\documentclass[
  aps,
  longbibliography,
  a4paper,
  amsfonts,
  amssymb,
  amsmath,
  reprint,
  showkeys,
  nofootinbib,
  twoside,
  superscriptaddress,
  footnotes
]{revtex4-2}

\DeclareUnicodeCharacter{2212}{\ensuremath{-}}
\usepackage[T1]{fontenc}
\usepackage{float}
\usepackage[version=4]{mhchem}
\usepackage{caption}
\usepackage{subcaption}
\usepackage{physics}
\usepackage{float}
\usepackage{mathtools}

\usepackage{adjustbox}
\usepackage{algorithm}
\usepackage{algpseudocode}
\usepackage{algcompatible}

\usepackage[unicode=true, colorlinks=true]{hyperref} % For hyperlinks in the PDF

\usepackage{tikz}
\usetikzlibrary{decorations,decorations.pathreplacing, shapes.geometric, arrows}
\tikzstyle{startstop} = [rectangle, rounded corners, minimum width=3cm, minimum height=1cm,text centered, draw=black, fill=red!30]
\tikzstyle{io} = [trapezium, trapezium left angle=70, trapezium right angle=110, minimum width=3cm, minimum height=1cm, text centered, draw=black, fill=blue!30]
\tikzstyle{process} = [rectangle, minimum width=3cm, minimum height=1cm, text centered, text width=3cm, draw=black, fill=orange!30]
\tikzstyle{decision} = [diamond, minimum width=3cm, minimum height=2cm, text centered, text width=2.5cm, draw=black, fill=green!30, aspect=1.75, inner sep=-0.5ex]
\tikzstyle{arrow} = [thick,->,>=stealth]
\tikzstyle{dotted-arrow} = [arrow, dotted]



\begin{document}
\title{Quantum-Accelerated Self-Consistent Field: A Hybrid Algorithm}

\author{Alexis Ralli*}
\affiliation{QMatter, Inc., 254 Chapman Rd, Office 109,  Suite 101-B, Newark, Delaware, 19702, USA}
\affiliation{Centre for Computational Science, Department of Chemistry, University College London, WC1H 0AJ, United Kingdom}
\email{alexis.ralli@qmatter.xyz}
\author{Tim Weaving}
\affiliation{QMatter, Inc., 254 Chapman Rd, Office 109,  Suite 101-B, Newark, Delaware, 19702, USA}
\affiliation{Centre for Computational Science, Department of Chemistry, University College London, WC1H 0AJ, United Kingdom}
%\email{timothy.weaving.20@ucl.ac.uk}
\author{Thomas M. Bickley}
\affiliation{Centre for Computational Science, Department of Chemistry, University College London, WC1H 0AJ, United Kingdom}
\author{Peter V. Coveney}
\affiliation{QMatter, Inc., 254 Chapman Rd, Office 109,  Suite 101-B, Newark, Delaware, 19702, USA}
\affiliation{Centre for Computational Science, Department of Chemistry, University College London, WC1H 0AJ, United Kingdom}
\affiliation{Advanced Research Computing Centre, University College London, WC1H 0AJ, United Kingdom}
% \email{p.v.coveney@ucl.ac.uk}
\author{Peter J. Love}
\affiliation{QMatter, Inc., 254 Chapman Rd, Office 109,  Suite 101-B, Newark, Delaware, 19702, USA}
\affiliation{Department of Physics and Astronomy, Tufts University, Medford, MA 02155, USA}
\affiliation{Computational Science Initiative, Brookhaven National Laboratory, Upton, NY 11973, USA}
\affiliation{Department of Computer Science, University of Toronto, 40 St. George Street, Toronto, ON M5S 3H6, Canada}
\affiliation{Department of Physics, University of Toronto, Toronto, Ontario M5S 1A7, Canada}
%\email{peter.love@tufts.edu}

\date{\today}

\begin{abstract}
\begin{center}
    \textbf{Abstract}
\end{center}
We present the Grover adaptive search self-consistent field (GAS-SCF) algorithm. GAS-SCF leverages quantum arithmetic to construct an efficient oracle that marks target states (Fock states) which improve upon some initial classical energy estimate. Amplitude amplification then increases the probability of measuring these states. This approach offers a theoretical quadratic speed-up for the optimization problem encountered in SCF quantum chemistry and establishes a baseline against which structured optimization algorithms, such as QAOA and DQI may be compared. In this work, we classically simulate three examples as proofs of concept of the algorithm, the largest consisting of $26$ qubits. We then extend our analysis to two larger systems, with \ce{O3} representing the largest case at $330$ qubits. These examples are chosen to probe classically challenging SCF regimes. Achieving chemically relevant applications of GAS-SCF will require large-scale, fault-tolerant quantum hardware.

%We conclude by examining the prospects of achieving quantum advantage by integration of this routine with classical SCF.
\end{abstract}

\maketitle

\section{Introduction} \label{sec:introduction}

Quantum chemistry is a widely studied application of quantum computing~\cite{aspuru2005simulated,cao2019quantum,santagati2024drug,chen2025framework}. Much effort has been devoted to studying the Full Configuration Interaction (FCI) problem on quantum computers~\cite{aspuru2005simulated,cao2019quantum}. In this case one is seeking exponential quantum speedups for problems out of reach of current on foreseeable classical computers due to the rapid increase in Hilbert space dimension of FCI with basis set size. However, making quantum algorithms for FCI realizable on reasonable sized quantum computers requires careful optimization and exploitation of problem structure~\cite{babbush2018low,babbush2018encoding,kivlichan2018quantum,babbush2019quantum,lee2021even,su2021fault,berry2025rapid,low2025fast}. This problem structure may be exploited by classical algorithms as well, leading to closer competition between classical and quantum approaches~\cite{lee2023evaluating, zhai2026classicalsolutionfemocofactormodel}. Furthermore, the importance of large-scale FCI calculations to chemistry is largely unknown, precisely because such calculations are classically intractable at present. Recently, these considerations have motivated the development of quantum algorithms that offer speedups over classical heuristics directly~\cite{chen2025framework,babbush2023quantum}. Such an approach is complementary to efforts to directly solve the FCI problem by quantum computation.

The self-consistent field (SCF) method is perhaps the oldest heuristic in quantum chemistry~\cite{Hartree_1928}. SCF methods, including Hartree–Fock (HF) and Kohn–Sham density functional theory (KS-DFT), provide approximate solutions to the time-independent Schr\"{o}dinger equation and are routinely used to define the ``canonical" molecular orbital basis in second-quantized formulations of the molecular Hamiltonian~\cite{szabo2012modern, helgaker2013molecular}. The difference between the HF energy and the FCI energy is the {\em correlation energy}. Recovery of the correlation energy is the goal of all post-Hartree-Fock methods. Post-Hartree-Fock methods are often defined in the basis of molecular orbitals obtained from HF.  Alternative molecular orbital (MO) bases, including localized and natural orbitals, are also commonly used. Improvements in the solutions to HF SCF problems can yield improved molecular orbitals, impacting all Post-Hartree-Fock methods.  

As shown in~\cite{whitfield2014np}, the Hartree-Fock method is NP-complete, and hence in the worst case one does not expect any classical or quantum polynomial time algorithm to provide a solution. This rather negative view is contradicted by the widespread success of HF in chemical problems, a fact which is partly explained by considering the approximability of the NP complete problem HF represents. In~\cite{ralli2025bridging} we showed that HF problems can be mapped to signed MaxCut problems, which admit performance guarantees in terms of the approximation ratio - the ratio of the approximate solution (the obtained HF energy from a particular SCF algorithm) to the true HF energy. Despite impressive classical advances - such as the HF calculation of $9,188$ water molecules reported by Barca \textit{et al.} (involving $119,444$ basis functions) \cite{barca2020scaling}, the optimality of such large-scale solutions remains unclear.

Hartree–Fock calculations have been performed on a quantum computer by Google AI Quantum and collaborators \cite{google2020hartree}. In that work the input (Fock) state was fixed and the orbital parameters were optimized with the variational quantum eigensolver \cite{peruzzo2014variational}. In contrast, the Hartree–Fock approach outlined in \cite{ralli2025bridging} (and here) differs in that a quantum algorithm is used to determine the reference Fock state. The orbital parameters are then updated using classical or quantum methods.

The HF formulation presented in \cite{ralli2025bridging} is related to the variational optimization of two-electron reduced density (2-RDM) matrices \cite{coleman2007reduced, mazziotti2011large, mazziotti2020dual}; however, in our approach \cite{ralli2025bridging}, both the 2-RDM and 1-RDM admit a significantly simpler structure, as they are derived from single Fock references. The specific optimization problem in our single reference SCF algorithm is composed of two steps: a continuous optimization over orbital parameters, and a discrete optimization to find the lowest energy Fock state for the new parameters. In order for the minimum state to change the eigenvalues (which are simply the diagonal entries in the second quantized molecular Hamiltonian) must cross. For sufficiently small changes in the orbital parameters, one can imagine that the states can be followed perturbatively from the initial orbitals without difficulty in identifying the updated state. Conversely, this suggests that tracking the ground state as the orbital parameters evolve is only reliable when those changes remain very small. However, the NP-completeness of the problem implies that such incremental variations are insufficient to guarantee finding the true minimum in the worst case~\cite{whitfield2014np}.
 
While quantum computers are not believed to be capable of solving NP complete problems exactly in polynomial time, there are many heuristic quantum approaches to combinatorial optimization problems~\cite{farhi2001quantum,farhi2014quantum,jordan2025optimization}. There is recent evidence that they may be able to obtain improvements in the approximation ratio obtained for some problems~\cite{jordan2025optimization,farhi2025lower}. Optimization of SCF approaches provides an interesting target for such algorithms, as the optimization problem itself arises in quantum mechanics and one may speculate optimistically that this will make it more amenable to quantum approaches. More specifically, classical performance guarantees on the approximation ratio for MaxCut with signed weights are weaker than in the unsigned case, making performance improvements easier to achieve. In the present paper we seek to exploit prior classical knowledge of approximate solutions and rigorous quantum speedups for unstructured problems to define a benchmark quantum algorithm, GAS-SCF, for improving SCF calculations. 

Grover’s search algorithm \cite{grover1996fast} and its generalization, amplitude amplification \cite{Brassard_2002}, are fundamental components underlying many quantum algorithms. Their use often provides quadratic speedups for many quantum algorithms over the best-known classical approaches. However, this advantage is asymptotic and constant prefactors can strongly influence practical performance. Prior work suggests that a modest fault-tolerant quantum computer is unlikely to realize a meaningful runtime benefit for quadratic speedups, as error-correction overheads can outweigh the gains for reasonable instance sizes \cite{babbush2021focus,hoefler2023disentangling}; similar conclusions are drawn elsewhere \cite{Campbell2019applyingquantum, sanders2020compilation}. Improvements in error correction can of course change these estimates~\cite{cain2026shor}. Grover speedups also remain an important goal for quantum algorithms: one would always wish to be able to obtain {\em at least} a Grover speedup. A Grover speedup sets a baseline against which more sophisticated quantum optimization techniques, such as the adiabatic algorithm~\cite{farhi2001quantum}, the quantum approximate optimization algorithm (QAOA)~\cite{farhi2014quantum}, or decoded quantum intreferometry (DQI)~\cite{jordan2025optimization} may be compared.

In this work, we present the implementation details of the recently proposed Grover Adaptive Search Self-Consistent Field (GAS-SCF) algorithm, which was previously described briefly by us in \cite{ralli2025bridging}. Here we define and analyze the algorithm in detail and provide an open source implementation \cite{gas_github} for the GAS \cite{durr1999quantumalgorithmfindingminimum, bulger2003implementing} subroutine to support other research into the algorithm. Theoretically, it has been proven that GAS provides a quadratic speedup over classical methods \cite{durr1999quantumalgorithmfindingminimum}, requiring $\mathcal{O}(\sqrt{N})$ operations compared to the classical $\mathcal{O}(N)$, where $N$ denotes the size of the search space. This quadratic speedup relies on an oracle. We show how to construct this oracle given a classical bound on the energy, which is always available for HF problems from variational classical approaches. Specifically, we construct an oracle that labels all Fock states that improve upon a classical reference energy. GAS-SCF is therefore a warm-started version of Grover search, in which all problem structure is assumed to have been fully utilized to obtain the classical bound, and therefore quantum unstructured search is used on the states that lie above this bound.  If another quantum optimization algorithm outperforms GAS-SCF it must do so by exploiting further problem structure not available to classical heuristics, and not used by GAS-SCF.

The remainder of the paper is organized as follows. Section \ref{sec:background} introduces the GAS-SCF algorithm and outlines its compilation into quantum circuits. Section \ref{sec:numerical} presents numerical examples, focusing in particular on \ce{H3-}, \ce{LiH}, \ce{OH-}, \ce{O2} and \ce{O3} molecular systems. Section \ref{sec:Q_ad} examines the potential of quantum advantage for the algorithm. Finally, the computational methodology for the numerical studies is given in Section \ref{sec:methodology}.

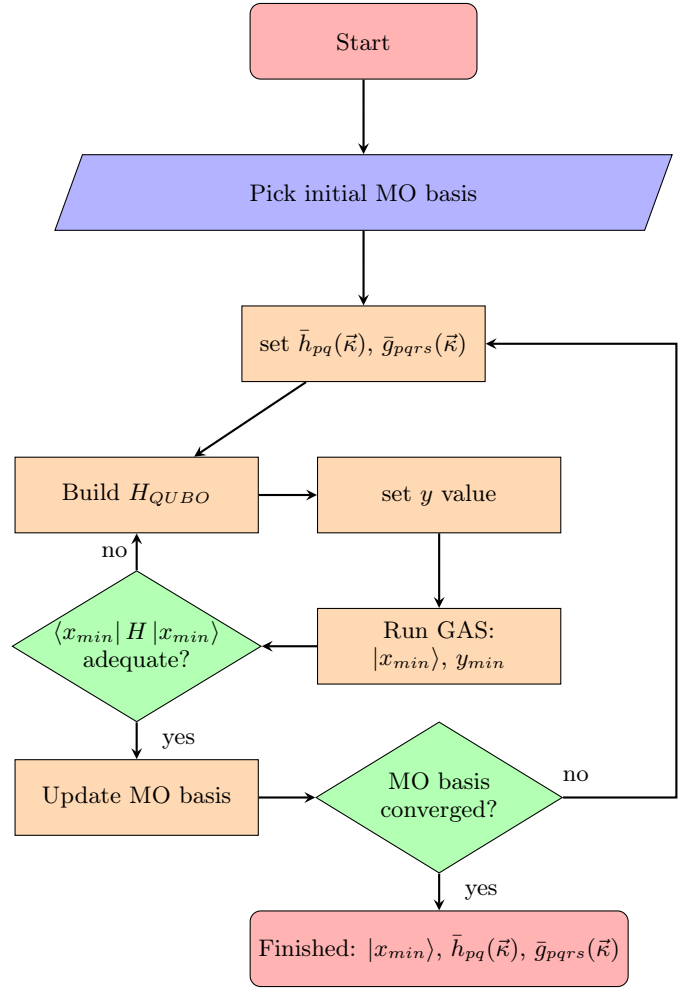
\begin{figure*}[ht]
    \begin{tikzpicture}[node distance=2cm]
    \node (start) [startstop] {Start};
    \node (in1) [io, below of=start, xshift=0.0cm] {Pick initial MO basis};
    \node (pro0) [process, below of=in1, xshift=0.0cm] {set $\bar{h}_{pq}(\vec{\kappa})$, $\bar{g}_{pqrs}(\vec{\kappa})$};
    \node (pro1) [process, below of=pro0, xshift=-3.0cm] {Build $H_{QUBO}$};
    \node (pro2) [process, right of=pro1, xshift=2.0cm] {set $y$ value};
    \node (pro3) [process, below of=pro2, xshift=0.0cm] {Run GAS: $\ket{x_{min}}$, $y_{min}$};    
    \node (dec1) [decision, left of=pro3, xshift=-2cm] {$\bra{x_{min}} H \ket{x_{min}}$ adequate?};
    \node (pro5) [process, below of=dec1, xshift=0.0cm] {Update MO basis};
    \node (dec2) [decision, right of=pro5, xshift=2cm] {MO basis converged?};
    \node (stop) [startstop, below of=dec2, xshift=0cm] {Finished: $\ket{x_{min}}$, $\bar{h}_{pq}(\vec{\kappa})$, $\bar{g}_{pqrs}(\vec{\kappa})$};

    \draw [arrow] (start) -- (in1);
    \draw [arrow] (in1) -- (pro0);
    \draw [arrow] (pro0) -- (pro1);
    \draw [arrow] (pro1) -- (pro2);
    % \draw [arrow, dashed] (pro2) -- (pro3);
    \draw [arrow] (pro2) -- (pro3);
    \draw [arrow] (pro3) -- (dec1);
    
    \draw [arrow] (dec1.north) -- node[anchor=east] {no} (pro1);
    \draw [arrow] (dec1.south) -- node[anchor=west] {{ }  yes} (pro5.north);

    \draw [arrow] (pro5) -- (dec2);
    \draw [arrow] (dec2.south) -- node[anchor=west] {{ }  yes} (stop.north);
    \draw[arrow] (dec2.east) -- +(1.5,0) |- (pro0);
    \draw [draw, xshift=-6.5cm, yshift=-9.5cm, xshift=13.5cm] -- node {no} (pro5);
    \end{tikzpicture}
\caption{Full outline of GAS-SCF algorithm (equation \ref{eq:QUBO-SCF}). Refer to Algorithms 1 and 2 in \cite{ralli2025bridging} for the procedure to update the MO basis. The initial MO basis may be set to any standard choice, such as L\"{o}wdin or canonical or similar. Here, ``adequate" refers to an energy that has converged, outperforms classical SCF solutions, or satisfies a predefined (possibly spectral-based) threshold for advancing the algorithm to the MO basis optimization stage.}
    \label{fig:GASSCF_outline}
\end{figure*}

\section{The GAS-SCF Algorithm}\label{sec:background}

In this section we introduce the GAS-SCF algorithm, first defining the SCF problem to be solved in section~\ref{sec:SCF}, then giving the GAS algorithm in section~\ref{sec:GAS}, section \ref{sec:non} then includes how to deal with SCF integrals by approximating them with integer values. 

\subsection{The SCF Problem}\label{sec:SCF}

The Quadratic Unconstrained Binary Optimization (QUBO) problem seeks the minimum of a function $f(\vec x)$ of the following form:
\begin{equation} \label{eq:QUBO}
f(\vec{x}) = \vec{x}^{T} Q \vec{x} + \vec{b}^{T}\vec{x} + c
= \sum_{i,j} Q_{ij}, x_{i} x_{j} + \sum_{i} b_{i}, x_{i} + c.
\end{equation}
Where $x_{i} \in \{0,1\}$ for $i = 1, \dots, n$ and $f(\vec x)\in \mathbb{R}$, so that $f(\vec x)$ is a pseudo-boolean function. The matrix $Q \in \mathbb{R}^{n \times n}$ encodes the quadratic coefficients, $\vec{b} \in \mathbb{R}^n$ represents the linear terms, and $c \in \mathbb{R}$ is a constant offset. Since $x_{i}^2 = x_{i}$, linear terms can be absorbed into the diagonal entries of $Q$, allowing the QUBO formulation to be written using only the $Q$ matrix. 

The QUBO problem arising from Hartree-Fock SCF was given in \cite{ralli2025bridging}: 
\begin{equation} \label{eq:qubo_full}
    \begin{aligned}
H_{QUBO}(\vec{\kappa}) 
        = &\sum_{w}^{M} \bar{h}_{ww}(\vec{\kappa})x_{w} + \\
        &\frac{1}{2} \sum_{\substack{m,w \\ m \neq w}}^{M} \bigg( \big[ \bar{g}_{mmww}(\vec{\kappa}) - \bar{g}_{mwwm}(\vec{\kappa})  \big] x_{m}x_{w} \bigg).
    \end{aligned}
\end{equation}
This problem is defined in the occupation number basis, so that the boolean variables represent fermionic occupancies of molecular orbitals. The coefficients $h_{pq}$ and $g_{pqrs}$ are one- and two-electron integrals, expressed in an orthonormal molecular orbital (MO) basis. Different choices of MO basis may be related via $H \mapsto H(\vec{\kappa}) = U^{\dagger} (\vec{\kappa})H U(\vec{\kappa})$, so the choice of MO basis is parametrized by $\vec{\kappa}$~\cite{ralli2025bridging}. %The mapping from translationally invariant problems to Ising spin glass QUBO problems was previously obtained in~\cite{whitfield2014np}. 

The SCF problem can be written as \cite{ralli2025bridging}:
\begin{equation} \label{eq:QUBO-SCF}
    \begin{aligned}
    E_{SCF} &= \underset{\vec{\kappa}}{\mathrm{min}}  \bigg[ \underset{\ket{x} \in \mathcal{V} \subset \mathcal{D}}{\mathrm{min}} \big[ \bra{x} H_{QUBO}(\vec{\kappa}) \ket{x} \big] \bigg].
    \end{aligned}
\end{equation}
Here, $\mathcal{D}$ denotes the set of all computational basis states or $n$-bit bitstrings. The size of this set is $2^{n}$, while $\mathcal{V}$ represents a subset consisting of  computational states that lie within the correct symmetry sector. For example states with the desired number of alpha/beta electrons, total spin angular momentum squared $S^{2}$, and molecular point group symmetries. These symmetries reduce the search space - e.g. particle number symmetry reduces the scaling from $2^{n}$ to ${n/2\choose n_{\alpha}} \cdot {n/2\choose n_{\beta}}$ for $n$ spin MOs and $(n_{\alpha}, n_{\beta})$ electrons. The dimension of $\vec{\kappa}$ is bounded by $|\vec{\kappa}| \leq \frac{n(n-1)}{2}$, reflecting the invariance of the SCF wavefunction under certain orbital rotations. For example, in HF closed-shell systems, only rotations between occupied and virtual orbitals contribute. This means that classical heuristics can both exploit the continuous/discrete nature of the problem and do not have to brute force search over all $2^{n}$ bitstrings nor redundant orbital rotations. 

To summarize the SCF problem defined in Equation~\ref{eq:QUBO-SCF}, first the inner optimization aims to find the best single bitstring that minimizes $H_{QUBO}(\vec{\kappa})$ for a fixed $\vec{\kappa}$ (the current MO basis). The outer optimization then updates the MO basis ($\vec{\kappa}$) with reference to the state obtained in the inner optimization. This is repeated until the energy converges. In this work, we give a quantum algorithm for the inner, discrete optimization, while the outer, continuous, optimization is handled by a classical routine. For the remainder of this paper, we focus on the discrete optimization step, as the classical routine has already been discussed in \cite{ralli2025bridging} where two approaches are given. Figure \ref{fig:GASSCF_outline} illustrates a high-level overview of the complete algorithm.

Classical solvers for the SCF optimization problem employ heuristic strategies rather than performing an exhaustive search of the solution space. The most widely used approach iteratively diagonalizes the Fock matrix until self-consistency is achieved, while the second most common class of methods uses second-order optimization techniques based on orbital rotations to directly minimize the SCF energy. In both cases, the reference determinant is held fixed. However, particularly in the latter approach, Thouless' theorem implies that any other single-determinant state can be generated from a given reference determinant through an orbital rotation, yielding a  (non-orthogonal) Fock state \cite[eq 2]{THOULESS1960225}. Consequently, this parameterization provides access to the complete single-determinant optimization space. This means that the optimal solution to the SCF problem can always be found by omitting the discrete optimization step in equation~\ref{eq:QUBO-SCF}. The discrete optimization step finds the minimum state  of SCF Hamiltonians along the continuous optimization path, which in general are distinct from the rotation of the initial minimum. Subsequent rotations proceed from this new starting point. The optimization of equation~\ref{eq:QUBO-SCF} can therefore be regarded as restarting the continuous optimization periodically, after discrete minimization of the intermediate SCF problem.  Although the theoretical optimum is the same for both GAS-SCF and traditional approaches, the use of different classical SCF heuristics complicates direct comparisons of runtime, convergence behavior, and solution quality. Nevertheless, our earlier work demonstrated improved convergence behavior \cite{ralli2025bridging}, a result further supported by numerical findings in this work for \ce{O2} and \ce{O3}, which indicate that improved solutions can be obtained relative to classical heuristics.

In the next subsection, we describe how $H_{QUBO}(\vec{\kappa})$ can be solved using GAS.

%Although the theoretical optimal solution is the same between GAS-SCF and traditional approaches, determining whether it provides a speedup over conventional implementations is challenging due to slightly different optimization heuristics being performed. This can result in a shorter path to the same or improved solution. Our earlier work demonstrated such shortcuts \cite{ralli2025bridging}. In the next subsection, we describe how $H_{QUBO}(\vec{\kappa})$ can be solved using GAS.

\subsection{Grover Adaptive Search}\label{sec:GAS}
\begin{figure*}
     \centering
     \begin{adjustbox}{width=1\textwidth}
\tikzset{every picture/.style={line width=0.75pt}} %set default line width to 0.75pt        

%! \usetikzlibrary{decorations.pathreplacing,decorations.pathmorphing}
\definecolor{mygreen}{RGB}{34,139,33}
\definecolor{myblue}{RGB}{157,220,229}
\definecolor{myred}{RGB}{255,99,98}
\begin{tikzpicture}[scale=1.500000,x=1pt,y=1pt]
\filldraw[color=white] (0.000000, -11.000000) rectangle (571.000000, 165.000000);
% Drawing wires
% Line 17: q0 W \ket{s}_{QUBO}
\draw[color=black] (0.000000,154.000000) -- (531.000000,154.000000);
\draw[color=black] (531.000000,153.500000) -- (557.500000,153.500000);
\draw[color=black] (531.000000,154.500000) -- (557.500000,154.500000);
\draw[color=black] (0.000000,154.000000) node[left] {$\ket{s}_{QUBO}$};
% Line 18: q1 W \ket{0}_{f(\vec{x})}
\draw[color=black] (0.000000,132.000000) -- (557.500000,132.000000);
\draw[color=black] (0.000000,132.000000) node[left] {$\ket{0}_{f(\vec{x})}$};
% Line 19: q2 W \ket{0}_{sign}
\draw[color=black] (0.000000,110.000000) -- (557.500000,110.000000);
\draw[color=black] (0.000000,110.000000) node[left] {$\ket{0}_{sign}$};
% Line 20: q3 W \ket{0}_{N_{\alpha}(x)}^{\otimes \mu}
\draw[color=black] (0.000000,88.000000) -- (557.500000,88.000000);
\draw[color=black] (0.000000,88.000000) node[left] {$\ket{0}_{N_{\alpha}(x)}^{\otimes \mu}$};
% Line 21: q4 W \ket{0}_{N_{\beta}(x)}^{\otimes \nu}
\draw[color=black] (0.000000,66.000000) -- (557.500000,66.000000);
\draw[color=black] (0.000000,66.000000) node[left] {$\ket{0}_{N_{\beta}(x)}^{\otimes \nu}$};
% Line 22: q5 W \ket{0}_{\alpha \text{-flag}}
\draw[color=black] (0.000000,44.000000) -- (557.500000,44.000000);
\draw[color=black] (0.000000,44.000000) node[left] {$\ket{0}_{\alpha \text{-flag}}$};
% Line 23: q6 W \ket{0}_{\beta \text{-flag}}
\draw[color=black] (0.000000,22.000000) -- (557.500000,22.000000);
\draw[color=black] (0.000000,22.000000) node[left] {$\ket{0}_{\beta \text{-flag}}$};
% Line 24: q7 W \ket{0}_{number \text{-flag}}
\draw[color=black] (0.000000,0.000000) -- (557.500000,0.000000);
\draw[color=black] (0.000000,0.000000) node[left] {$\ket{0}_{number \text{-flag}}$};
% Done with wires; drawing gates
% Line 26: q0 / n
\draw (8.833333, 147.000000) -- (18.166667, 161.000000);
\draw (15.833333, 157.500000) node[right] {$\scriptstyle{n}$};
% Line 27: q1 / m
\draw (8.833333, 125.000000) -- (18.166667, 139.000000);
\draw (15.833333, 135.500000) node[right] {$\scriptstyle{m}$};
% Line 28: q2 LABEL
% Line 29: q3 / \mu
\draw (8.833333, 81.000000) -- (18.166667, 95.000000);
\draw (15.833333, 91.500000) node[right] {$\scriptstyle{\mu}$};
% Line 30: q4 / \nu
\draw (8.833333, 59.000000) -- (18.166667, 73.000000);
\draw (15.833333, 69.500000) node[right] {$\scriptstyle{\nu}$};
% Line 48: q5 LABEL
% Line 49: q6 LABEL
% Line 50: q7 LABEL
% Line 35: q0 LABEL
% Line 37: q1 G:state width=23 $\hat H^{\otimes m}$
\begin{scope}[rounded corners=3pt]
\begin{scope}
\draw[fill=myblue] (44.500000, 132.000000) +(-45.000000:16.263456pt and 9.899495pt) -- +(45.000000:16.263456pt and 9.899495pt) -- +(135.000000:16.263456pt and 9.899495pt) -- +(225.000000:16.263456pt and 9.899495pt) -- cycle;
\clip (44.500000, 132.000000) +(-45.000000:16.263456pt and 9.899495pt) -- +(45.000000:16.263456pt and 9.899495pt) -- +(135.000000:16.263456pt and 9.899495pt) -- +(225.000000:16.263456pt and 9.899495pt) -- cycle;
\draw (44.500000, 132.000000) node {$\hat H^{\otimes m}$};
\end{scope}
\end{scope}
% Line 38: q2 G:state width=23 $\hat H$
\begin{scope}[rounded corners=3pt]
\begin{scope}
\draw[fill=myblue] (44.500000, 110.000000) +(-45.000000:16.263456pt and 9.899495pt) -- +(45.000000:16.263456pt and 9.899495pt) -- +(135.000000:16.263456pt and 9.899495pt) -- +(225.000000:16.263456pt and 9.899495pt) -- cycle;
\clip (44.500000, 110.000000) +(-45.000000:16.263456pt and 9.899495pt) -- +(45.000000:16.263456pt and 9.899495pt) -- +(135.000000:16.263456pt and 9.899495pt) -- +(225.000000:16.263456pt and 9.899495pt) -- cycle;
\draw (44.500000, 110.000000) node {$\hat H$};
\end{scope}
\end{scope}
% Line 41: q3 LABEL
% Line 42: q4 LABEL
% Line 39: q1 q2 G:state width=28 $\hat{A}(-y)$
\draw[rounded corners=3pt] (82.000000,132.000000) -- (82.000000,110.000000);
\begin{scope}[rounded corners=3pt]
\begin{scope}
\draw[fill=myblue] (82.000000, 121.000000) +(-45.000000:19.798990pt and 25.455844pt) -- +(45.000000:19.798990pt and 25.455844pt) -- +(135.000000:19.798990pt and 25.455844pt) -- +(225.000000:19.798990pt and 25.455844pt) -- cycle;
\clip (82.000000, 121.000000) +(-45.000000:19.798990pt and 25.455844pt) -- +(45.000000:19.798990pt and 25.455844pt) -- +(135.000000:19.798990pt and 25.455844pt) -- +(225.000000:19.798990pt and 25.455844pt) -- cycle;
\draw (82.000000, 121.000000) node {$\hat{A}(-y)$};
\end{scope}
\end{scope}
% Line 43: q3 LABEL
% Line 44: q4 LABEL
% Line 45: q3 LABEL
% Line 46: q4 LABEL
% Line 63: q1 q2 G:state width=20 $f(\vec{x})$ q0:shape=4
\draw[rounded corners=3pt] (118.000000,154.000000) -- (118.000000,110.000000);
\begin{scope}[rounded corners=3pt]
\begin{scope}
\draw[fill=myblue] (118.000000, 121.000000) +(-45.000000:14.142136pt and 25.455844pt) -- +(45.000000:14.142136pt and 25.455844pt) -- +(135.000000:14.142136pt and 25.455844pt) -- +(225.000000:14.142136pt and 25.455844pt) -- cycle;
\clip (118.000000, 121.000000) +(-45.000000:14.142136pt and 25.455844pt) -- +(45.000000:14.142136pt and 25.455844pt) -- +(135.000000:14.142136pt and 25.455844pt) -- +(225.000000:14.142136pt and 25.455844pt) -- cycle;
\draw (118.000000, 121.000000) node {$f(\vec{x})$};
\end{scope}
\end{scope}
\begin{scope}
\draw[fill=white] (118.000000, 154.000000) +(-45.000000:3.000000pt) -- +(45.000000:3.000000pt) -- +(135.000000:3.000000pt) -- +(225.000000:3.000000pt) -- cycle;
\clip (118.000000, 154.000000) +(-45.000000:3.000000pt) -- +(45.000000:3.000000pt) -- +(135.000000:3.000000pt) -- +(225.000000:3.000000pt) -- cycle;
\draw (115.000000, 154.000000) -- (121.000000, 154.000000);
\draw (118.000000, 151.000000) -- (118.000000, 157.000000);
\end{scope}
% Line 53: q3 G:state width=23 $\hat H^{\otimes \mu}$
\begin{scope}[rounded corners=3pt]
\begin{scope}
\draw[fill=myblue] (151.500000, 88.000000) +(-45.000000:16.263456pt and 9.899495pt) -- +(45.000000:16.263456pt and 9.899495pt) -- +(135.000000:16.263456pt and 9.899495pt) -- +(225.000000:16.263456pt and 9.899495pt) -- cycle;
\clip (151.500000, 88.000000) +(-45.000000:16.263456pt and 9.899495pt) -- +(45.000000:16.263456pt and 9.899495pt) -- +(135.000000:16.263456pt and 9.899495pt) -- +(225.000000:16.263456pt and 9.899495pt) -- cycle;
\draw (151.500000, 88.000000) node {$\hat H^{\otimes \mu}$};
\end{scope}
\end{scope}
% Line 54: q4 G:state width=23 $\hat H^{\otimes \nu}$
\begin{scope}[rounded corners=3pt]
\begin{scope}
\draw[fill=myblue] (151.500000, 66.000000) +(-45.000000:16.263456pt and 9.899495pt) -- +(45.000000:16.263456pt and 9.899495pt) -- +(135.000000:16.263456pt and 9.899495pt) -- +(225.000000:16.263456pt and 9.899495pt) -- cycle;
\clip (151.500000, 66.000000) +(-45.000000:16.263456pt and 9.899495pt) -- +(45.000000:16.263456pt and 9.899495pt) -- +(135.000000:16.263456pt and 9.899495pt) -- +(225.000000:16.263456pt and 9.899495pt) -- cycle;
\draw (151.500000, 66.000000) node {$\hat H^{\otimes \nu}$};
\end{scope}
\end{scope}
% Line 69: q1 LABEL
% Line 70: q2 LABEL
% Line 65: q3 G:state width=20 $\hat{N}_{\alpha}(x)$ q0:shape=4
\draw[rounded corners=3pt] (185.000000,154.000000) -- (185.000000,88.000000);
\begin{scope}[rounded corners=3pt]
\begin{scope}
\draw[fill=myblue] (185.000000, 88.000000) +(-45.000000:14.142136pt and 9.899495pt) -- +(45.000000:14.142136pt and 9.899495pt) -- +(135.000000:14.142136pt and 9.899495pt) -- +(225.000000:14.142136pt and 9.899495pt) -- cycle;
\clip (185.000000, 88.000000) +(-45.000000:14.142136pt and 9.899495pt) -- +(45.000000:14.142136pt and 9.899495pt) -- +(135.000000:14.142136pt and 9.899495pt) -- +(225.000000:14.142136pt and 9.899495pt) -- cycle;
\draw (185.000000, 88.000000) node {$\hat{N}_{\alpha}(x)$};
\end{scope}
\end{scope}
\begin{scope}
\draw[fill=white] (185.000000, 154.000000) +(-45.000000:3.000000pt) -- +(45.000000:3.000000pt) -- +(135.000000:3.000000pt) -- +(225.000000:3.000000pt) -- cycle;
\clip (185.000000, 154.000000) +(-45.000000:3.000000pt) -- +(45.000000:3.000000pt) -- +(135.000000:3.000000pt) -- +(225.000000:3.000000pt) -- cycle;
\draw (182.000000, 154.000000) -- (188.000000, 154.000000);
\draw (185.000000, 151.000000) -- (185.000000, 157.000000);
\end{scope}
% Line 71: q1 LABEL
% Line 72: q2 LABEL
% Line 66: q4 G:state width=20 $\hat{N}_{\beta}(x)$  q0:shape=4
\draw[rounded corners=3pt] (232.000000,154.000000) -- (232.000000,66.000000);
\begin{scope}[rounded corners=3pt]
\begin{scope}
\draw[fill=myblue] (232.000000, 66.000000) +(-45.000000:14.142136pt and 9.899495pt) -- +(45.000000:14.142136pt and 9.899495pt) -- +(135.000000:14.142136pt and 9.899495pt) -- +(225.000000:14.142136pt and 9.899495pt) -- cycle;
\clip (232.000000, 66.000000) +(-45.000000:14.142136pt and 9.899495pt) -- +(45.000000:14.142136pt and 9.899495pt) -- +(135.000000:14.142136pt and 9.899495pt) -- +(225.000000:14.142136pt and 9.899495pt) -- cycle;
\draw (232.000000, 66.000000) node {$\hat{N}_{\beta}(x)$};
\end{scope}
\end{scope}
\begin{scope}
\draw[fill=white] (232.000000, 154.000000) +(-45.000000:3.000000pt) -- +(45.000000:3.000000pt) -- +(135.000000:3.000000pt) -- +(225.000000:3.000000pt) -- cycle;
\clip (232.000000, 154.000000) +(-45.000000:3.000000pt) -- +(45.000000:3.000000pt) -- +(135.000000:3.000000pt) -- +(225.000000:3.000000pt) -- cycle;
\draw (229.000000, 154.000000) -- (235.000000, 154.000000);
\draw (232.000000, 151.000000) -- (232.000000, 157.000000);
\end{scope}
% Line 73: q3 LABEL
% Line 82: q1 LABEL
% Line 83: q2 LABEL
% Line 84: q1 q2 G:env width=25 $QFT^{\dagger}$
\draw[rounded corners=3pt] (281.500000,132.000000) -- (281.500000,110.000000);
\begin{scope}[rounded corners=3pt]
\begin{scope}
\draw[fill=myred] (281.500000, 121.000000) +(-45.000000:17.677670pt and 25.455844pt) -- +(45.000000:17.677670pt and 25.455844pt) -- +(135.000000:17.677670pt and 25.455844pt) -- +(225.000000:17.677670pt and 25.455844pt) -- cycle;
\clip (281.500000, 121.000000) +(-45.000000:17.677670pt and 25.455844pt) -- +(45.000000:17.677670pt and 25.455844pt) -- +(135.000000:17.677670pt and 25.455844pt) -- +(225.000000:17.677670pt and 25.455844pt) -- cycle;
\draw (281.500000, 121.000000) node {$QFT^{\dagger}$};
\end{scope}
\end{scope}
% Line 85: q3 G:env width=25 $QFT^{\dagger}$
\begin{scope}[rounded corners=3pt]
\begin{scope}
\draw[fill=myred] (281.500000, 88.000000) +(-45.000000:17.677670pt and 9.899495pt) -- +(45.000000:17.677670pt and 9.899495pt) -- +(135.000000:17.677670pt and 9.899495pt) -- +(225.000000:17.677670pt and 9.899495pt) -- cycle;
\clip (281.500000, 88.000000) +(-45.000000:17.677670pt and 9.899495pt) -- +(45.000000:17.677670pt and 9.899495pt) -- +(135.000000:17.677670pt and 9.899495pt) -- +(225.000000:17.677670pt and 9.899495pt) -- cycle;
\draw (281.500000, 88.000000) node {$QFT^{\dagger}$};
\end{scope}
\end{scope}
% Line 86: q4 G:env width=25 $QFT^{\dagger}$
\begin{scope}[rounded corners=3pt]
\begin{scope}
\draw[fill=myred] (281.500000, 66.000000) +(-45.000000:17.677670pt and 9.899495pt) -- +(45.000000:17.677670pt and 9.899495pt) -- +(135.000000:17.677670pt and 9.899495pt) -- +(225.000000:17.677670pt and 9.899495pt) -- cycle;
\clip (281.500000, 66.000000) +(-45.000000:17.677670pt and 9.899495pt) -- +(45.000000:17.677670pt and 9.899495pt) -- +(135.000000:17.677670pt and 9.899495pt) -- +(225.000000:17.677670pt and 9.899495pt) -- cycle;
\draw (281.500000, 66.000000) node {$QFT^{\dagger}$};
\end{scope}
\end{scope}
% Line 89: q3 P:width=28 $| n_{\alpha} \rangle \langle n_{\alpha} |$ +q5
\draw (320.000000,88.000000) -- (320.000000,44.000000);
\begin{scope}
\draw[fill=white] (320.000000, 88.000000) circle(14.000000pt);
\clip (320.000000, 88.000000) circle(14.000000pt);
\draw (320.000000, 88.000000) node {$| n_{\alpha} \rangle \langle n_{\alpha} |$};
\end{scope}
\begin{scope}
\draw[fill=white] (320.000000, 44.000000) circle(3.000000pt);
\clip (320.000000, 44.000000) circle(3.000000pt);
\draw (317.000000, 44.000000) -- (323.000000, 44.000000);
\draw (320.000000, 41.000000) -- (320.000000, 47.000000);
\end{scope}
% Line 90: q4 P:width=28 $| n_{\beta} \rangle \langle n_{\beta} |$ +q6
\draw (348.000000,66.000000) -- (348.000000,22.000000);
\begin{scope}
\draw[fill=white] (348.000000, 66.000000) circle(14.000000pt);
\clip (348.000000, 66.000000) circle(14.000000pt);
\draw (348.000000, 66.000000) node {$| n_{\beta} \rangle \langle n_{\beta} |$};
\end{scope}
\begin{scope}
\draw[fill=white] (348.000000, 22.000000) circle(3.000000pt);
\clip (348.000000, 22.000000) circle(3.000000pt);
\draw (345.000000, 22.000000) -- (351.000000, 22.000000);
\draw (348.000000, 19.000000) -- (348.000000, 25.000000);
\end{scope}
% Line 93: q7 T q5 q6
\draw (377.000000,44.000000) -- (377.000000,0.000000);
\begin{scope}
\draw[fill=white] (377.000000, 0.000000) circle(3.000000pt);
\clip (377.000000, 0.000000) circle(3.000000pt);
\draw (374.000000, 0.000000) -- (380.000000, 0.000000);
\draw (377.000000, -3.000000) -- (377.000000, 3.000000);
\end{scope}
\filldraw (377.000000, 44.000000) circle(1.500000pt);
\filldraw (377.000000, 22.000000) circle(1.500000pt);
% Line 96: q0 G:state $-\hat I$ q2 q7
\draw[rounded corners=3pt] (399.000000,154.000000) -- (399.000000,0.000000);
\begin{scope}[rounded corners=3pt]
\begin{scope}
\draw[fill=myblue] (399.000000, 154.000000) +(-45.000000:9.899495pt and 9.899495pt) -- +(45.000000:9.899495pt and 9.899495pt) -- +(135.000000:9.899495pt and 9.899495pt) -- +(225.000000:9.899495pt and 9.899495pt) -- cycle;
\clip (399.000000, 154.000000) +(-45.000000:9.899495pt and 9.899495pt) -- +(45.000000:9.899495pt and 9.899495pt) -- +(135.000000:9.899495pt and 9.899495pt) -- +(225.000000:9.899495pt and 9.899495pt) -- cycle;
\draw (399.000000, 154.000000) node {$-\hat I$};
\end{scope}
\end{scope}
\filldraw (399.000000, 110.000000) circle(1.500000pt);
\filldraw (399.000000, 0.000000) circle(1.500000pt);
% Line 99: q0 q1 q2 q3 q4 q5 q6 q7 G:state width=20 $\hat W^{\dagger}$
\draw[rounded corners=3pt] (428.000000,154.000000) -- (428.000000,0.000000);
\begin{scope}[rounded corners=3pt]
\begin{scope}
\draw[fill=myblue] (428.000000, 77.000000) +(-45.000000:14.142136pt and 118.793939pt) -- +(45.000000:14.142136pt and 118.793939pt) -- +(135.000000:14.142136pt and 118.793939pt) -- +(225.000000:14.142136pt and 118.793939pt) -- cycle;
\clip (428.000000, 77.000000) +(-45.000000:14.142136pt and 118.793939pt) -- +(45.000000:14.142136pt and 118.793939pt) -- +(135.000000:14.142136pt and 118.793939pt) -- +(225.000000:14.142136pt and 118.793939pt) -- cycle;
\draw (428.000000, 77.000000) node {$\hat W^{\dagger}$};
\end{scope}
\end{scope}
% Line 102: q0 G:env width=35 $2 | s \rangle \langle s | - I $
\begin{scope}[rounded corners=3pt]
\begin{scope}
\draw[fill=myred] (467.500000, 154.000000) +(-45.000000:24.748737pt and 9.899495pt) -- +(45.000000:24.748737pt and 9.899495pt) -- +(135.000000:24.748737pt and 9.899495pt) -- +(225.000000:24.748737pt and 9.899495pt) -- cycle;
\clip (467.500000, 154.000000) +(-45.000000:24.748737pt and 9.899495pt) -- +(45.000000:24.748737pt and 9.899495pt) -- +(135.000000:24.748737pt and 9.899495pt) -- +(225.000000:24.748737pt and 9.899495pt) -- cycle;
\draw (467.500000, 154.000000) node {$2 | s \rangle \langle s | - I $};
\end{scope}
\end{scope}
% Line 108: q1 LABEL
% Line 109: q2 LABEL
% Line 110: q3 LABEL
% Line 111: q4 LABEL
% Line 112: q5 LABEL
% Line 113: q6 LABEL
% Line 114: q7 LABEL
% Line 107: q0 LABEL
% Line 116: q0 M
\draw[fill=white] (524.000000, 147.000000) rectangle (538.000000, 161.000000);
\draw[very thin] (531.000000, 154.700000) arc (90:150:7.000000pt);
\draw[very thin] (531.000000, 154.700000) arc (90:30:7.000000pt);
\draw[->,>=stealth] (531.000000, 147.700000) -- +(80:12.124356pt);
% Line 118: q0 END
% Line 119: q1 END
% Line 120: q2 END
% Line 121: q3 END
% Line 122: q4 END
% Line 123: q5 END
% Line 124: q6 END
% Line 125: q7 END
% Done with gates; drawing ending labels
% Done with ending labels; drawing cut lines and comments
% Line 127: @ 1 9 %% $\hat W$
\draw[decorate,decoration={brace,mirror,amplitude = 4.666667pt},very thick] (30.000000,-11.000000) -- (383.000000,-11.000000);
\draw (206.500000, -15.666667) node[text width=144pt,below,text centered] {$\hat W$};
% Line 130: @ 1 13 % repeat $L$ times
\draw[decorate,decoration={brace,amplitude = 4.666667pt},very thick] (30.000000,165.000000) -- (515.000000,165.000000);
\draw (272.500000, 169.666667) node[text width=144pt,above,text centered] {repeat $L$ times};
% Done with comments
\end{tikzpicture}

\end{adjustbox}
        \caption{Outline of the full quantum circuit to implement the Grover Adaptive Search Self-Consistent Field (GAS-SCF) algorithm. Here the search space is over the uniform superposition of computational states: $\ket{s} = H^{\otimes n}\ket{0}^{\otimes n}$. Figure~\ref{fig:GAS-Dicke} provides an alternate construction, where the state on the first $n$-qubit register is different. The $\hat{A}$ and $f(\vec{x})$ gates are defined in Figure~\ref{fig:adder} and Figure~\ref{fig:QUBO-adder} in the supporting information. The former encodes the negative reference value ($-y$) setting a threshold, while the latter encodes the cost function for different binary states in the QUBO register state, as observed by the controls. The gates defined with clear circular controls, with projectors inside the controls, represent multi-controlled $X$ gates whose control settings correspond to the target occupation numbers expressed in binary, see Figure~\ref{fig:QUBO-proj} for further detail.  An alternative compilation is provided in the Supplemental Information \ref{sec:FT_comp}, which gives the corresponding $T$-gate cost. In this realization no $QFT^{\dagger}$ is required.} 
        \label{fig:GAS-fig}
\end{figure*}
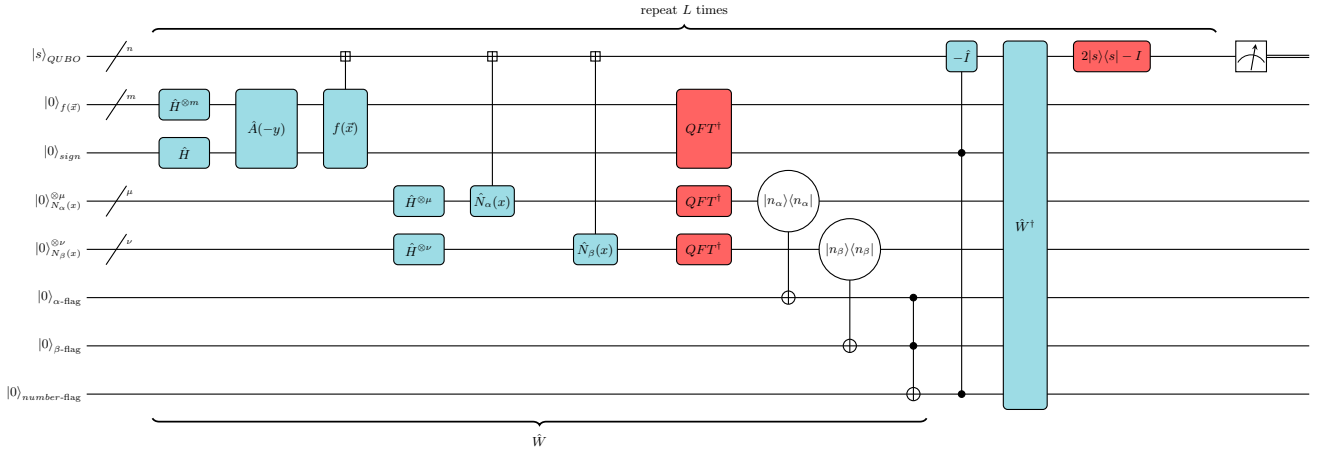

Gilliam, Woerner, and Gonciulea \cite{gilliam2021grover} demonstrated how Grover Adaptive Search (GAS) \cite{durr1999quantumalgorithmfindingminimum, bulger2003implementing, baritompa2005grover} can be applied to binary optimization problems. We build on their framework by incorporating additional constraints arising from spin and particle number symmetries, and restrict the search to solutions of particular Hamming weights . Figure~\ref{fig:GAS-fig} illustrates the quantum circuit used in the Grover Adaptive Search routine of GAS-SCF, which we summarize below.

The first register contains the $n$ qubits on which the QUBO problem is defined. Each term in the QUBO can be represented using a single- or doubly-controlled phase gate, with the control qubits determined by the structure of each QUBO term. These gates are applied to a secondary register of $(m+1)$ qubits, which encodes the value of the QUBO cost function in the Fourier basis \cite{draper2000additionquantumcomputer}.  This operation is represented by the $f(\vec{x})$ gate in this work. Higher-order polynomial terms, such as cubic or quartic interactions, can be represented by increasing the number of control qubits, but that is not relevant for the QUBO setting of the present work.

The minimum number of qubits $m$ required to store the maximum value of the QUBO cost function is given by:
\begin{equation} \label{eq:m_reg}
m = \left\lceil \log_{2}\left( \max[f(\vec{x})] \right) \right\rceil \leq \left\lceil \log_{2}\left( \sum_{ij}|Q_{ij}| + \sum_{i}|b_{i}| \right) \right\rceil .
\end{equation}
A single additional qubit is included to allow the representation of negative values, enabling the encoding of integers in the range $[-2^m, 2^m)$. The $(m+1)$-qubit register stores the cost function in the Fourier basis, which is then transformed into the computational basis using the inverse Quantum Fourier Transform ($QFT^{\dagger}$) \cite{draper2000additionquantumcomputer, gilliam2021grover}. The single qubit below the $m$ register stores the sign of the cost function. We give details of the necessary components in the supporting information \ref{sec:Fourier_add} and \ref{sec:twosC}.

To adapt this algorithm for chemical problems, constraints on the number of alpha (spin-up) and beta (spin-down) electrons must be enforced. These constraints reflect number and spin symmetry, thereby reducing the optimization space. To incorporate these constraints, a similar procedure is used as for encoding the QUBO cost function  onto an ancilla register. However, since the number operator has only non-negative expectation values, the additional qubit required for negative values is unnecessary. The number of qubits needed is:
\begin{equation}
\mu = \nu = \lceil \log_{2}\left(\frac{n}{2}\right) \rceil.
\end{equation}
Here, $n \in 2\mathbb{Z}^{+}$ is the total number of spin-orbitals (or qubits), which is an even positive integer for chemical problems. Thus, $n/2 \in \mathbb{Z}^{+}$ corresponds to the number of spatial orbitals. 

%%%
The number of alpha ($n_{\alpha}$) and beta ($n_{\beta}$) electrons are also stored in ancillary registers using the Fourier basis, and subsequently transformed into the computational basis via the inverse Quantum Fourier Transform ($QFT^{\dagger}$). Alternative quantum adder circuits could also be used here; however, given their variety \cite{ORTS2020102810}, a detailed discussion is beyond the scope of this work.

\begin{figure*}
     \centering
     \begin{adjustbox}{width=1\textwidth}
\tikzset{every picture/.style={line width=0.75pt}} %set default line width to 0.75pt        

%! \usetikzlibrary{decorations.pathreplacing,decorations.pathmorphing}
\definecolor{mygreen}{RGB}{34,139,33}
\definecolor{myblue}{RGB}{157,220,229}
\definecolor{myred}{RGB}{255,99,98}
\begin{tikzpicture}[scale=1.500000,x=1pt,y=1pt]
\filldraw[color=white] (0.000000, -11.000000) rectangle (356.000000, 55.000000);
% Drawing wires
% Line 17: q0 W \ket{d}_{QUBO}
\draw[color=black] (0.000000,44.000000) -- (316.000000,44.000000);
\draw[color=black] (316.000000,43.500000) -- (342.500000,43.500000);
\draw[color=black] (316.000000,44.500000) -- (342.500000,44.500000);
\draw[color=black] (0.000000,44.000000) node[left] {$\ket{d}_{QUBO}$};
% Line 18: q1 W \ket{0}_{f(\vec{x})}
\draw[color=black] (0.000000,22.000000) -- (342.500000,22.000000);
\draw[color=black] (0.000000,22.000000) node[left] {$\ket{0}_{f(\vec{x})}$};
% Line 19: q2 W \ket{0}_{sign}
\draw[color=black] (0.000000,0.000000) -- (342.500000,0.000000);
\draw[color=black] (0.000000,0.000000) node[left] {$\ket{0}_{sign}$};
% Done with wires; drawing gates
% Line 21: q0 / n
\draw (8.833333, 37.000000) -- (18.166667, 51.000000);
\draw (15.833333, 47.500000) node[right] {$\scriptstyle{n}$};
% Line 22: q1 / m
\draw (8.833333, 15.000000) -- (18.166667, 29.000000);
\draw (15.833333, 25.500000) node[right] {$\scriptstyle{m}$};
% Line 23: q2 LABEL
% Line 28: q0 LABEL
% Line 29: q1 G:state width=23 $\hat H^{\otimes m}$
\begin{scope}[rounded corners=3pt]
\begin{scope}
\draw[fill=myblue] (44.500000, 22.000000) +(-45.000000:16.263456pt and 9.899495pt) -- +(45.000000:16.263456pt and 9.899495pt) -- +(135.000000:16.263456pt and 9.899495pt) -- +(225.000000:16.263456pt and 9.899495pt) -- cycle;
\clip (44.500000, 22.000000) +(-45.000000:16.263456pt and 9.899495pt) -- +(45.000000:16.263456pt and 9.899495pt) -- +(135.000000:16.263456pt and 9.899495pt) -- +(225.000000:16.263456pt and 9.899495pt) -- cycle;
\draw (44.500000, 22.000000) node {$\hat H^{\otimes m}$};
\end{scope}
\end{scope}
% Line 30: q2 G:state width=23 $\hat H$
\begin{scope}[rounded corners=3pt]
\begin{scope}
\draw[fill=myblue] (44.500000, -0.000000) +(-45.000000:16.263456pt and 9.899495pt) -- +(45.000000:16.263456pt and 9.899495pt) -- +(135.000000:16.263456pt and 9.899495pt) -- +(225.000000:16.263456pt and 9.899495pt) -- cycle;
\clip (44.500000, -0.000000) +(-45.000000:16.263456pt and 9.899495pt) -- +(45.000000:16.263456pt and 9.899495pt) -- +(135.000000:16.263456pt and 9.899495pt) -- +(225.000000:16.263456pt and 9.899495pt) -- cycle;
\draw (44.500000, -0.000000) node {$\hat H$};
\end{scope}
\end{scope}
% Line 31: q1 q2 G:state width=28 $\hat{A}(-y)$
\draw[rounded corners=3pt] (82.000000,22.000000) -- (82.000000,0.000000);
\begin{scope}[rounded corners=3pt]
\begin{scope}
\draw[fill=myblue] (82.000000, 11.000000) +(-45.000000:19.798990pt and 25.455844pt) -- +(45.000000:19.798990pt and 25.455844pt) -- +(135.000000:19.798990pt and 25.455844pt) -- +(225.000000:19.798990pt and 25.455844pt) -- cycle;
\clip (82.000000, 11.000000) +(-45.000000:19.798990pt and 25.455844pt) -- +(45.000000:19.798990pt and 25.455844pt) -- +(135.000000:19.798990pt and 25.455844pt) -- +(225.000000:19.798990pt and 25.455844pt) -- cycle;
\draw (82.000000, 11.000000) node {$\hat{A}(-y)$};
\end{scope}
\end{scope}
% Line 37: q1 q2 G:state width=20 $f(\vec{x})$ q0:shape=4
\draw[rounded corners=3pt] (118.000000,44.000000) -- (118.000000,0.000000);
\begin{scope}[rounded corners=3pt]
\begin{scope}
\draw[fill=myblue] (118.000000, 11.000000) +(-45.000000:14.142136pt and 25.455844pt) -- +(45.000000:14.142136pt and 25.455844pt) -- +(135.000000:14.142136pt and 25.455844pt) -- +(225.000000:14.142136pt and 25.455844pt) -- cycle;
\clip (118.000000, 11.000000) +(-45.000000:14.142136pt and 25.455844pt) -- +(45.000000:14.142136pt and 25.455844pt) -- +(135.000000:14.142136pt and 25.455844pt) -- +(225.000000:14.142136pt and 25.455844pt) -- cycle;
\draw (118.000000, 11.000000) node {$f(\vec{x})$};
\end{scope}
\end{scope}
\begin{scope}
\draw[fill=white] (118.000000, 44.000000) +(-45.000000:3.000000pt) -- +(45.000000:3.000000pt) -- +(135.000000:3.000000pt) -- +(225.000000:3.000000pt) -- cycle;
\clip (118.000000, 44.000000) +(-45.000000:3.000000pt) -- +(45.000000:3.000000pt) -- +(135.000000:3.000000pt) -- +(225.000000:3.000000pt) -- cycle;
\draw (115.000000, 44.000000) -- (121.000000, 44.000000);
\draw (118.000000, 41.000000) -- (118.000000, 47.000000);
\end{scope}
% Line 49: q1 q2 G:env width=25 $QFT^{\dagger}$
\draw[rounded corners=3pt] (152.500000,22.000000) -- (152.500000,0.000000);
\begin{scope}[rounded corners=3pt]
\begin{scope}
\draw[fill=myred] (152.500000, 11.000000) +(-45.000000:17.677670pt and 25.455844pt) -- +(45.000000:17.677670pt and 25.455844pt) -- +(135.000000:17.677670pt and 25.455844pt) -- +(225.000000:17.677670pt and 25.455844pt) -- cycle;
\clip (152.500000, 11.000000) +(-45.000000:17.677670pt and 25.455844pt) -- +(45.000000:17.677670pt and 25.455844pt) -- +(135.000000:17.677670pt and 25.455844pt) -- +(225.000000:17.677670pt and 25.455844pt) -- cycle;
\draw (152.500000, 11.000000) node {$QFT^{\dagger}$};
\end{scope}
\end{scope}
% Line 53: q0 G:state $-\hat I$ q2
\draw[rounded corners=3pt] (184.000000,44.000000) -- (184.000000,0.000000);
\begin{scope}[rounded corners=3pt]
\begin{scope}
\draw[fill=myblue] (184.000000, 44.000000) +(-45.000000:9.899495pt and 9.899495pt) -- +(45.000000:9.899495pt and 9.899495pt) -- +(135.000000:9.899495pt and 9.899495pt) -- +(225.000000:9.899495pt and 9.899495pt) -- cycle;
\clip (184.000000, 44.000000) +(-45.000000:9.899495pt and 9.899495pt) -- +(45.000000:9.899495pt and 9.899495pt) -- +(135.000000:9.899495pt and 9.899495pt) -- +(225.000000:9.899495pt and 9.899495pt) -- cycle;
\draw (184.000000, 44.000000) node {$-\hat I$};
\end{scope}
\end{scope}
\filldraw (184.000000, 0.000000) circle(1.500000pt);
% Line 56: q0 q1 q2 G:state width=20 $\hat W^{\dagger}$
\draw[rounded corners=3pt] (213.000000,44.000000) -- (213.000000,0.000000);
\begin{scope}[rounded corners=3pt]
\begin{scope}
\draw[fill=myblue] (213.000000, 22.000000) +(-45.000000:14.142136pt and 41.012193pt) -- +(45.000000:14.142136pt and 41.012193pt) -- +(135.000000:14.142136pt and 41.012193pt) -- +(225.000000:14.142136pt and 41.012193pt) -- cycle;
\clip (213.000000, 22.000000) +(-45.000000:14.142136pt and 41.012193pt) -- +(45.000000:14.142136pt and 41.012193pt) -- +(135.000000:14.142136pt and 41.012193pt) -- +(225.000000:14.142136pt and 41.012193pt) -- cycle;
\draw (213.000000, 22.000000) node {$\hat W^{\dagger}$};
\end{scope}
\end{scope}
% Line 59: q0 G:env width=35 $2 | d \rangle \langle d | - I $
\begin{scope}[rounded corners=3pt]
\begin{scope}
\draw[fill=myred] (252.500000, 44.000000) +(-45.000000:24.748737pt and 9.899495pt) -- +(45.000000:24.748737pt and 9.899495pt) -- +(135.000000:24.748737pt and 9.899495pt) -- +(225.000000:24.748737pt and 9.899495pt) -- cycle;
\clip (252.500000, 44.000000) +(-45.000000:24.748737pt and 9.899495pt) -- +(45.000000:24.748737pt and 9.899495pt) -- +(135.000000:24.748737pt and 9.899495pt) -- +(225.000000:24.748737pt and 9.899495pt) -- cycle;
\draw (252.500000, 44.000000) node {$2 | d \rangle \langle d | - I $};
\end{scope}
\end{scope}
% Line 65: q1 LABEL
% Line 66: q2 LABEL
% Line 64: q0 LABEL
% Line 68: q0 M
\draw[fill=white] (309.000000, 37.000000) rectangle (323.000000, 51.000000);
\draw[very thin] (316.000000, 44.700000) arc (90:150:7.000000pt);
\draw[very thin] (316.000000, 44.700000) arc (90:30:7.000000pt);
\draw[->,>=stealth] (316.000000, 37.700000) -- +(80:12.124356pt);
% Line 70: q0 END
% Line 71: q1 END
% Line 72: q2 END
% Done with gates; drawing ending labels
% Done with ending labels; drawing cut lines and comments
% Line 74: @ 1 4 %% $\hat W$
\draw[decorate,decoration={brace,mirror,amplitude = 4.666667pt},very thick] (30.000000,-11.000000) -- (168.000000,-11.000000);
\draw (99.000000, -15.666667) node[text width=144pt,below,text centered] {$\hat W$};
% Line 76: @ 1 7 % repeat $L$ times
\draw[decorate,decoration={brace,amplitude = 4.666667pt},very thick] (30.000000,55.000000) -- (273.000000,55.000000);
\draw (151.500000, 59.666667) node[text width=144pt,above,text centered] {repeat $L$ times};
% Done with comments
\end{tikzpicture}
\end{adjustbox}
        \caption{Outline of the full quantum circuit to implement Grover Adaptive Search Self-Consistent Field (GAS-SCF) algorithm searching over Dicke states. Here, a Dicke state $\ket{d}$ is prepared on the $n$-qubit register. This restricts the search space to bitstrings with the correct number of alpha and beta electrons and therefore the symmetry flagging step in Figure~\ref{fig:GAS-fig} is no longer needed. The $\hat{A}$ and $f(\vec{x})$ gates are defined in Figure~\ref{fig:adder} and Figure~\ref{fig:QUBO-adder} in the supporting information. An alternative compilation is provided in the Supplemental Information \ref{sec:FT_comp}, which gives the corresponding $T$-gate cost. In this realization no $QFT^{\dagger}$ is required.} 
        \label{fig:GAS-Dicke}
\end{figure*}
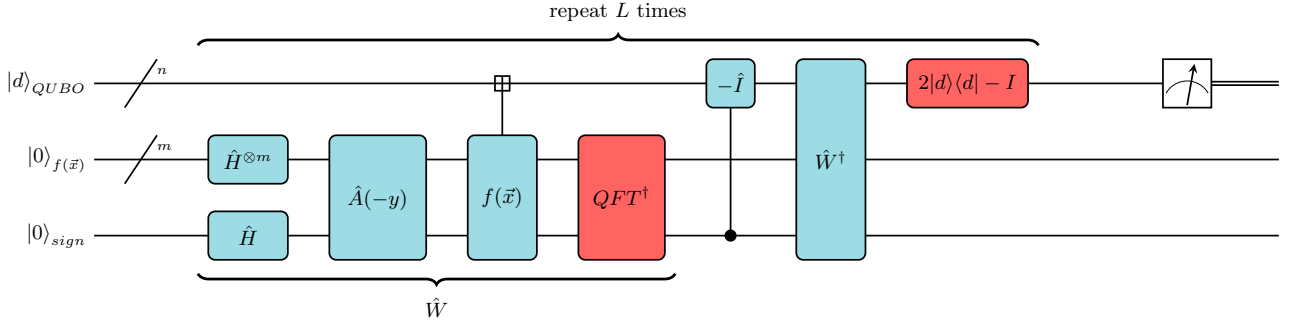

The next step of the algorithm is to mark states that both:
\begin{enumerate}
    \item Have a negative value in the QUBO cost function.
    \item Satisfy the correct number of alpha and beta electrons.
\end{enumerate}

To achieve this, a multi-controlled X gate is used, with controls defined by the bitstring representation of the desired binary occupation state. In Figure~\ref{fig:GAS-fig} this is represented by the projectors in each control. See Figure~\ref{fig:QUBO-proj} in the supporting information for further details. This gate acts on a single ancillary qubit, effectively flagging valid configurations. In the language of second quantization, these correspond to Fock states (occupation number states) in the correct particle-number symmetry sector. A Toffoli gate is then applied to store a $\ket{1}$ state which acts as a flag if and only if both the alpha and beta electron count registers match their target values.

To mark the resulting states, a multi-controlled $-I$ gate is applied. This gate is controlled by:

\begin{itemize}
    \item The number-flag qubit (indicating the correct electron counts)
    \item The \emph{sign} qubit from the $(m+1)$ register (indicating a negative QUBO value).
\end{itemize}

We note that the \emph{sign} qubit serves as a flag indicating whether the $(m+1)$-qubit register encodes a negative value of the (QUBO) cost function. Specifically, this single qubit emerges from the Two’s-complement representation of integers, in which the most significant (leftmost) bit determines the sign of the encoded value. For readers unfamiliar with Two’s‑complement arithmetic, additional background is provided in the supporting information \ref{sec:twosC}.

Overall, the controlled $-I$ operation, implemented via a controlled $R_{z}(- 2\pi)$, selectively applies a phase flip to all bitstrings $\vec{x}$ with value $f(\vec{x})-y<0$ and with the correct number of alpha and beta electrons. Here $y$ is a constant shift representing the best value found so far (a random bitstring with the correct symmetries can be used to initialize this value or the output of a classical SCF calculation). Undoing the circuit which generated the marked states followed by a reflection around $\ket{s}$ (the input state for Grover) then amplifies all marked states.

\begin{table*}[t]
    \centering
    \begin{adjustbox}{width=\textwidth}
    \begin{tabular}{c@{\hskip 20pt}c@{\hskip 20pt}c@{\hskip 20pt}c@{\hskip 20pt}c}
        \hline \hline
            & \textbf{Gate} & \textbf{Count} & \textbf{Asymptotic} & \textbf{Note} \\ 
        \hline
        (1) & $H$ & $2(m+1)$& $\mathcal{O}(m)$ & Preparation for storing QUBO value in Fourier basis \\
            & $P$\footnote{Phase gate.} & $2(m+1)$& $\mathcal{O}(m)$ & $\hat{A}(-y)$ \\
            & Singly-controlled $P$ & At most $2\big(n[m+1]\big)$& $\mathcal{O}(nm)$ & $f(\vec{x})$ linear terms \\
            & Doubly-controlled $P$ & At most $2\big(\binom{n}{2} [m+1] \big)$& $\mathcal{O}(n^{2}m)$ & $f(\vec{x})$ quadratic terms \\
            & $m$-qubit $QFT^{\dag}$ & $2$ & $\mathcal{O}(1)$ & To convert QUBO value to decimal basis \\
            & $(n-1)$-controlled $Z$ & $1$& $\mathcal{O}(1)$ & For $2|0^{\otimes n}\rangle \langle 0^{\otimes n}|-I^{\otimes n}$  reflection \\
            & $X$ & $2n$& $\mathcal{O}(n)$ & For $2|0^{\otimes n}\rangle \langle 0^{\otimes n}|-I^{\otimes n}$ reflection  \\
            & $CNOT$ & $2(n-1)$& $\mathcal{O}(n)$ & For marking operation \\
        \hline
        (2) & $H$ & $2(\mu + \nu)$& $\mathcal{O}(\log n)$ & Preparation for storing occupation value in Fourier basis \\
            & Singly-controlled $P$ & $2n$& $\mathcal{O}(n)$ & $\hat{N}_{\alpha / \beta}(x)$ linear terms \\
            & $\mu$-controlled $X$ & $2$& $\mathcal{O}(1)$ & To mark correct Hamming weight states ($\alpha$ electrons)\\
            & $\mu$-qubit $QFT^{\dag}$ & $2$& $\mathcal{O}(1)$ & To convert occupation values to decimal basis \\
            & $\nu$-controlled $X$ & $2$& $\mathcal{O}(1)$ & To mark correct Hamming weight states ($\beta$ electrons)\\
            & $\nu$-qubit $QFT^{\dag}$ & $2$& $\mathcal{O}(1)$ & To convert Fourier occupation values to decimal basis \\
            & Toffoli & $2$& $\mathcal{O}(1)$ & Check for correct number of  ${\alpha}$ \& ${\beta}$ electrons \\
        \hline
        (3) & $H$ & $2n$& $\mathcal{O}(n)$ & To reflect around $\ket{s}$ on $n$-qubits \\
            & Doubly-controlled $R_{Z}(-2\pi)$ & $1$ & $\mathcal{O}(1)$ & For marking operation \\
        \hline
        (4) & Dicke state construction. See \textit{e.g.} \cite{bartschi2019deterministic, bartschi2022short} & 
        $2 \big( w(n/2,n_{\alpha}) + w(n/2,n_{\beta}) \big)$\footnote{$w(n,k)$  used as a placeholder for quantum circuit cost to generate $\binom{n}{k}$ Dicke state.} & $\mathcal{O}(kn)\leq \mathcal{O}(n^{2})$\footnote{Scaling for the Dicke circuit implementation we used is at worst quadratic when $k = \mathcal{O}(n)$ \cite{bartschi2019deterministic}; other approaches may differ.} & To reflect around $\ket{d}$ on $n$-qubits \\
            & Singly-controlled $R_{Z}(-2\pi)$ & $1$& $\mathcal{O}(1)$ & For marking operation \\
        \hline \hline
    \end{tabular}
    \end{adjustbox}
    \caption{Gate requirements for one repetition  of GAS-SCF (Figure~\ref{fig:GAS-fig} / Figure~\ref{fig:GAS-Dicke}) that is repeated $L$ times. The table is broken into four sections as follows: (1), gates required for both $\ket{s}$ and $\ket{d}$ algorithms; (2), extra gates required to enforce occupation numbers (for $\ket{s}$ version only); (3), extra gates required for reflection (for $\ket{s}$ version only); (4), extra gates required for reflection (for $\ket{d}$ version only). Here $\alpha$ and $\beta$ denote the number of spin-up and spin-down electrons respectively. An alternative compilation is provided in the Supplemental Information \ref{sec:FT_comp}, which gives the corresponding $T$-gate cost for both realizations. In this approach no $QFT^{\dagger}$ is required.}
    \label{tab:circuit_cost}
\end{table*}

In summary, a high-level rundown of the GAS (inner optimization of Equation~\ref{eq:QUBO-SCF}) is as follows:

\begin{enumerate}
    % \item Initialize a uniform superposition over all $N$ bitstrings in the QUBO register in the $n$-qubit register.
    \item Initialize a uniform superposition over all (or a selected subset of) bitstrings in the $n$-qubit QUBO register; in this work, this state is denoted by $\ket{s}$ or $\ket{d}$.
    \item Initialize the register of $(m+1)$ -qubits, which stores the current value of the QUBO function, with the chosen starting value: $-y$ (in the Fourier basis).
    \item Conditioned on the $n$-qubit QUBO  register, which encodes the bitstring inputs for the QUBO cost function, compute and store the cost function value in the $(m+1)$-qubit register using quantum arithmetic.
    \item Mark all bitstrings (in the QUBO register) with a negative sign if their value is less than $y$ \emph{and} if they have the correct number of electrons.
    \item Undo steps 2 and 3 ($W^{\dagger}$ in circuit).
    \item In the $n$-qubit QUBO register, apply a reflection around the input state ($\ket{s}$ or $\ket{d}$) to increase the amplitude of marked states.
    \item Repeat the amplification routine (steps 2-6) $L$ times to increase the probability of measuring marked bitstrings.
    \item Measure the QUBO register. With high probability, this yields a marked bitstring, representing an improved QUBO solution.
    \item Set $y$ to be the new QUBO value of this improved bitstring and repeat from step 1 until convergence.
\end{enumerate}

The amplification routine should be repeated $L$ times where \cite{boyer1998tight, nielsen2010quantum}:

% \begin{equation}
%     L  \leq \Big\lceil \frac{\pi}{4} \sqrt{\frac{N}{T}} \Big\rceil, 
% \end{equation}

\begin{equation} \label{eq:Lrepeats}
\begin{aligned}
    L  &=\frac{\pi}{4}\Bigg[\frac{1}{\arcsin \bigg( \sqrt{\frac{T}{N}} \bigg)} \Bigg]-\frac{1}{2} \\
    &< \frac{\pi}{4} \sqrt{\frac{N}{T}}-\frac{1}{2}, 
    \end{aligned}
\end{equation}
$L$ should be rounded to the nearest integer based on the top line of this equation (we assume $0<T<\frac{N}{2}$). Here, $T$ is the number of marked states and $N\leq2^{n}$ is the size of the search space. However, as noted in \cite{gilliam2021grover}, the number of marked states, $T$, is generally unknown, and thus they define a randomized strategy to select $L$. An alternate approach has also been presented in \cite{ominato2024grover} that has been shown to require fewer queries.  On this issue, we appreciate Brassard’s reference to Kristen Fuchs, who likens Grover’s search algorithm to cooking a souffl{\'e}: if you use too many (overcooked) or too few (undercooked) queries, the outcome won’t yield good results \cite{brassard1997searching}. Alternatively, fixed-point methods \cite{Grover2005Fixedpoint} ensure that at each step (increasing $L$), the probability of measuring a marked state monotonically increases. However, this guarantee comes at the cost of losing the quadratic speedup. Yoder \textit{et al.} later introduced an approach that achieves fixed-point behavior without sacrificing quantum speedup \cite{Chuang2014FixedPoint}.

For electronic structure SCF applications, the total number of states with the correct number of spin up and down electrons ($n_{\alpha}$, $n_{\beta}$) is given by:
\begin{equation} \label{eq:S_bound}
\binom{n_{spatial}}{n_{\alpha}} \cdot \binom{n_{spatial}}{n_{\beta}} = \binom{n/2}{n_{\alpha}} \cdot \binom{n/2}{n_{\beta}} \geq T
\end{equation}
which bounds the number of marked states, $T$, and thus the maximum value $L$ can take for GAS-SCF. Weyl's formula can also be utilize to give a better bound \cite[eq 8]{cremer2013configuration} \cite[eq 92]{szalay2012multiconfiguration}, but requires a specified spin sector. Note in practice $T$ will be smaller than this, as the marked states must also have a negative value in the $(m+1)$ register. If a good classical reference (corresponding to an appropriate $y$-value) is used, it can significantly reduce the number of relevant states and, consequently, the number of repetitions required.

This structure can be extended to construct an alternative quantum circuit for GAS-SCF. Instead of preparing an equal superposition over all computational basis states, we generate an equal superposition over states with the correct number of electrons. This corresponds to a Dicke state $\ket{d}$, which is an equal superposition of computational basis states of a fixed Hamming weight. The overall circuit is illustrated in Figure~\ref{fig:GAS-Dicke}. While this algorithm introduces additional depth, due to Dicke state preparation and the Grover reflection step, it eliminates the need for ancilla qubits used to track electron count. In essence, the search space is inherently restricted to valid states within the correct number symmetry sector. This structure is exactly what classical simulation techniques also take advantage of to simplify the optimization problem. Table~\ref{tab:circuit_cost} summarizes the circuit costs for both circuit  implementations. 

In the Supplemental Information (Section~\ref{sec:mol_syms}), we describe how $S^2$ and molecular point-group symmetries can be enforced within the GAS-SCF algorithm. The underlying methodology is analogous to that used to preserve the number of spin-up and spin-down electrons, keeping the optimization to the desired symmetry sector. We note enforcing more symmetries will reduce the search space further. 

\subsection{Non-Integers}\label{sec:non}

A challenge with the GAS routine is that it requires the QUBO formulation (Equation~\ref{eq:qubo_full}) to have integer coefficients, whereas in the SCF problem the coefficients (integrals) are real (floating point) numbers. Gilliam \textit{et al.} explored two encoding strategies: integer-approximation and direct (Fej\'{e}r) encoding  schemes for treating non-integer functions \cite{gilliam2021grover}. In \cite[Section III.C]{norimoto2023quantum}, Norimoto and Ishikawa  discuss the Fej\'{e}r encoding scheme in more detail.

In this work, we employ the integer-approximation scheme to enable the implementation of GAS-SCF. We note that multiplying the Hamiltonian’s coefficients by a constant $\Lambda$ preserves the eigenspectrum, only rescaling the eigenvalues. Operationally for GAS-SCF this merely changes the number of qubits required in the 
$(m+1)$-qubit register by $\log \Lambda$. If we multiply by a factor of $2^{\epsilon}$ ($\epsilon > 0, \; \epsilon \in \mathbb{R}$) we find Equation~\ref{eq:m_reg} is modified as:
\begin{equation}
\begin{aligned}
    m_{\epsilon} &= \left\lceil \log_{2}\left( 2^{\epsilon}\max[ f(\vec{x})] \right) \right\rceil =\epsilon + \left\lceil \log_{2}\left( \max[ f(\vec{x})] \right) \right\rceil 
    \\
    &\leq  \epsilon +  \left\lceil \log_{2} \left(  \sum_{ij}|Q_{ij}| + \sum_{i}|b_{i}| \right) \right\rceil.
\end{aligned}
\end{equation}

For $\epsilon=50$, we have $2^{50} \approx 1\times 10^{15}$. This increases the number of ancilla qubits by fifty but scales all coefficients by roughly $1 \times 10^{15}$. Given that most chemistry libraries only calculate the Hamiltonian coefficients (integrals) to fifteen decimal places \cite{sun2015libcint}, this $\epsilon$ will round all the coefficients to integers as required. In the fault-tolerant regime, this additional qubit overhead is negligible. However, the number of non-Clifford rotations is increased, which may incur additional costs. If this becomes a limitation, alternative approaches to quantum arithmetic may help mitigate it \cite{ORTS2020102810} \cite[Appendix D]{Simon2025ladderoperatorblock}. Moreover, for most practical applications, rounding to fewer decimal places is sufficient, as demonstrated by the numerical results presented in this work. Nonetheless, in the Supplemental information \ref{sec:FT_comp} we compile the GAS circuit using adder circuits derived in \cite{Simon2025ladderoperatorblock} and provide the overall T-gate cost. This alternative approach removes the need for $QFT^{\dagger}$.

In Section~\ref{sec:numerical}, we present three numerical examples. The first two illustrate the inner workings of the algorithm and demonstrate the feasibility of rounding the Hamiltonian to integer coefficients. The third, and largest, example highlights the advantages of the algorithm, albeit for a classically tractable problem.
\section{Numerical Study\label{sec:numerical}} 

\begin{figure*}[ht]
    \centering
    \begin{subfigure}[b]{0.46\textwidth}
\includegraphics[width=0.95\linewidth]{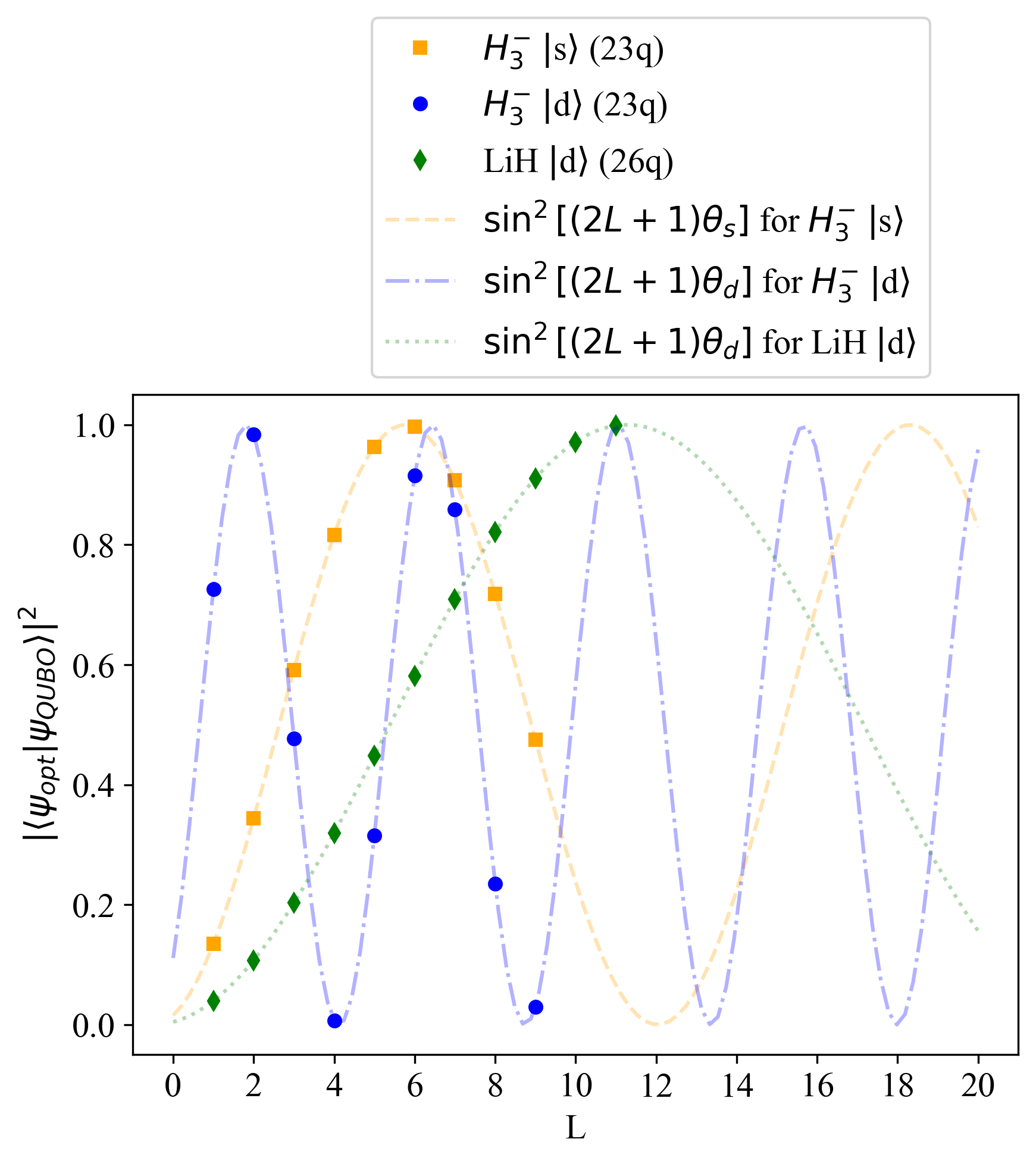}
\caption{} \label{fig:L_results_left}
\end{subfigure}
    \hspace{0.05\textwidth}
    \begin{subfigure}[b]{0.46\textwidth}
\includegraphics[width=0.95\linewidth]{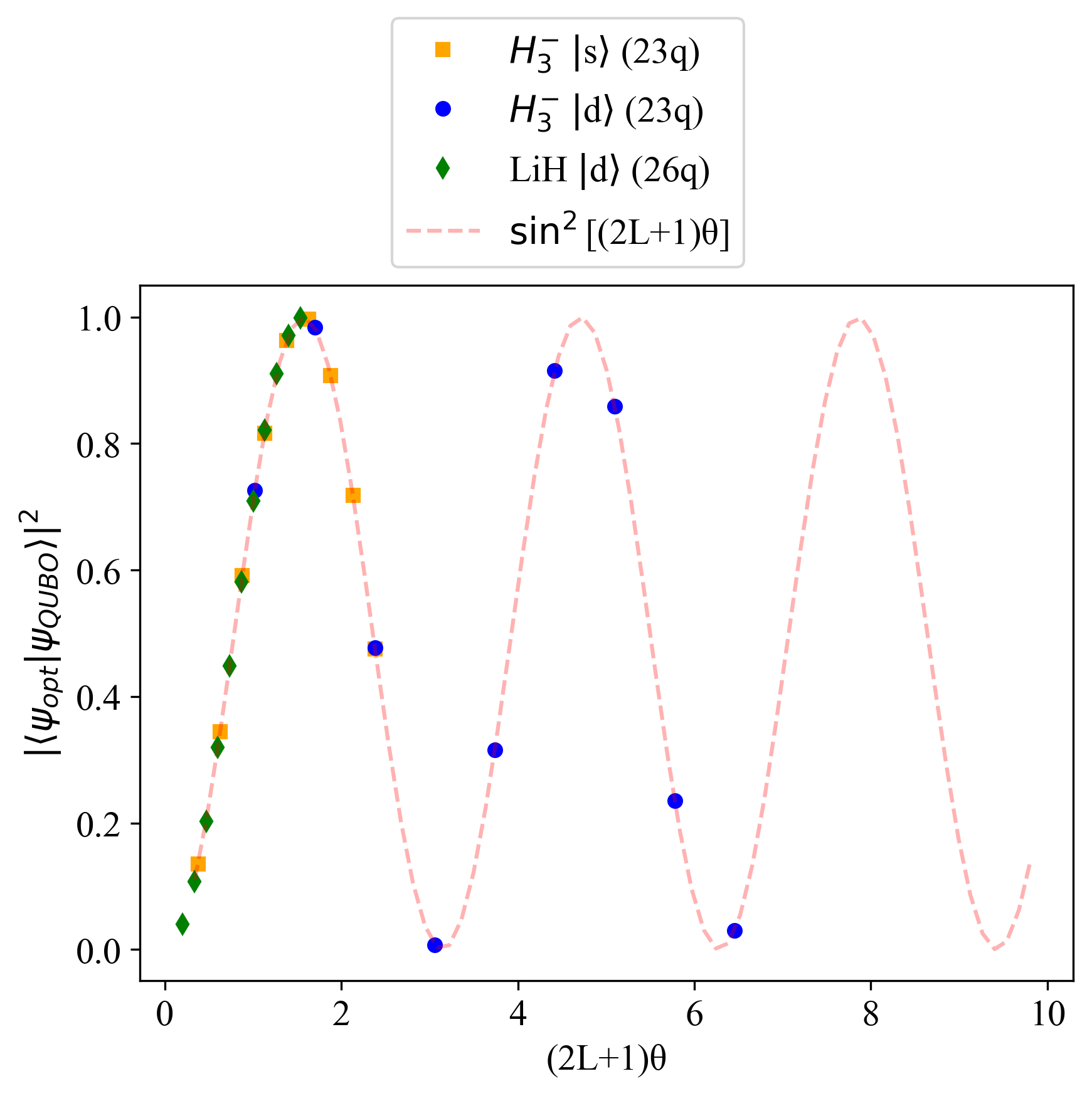}
\caption{} \label{fig:L_results_right}
    \end{subfigure}
    \caption{ (a) GAS-SCF results for increasing number of iterations $L$ and (b) $(2L+1) \theta$ where initial $y$ was selected so that GAS-SCF would only amplify the ground state. The overlap between the true ground state $\ket{\psi_{opt}}$ and the circuit statevector $\ket{\psi_{QUBO}}$ is plotted for \ce{H3-} $\blacksquare$ (uniform input $\ket{s}$), \ce{H3-} $\bullet$ (Dicke input $\ket{d}$) and \ce{LiH} $\blacklozenge$ (Dicke input $\ket{d}$). Qubit counts in the legend parenthesis provide the total number of qubits including all ancilla registers. Note $\theta$ for each system is $arcsin(1/\sqrt{2^{6}})$, $arcsin(1/\sqrt{9})$ and $arcsin(1/\sqrt{225})$ respectively (equation \ref{eq:Lrepeats}).}
    \label{fig:L_results}
\end{figure*}

% \begin{figure*}[ht]
%     \centering
% \includegraphics[width=0.75\linewidth]{figures/combined_figure.png}
%     \caption{GAS-SCF results for increasing number of iterations ($L$) where initial $y$ was selected so that GAS-SCF would only amplify the ground state. The overlap between the true ground state $\ket{\psi_{opt}}$ and the circuit statevector $\ket{\psi_{QUBO}}$ is plotted for \ce{H3-} $\blacksquare$ (uniform input $\ket{s}$), \ce{H3-} $\bullet$ (Dicke input $\ket{d}$) and \ce{LiH} $\blacklozenge$ (Dicke input $\ket{d}$). Qubit counts in the legend parenthesis provide the total number of qubits including all ancilla registers.}
%     \label{fig:L_results}
% \end{figure*}

\subsection{GAS simulation\label{sec:gas_numerical}}

To study the performance of the GAS routine in GAS-SCF, we investigate a single inner optimization step within Equation~\ref{eq:qubo_full} for the MO optimization of \ce{H3-} and \ce{LiH} in the STO-3G basis set. The goal is to find the bitstring with the lowest energy in the correct symmetry sector. Neither of these systems exhibits classical convergence issues and are small test examples, not as systems with potential practical advantage. Applying Eq~\ref{eq:S_bound} we find for \ce{H3-} there are  $\binom{3}{2} \cdot \binom{3}{2}=9$ and for \ce{LiH} there are $\binom{6}{2} \cdot \binom{6}{2}=225$ valid bitstrings. 

The ground states of \ce{H3-} and \ce{LiH} in the STO-3G basis have energy $-24$ Ha and $-888$ Ha respectively (in the integer coefficient problem). We first consider setting the threshold for GAS sufficiently low that the algorithm marks only the ground state. In this case GAS reduces to Grover search for the ground state in the symmetry subspace. To accomplish this, the GAS routine was configured with thresholds of $y=-20$ and $y=-880$ for \ce{H3-} and \ce{LiH} respectively. We initialized the system in the uniform state $\ket{s}$ and the Dicke state $\ket{d}$ for $\ce{H3-}$ and only in the Dicke state for \ce{LiH}. We simulated quantum circuits with increasing number of repetitions $L$. We plot the probability of the marked state as a function of $L$ in Figure~\ref{fig:L_results}. In each case, as $L$ increases, the probability of measuring the optimal bitstring is observed to vary periodically as expected in Grover's search. The optimal number of iterations can be determined by rounding Equation~\ref{eq:Lrepeats} to the closest integer. For \ce{H3-} in the STO-3G basis with the uniform initialization $\ket{s}$ and $y=-20$ we have $N=2^6$ and $T=1$ and  the optimal number of repetitions is $L \approx 5.77 \mapsto 6$.  For \ce{H3-} in the STO-3G basis with the Dicke state initialization $\ket{d}$ and $y=-20$ we have $N=9$, $T=1$ and $L  \approx 1.81 \mapsto 2$.

% comment on H3- amplifying the wrong state
If the classical estimate $y$ used in GAS is so poor that more than half the states are in the search space are marked, GAS fails due to a well understood failure mode for Grover's search when more than half the states are marked - called {\em overbalancing}. For \ce{H3-} in the STO-3G basis with the Dicke state $\ket{d}$ initialization, we found that setting $y=-4$ as the initial starting point would cause the GAS-SCF routine to fail. In this case, the amplitude of the $y=-4$ eigenstate increased, rather than the eigenstates with lower eigenvalues. As the problem is small (nine valid eigenstates) the allowed eigenvalues are: $\{-24,-20,-16,-20,-12,-8,-16,-8,-4\}$. The algorithm would therefore mark all the states below $y=-4$, meaning that there are more marked states than unmarked ones, thus causing the unmarked (incorrect) states to be amplified instead. Additional details for this example are provided in Supplemental Information~\ref{sec:H3_data}. As discussed in Supplemental Information~\ref{sec:mol_syms}, the incorporation of additional symmetry constraints, such as $S^{2}$, can eliminate the overbalancing issue. In particular, enforcing the singlet-spin constraint, $\langle S^{2}\rangle=0$, completely removes overbalancing in this example, as the number of marked states can no longer exceed half of the search space.

\begin{figure*}[ht]
    \centering
    \begin{subfigure}[b]{0.45\textwidth}
\includegraphics[width=0.95\linewidth]{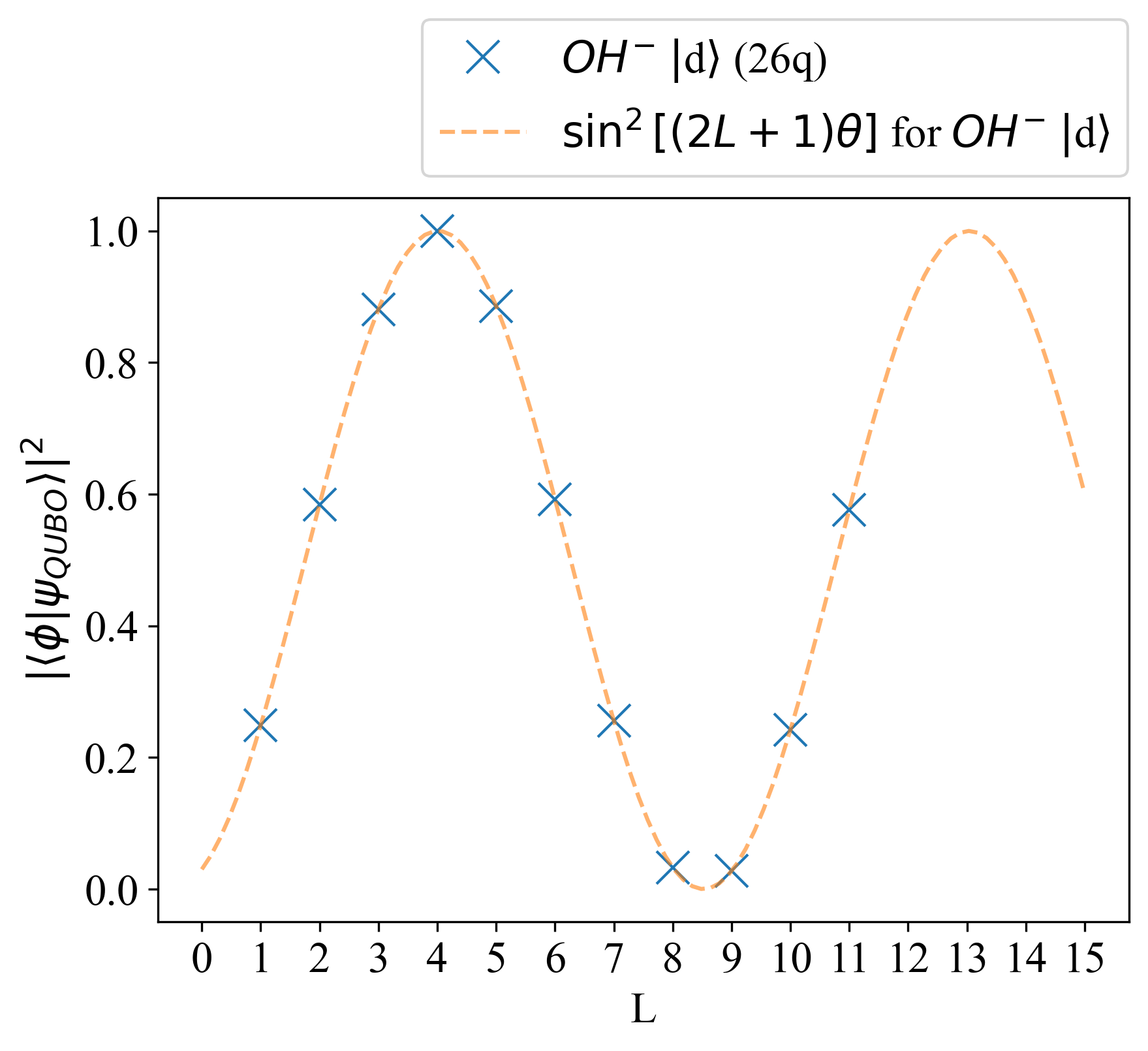}
\caption{\ce{OH-} (search space $\ket{d}$, total qubits $26$)} \label{fig:OH_dicke}
\end{subfigure}
    \hspace{0.05\textwidth}
    \begin{subfigure}[b]{0.45\textwidth}
\includegraphics[width=0.95\linewidth]{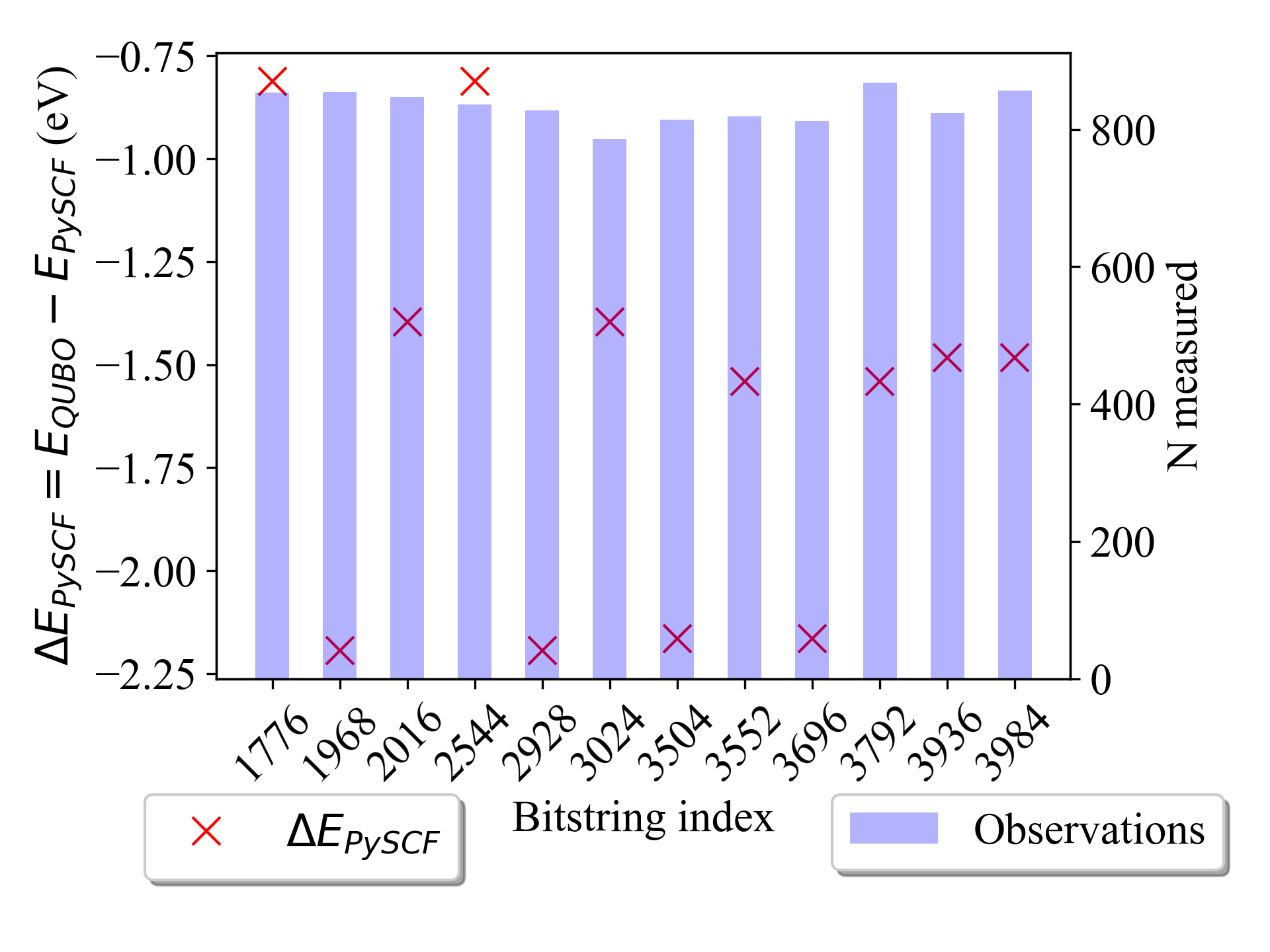}
\caption{Results for $L=4$ repeats, showing the statevector sampled $10,000$ times.} \label{fig:OH_sample_results}
    \end{subfigure}
    \caption{GAS-SCF Dicke simulation result for \ce{OH-}. The $y=-780$ (classical \texttt{PySCF} value)  was used to seed the GAS-SCF routine. (a) Illustrates the overlap of the QUBO statevector with $\ket{\phi}$ that is an equal superposition of the $12$ Fock states with lower energy than the \texttt{PySCF} solution. (b) Results of the $L=4$ statevector sampling simulation with $10,000$ shots. The $x$-axis gives the index of the sampled bitstring in base-$10$ (converting to binary gives the Fock state). The left $y$-axis gives the error between the GAS-SCF and \texttt{PySCF} RHF solution (scatter), where a negative value indicates GAS-SCF finding a better result. We see that in all cases a better bitstring over \texttt{PySCF} is obtained. The right $y$-axis gives the number of times each bitstring was sampled (bar-chart). Here $\theta = arcsin(\sqrt{12/400})$ (equation \ref{eq:Lrepeats}).}
    \label{fig:OH-2-result}
\end{figure*}

We propose the use of GAS-SCF as a method to improve upon the best available classical SCF solution, using the best classical reference to identify the states to mark. Avoiding overbalancing in GAS requires that the density of states of the molecule is such that the classical reference has lower energy than more than half of the states in the search space. Large numbers of configurations can have very similar energies if the ground state has large static correlation. If this causes overbalancing, Faro and Marino’s method of searching in overbalanced domains, applicable when valid solutions exceed half of the input space, can be used \cite{faro2025scaling}. 

Overbalancing can never happen for the uniform superposition input $\ket{s}$ state. This is because states belonging to incorrect symmetry sectors remain within the search space but are not marked, due to the constraints imposed by the number-symmetry registers. In particular, for $n_{s}$ spatial orbitals the  search space has dimension $2^{2n_{s}}$, whereas the maximum possible number of marked states is $\binom{n_{s}}{n_{\alpha}} \cdot \binom{n_{s}}{n_{\beta}}$. The worst-case subspace dimension occurs when: $\binom{n_{s}}{n_{s}/2} ^{2}$. The corresponding fraction of the total search space can be written as: $\tau ={\binom{n_{s}}{{n_{s}} /{2}} ^{2}} / {2^{2n_{s}}}$ which we can bound as: $ \tau < \frac{2}{\pi n_{s}}$. For $n_{s}>1$, this implies $\tau <0.5$, so the marked subspace always occupies less than half of the full search space. For the $n_{s}=1$ case, the solution is trivial.

Next, we focus on the more complex \ce{OH-} SCF optimization problem, which we previously showed had convergence issues \cite{ralli2025bridging}. Running RHF in \texttt{PySCF} returned a value of $-75.0730$~Ha ($y=-780$ for the rounded problem), which is a local minimum. This result could be improved by running more expensive second-order and stability methods in \texttt{PySCF} \cite{ralli2025bridging}. However, in the context of this work we use this result as a seed to show that GAS-SCF can improve a classical solution that has not converged properly; Figure~\ref{fig:OH-2-result} summarizes the results. In the chosen active space, there are $\binom{6}{3} \cdot \binom{6}{3}=400$ valid bitstring, a twelve qubit problem. 

Figure~\ref{fig:OH_dicke} shows the overlap squared of the statevector from our GAS-SCF circuit, for different repetitions $L$, with the equal superposition of the $12$ eigenstates with lower eigenvalues relative to the \texttt{PySCF} SCF solution. This demonstrates the probability of measuring a state that surpasses the \texttt{PySCF} solution. Using Equation~\ref{eq:Lrepeats}, $N=400$ and $T=12$, we see that  $L \approx 4.01 \mapsto 4$. As expected, we observe that at $L=4$, the probability of measuring a marked state is maximized. We performed a sampling statevector simulation for this scenario, which is summarized in Figure~\ref{fig:OH_sample_results}.  We observe that, across the $10,000$ samples taken, all the bitstrings obtained result in a lower energy Fock state than the \texttt{PySCF} solution. In this case, the largest improvement over the \texttt{PySCF} solution is $\approx 2.2$~eV. This illustrates how GAS-SCF can be used to search for solutions that are better than the best classical reference available. 

Finally, we note that this work does not implement a strategy for selecting the number of Grover iterations, $L$. Our objective was instead to highlight the key components of the algorithm and assess its performance in representative scenarios without introducing case-specific tuning of $L$, which could obscure the underlying algorithmic behavior. The problem of choosing $L$ has been extensively studied in the literature such as in \cite{gilliam2021grover, ominato2024grover, Grover2005Fixedpoint, Chuang2014FixedPoint}. Accordingly, a detailed analysis of iteration-selection strategies lies beyond the scope of this work.

\subsection{Examples of SCF Convergence Issues\label{sec:hard_numerical}}

\begin{figure*}[ht]
    \centering
\includegraphics[width=0.95\linewidth]{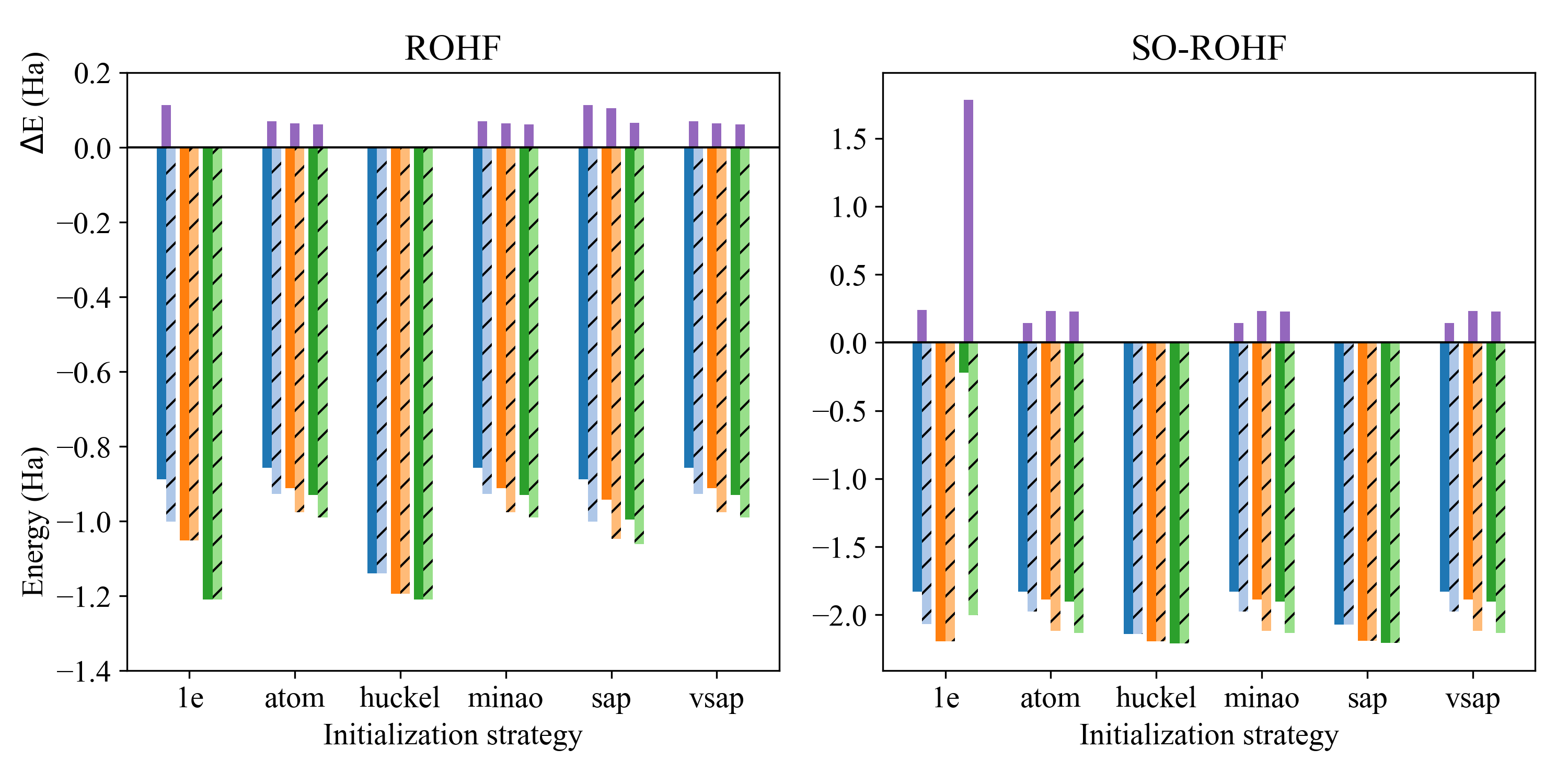}
    \caption{Comparison of SCF initialization strategies for triplet \ce{O3}, showing the PySCF solution and alternative single Fock-state solutions when found. The left panels show ROHF results, and the right panels show second-order ROHF (SO-ROHF) results. In the lower panels, results are grouped into cc-pVDZ (blue), cc-pVTZ (orange), and cc-pVQZ (green) basis sets (from left to right); within each basis-set pair, the left bar corresponds to the PySCF solution and the right bar (hatched) to the alternative Fock-state solution $\ket{x}$. The upper panels show the corresponding energy difference, $\Delta E = E_{PySCF} - E_{\ket{x}} $, where positive values indicate that $\ket{x}$ is lower in energy than the PySCF solution. The lower panels show the total energies obtained from each calculation, shifted by $223$ Ha (left) and $222$ Ha (right) for clarity. The initialization strategies presented are superposition of atomic densities (MINAO, atom) \cite{almlof1982principles, van2006starting}, the core-Hamiltonian guess (1e) \cite{lehtola2019assessment}, the parameter-free H\"uckel guess \cite{lehtola2019assessment}, and superposition of atomic potentials (VSAP, SAP) \cite{lehtola2019assessment}. Additional details, including the improved Fock-state solutions $\ket{x}$, are available in the online repository \cite{gas_github}. The cc-pVDZ, cc-pVTZ, and cc-pVQZ calculations correspond to systems with $84$, $180$, and $330$ qubits, respectively, excluding ancillary qubits.}
    \label{fig:O3_linear}
\end{figure*}

In this section, no GAS-SCF simulations are performed. Instead, conventional quantum chemistry calculations on classical hardware are used to identify example systems in which a GAS-SCF approach would provide benefits. Although GAS-SCF can, in principle, be simulated classically, the computational cost increases rapidly with system size, limiting classical implementations to relatively small benchmark problems, as demonstrated in the previous section. The method is ultimately intended for execution on quantum computers, where it has the potential to address practically relevant problems that lie beyond the reach of classical computational resources.

A motivating example for GAS-SCF is triplet \ce{O2} at a bond length of 2.55~{\AA}. Rubin \textit{et al} in ~\cite{rubin2022compressing} reported a low overlap between the restricted open-shell Hartree-Fock (ROHF) reference and the full configuration interaction (FCI) ground state in the STO-3G basis. We investigate this system in the larger 6-31G basis, FCI is not feasible, so we performed a configuratin interaction with singles and doubles (CISD) and a heat bath configuration interaction (HCI) \cite{dobrautz2019efficient} calculation. The ROHF overlaps are $|\langle \mathrm{CISD}|\mathrm{ROHF}\rangle|^{2} = 0.8958$ and $|\langle \mathrm{HCI}|\mathrm{ROHF}\rangle|^{2} = 0.0088$. We attribute the large CISD overlap value largely to truncation: in the CISD space the reference is directly coupled to all single and double excitations (Slater-Condon rules~\cite{slater1929theory, condon1930theory}), whereas any other determinant couples to only a subset of the space. This can inflate the apparent weight of the reference. Including higher excitations removes this bias, and the overlap can drops sharply, as seen here.

We stress that overlap with the ground state is a diagnostic of single-reference character, not a criterion for the HF solution. For the 6-31G result, we note other Fock states have larger amplitude than ROHF, for example $ |\langle \mathrm{HCI}_{\text{6-31G}}|68701388800_{10}\rangle |^{2}= 0.043$, where the integer is the bitstring of the Slater determinant in base 10. Since a larger amplitude does not imply lower energy (see below), we compare energies directly: $ \bra{ROHF} H_{\text{6-31G}}\ket{ROHF} = -149.4423$ Ha, $ \bra{68701388800_{10}} H_{\text{6-31G}}\ket{68701388800_{10}} = -148.9665$ Ha. This alternate single Fock state lies $0.476$~Ha below the ROHF state while retaining $\langle S^2 \rangle = 2$, corresponding to a triplet state.  The ROHF state is therefore not the lowest-energy determinant in this particular molecular orbital basis. The 6-31G search space is too large to  establish by brute force whether this determinant is optimal, so lower-energy solutions may remain undiscovered, and for larger problems this issue will be harder still. 

From these results we make two remarks. (1) Firstly, Hartree-Fock finds the molecular orbital basis that minimizes the energy of a single Fock state with the correct symmetries; equivalently, it minimizes one diagonal element of the second-quantized Hamiltonian (represented by the single reference state). This is defined independently of the ground state. How well that determinant approximates the ground state depends on the off-diagonal couplings and if diagonal elements are equal or similar in weight (no individual term is energetically favored) where the ground state is therefore expected to have comparable support across them and low overlap signals multi-reference character. This feature is common, for example in bond dissociation events. A multi-reference method such as CASSCF is typically required in these situations. (2) Secondly, a large amplitude (from an approximate ground state wavefunction) on a single determinant does not imply low energy for that determinant. Consider a toy CI matrix:
\begin{equation}
H = \begin{pmatrix}
-1.0 & 0.1 & 0.1 \\
0.1 & 0.5 & -2.0 \\
0.1 & -2.0 & 0.1
\end{pmatrix}.
\end{equation}
The single-determinant state of lowest energy is $|0\rangle$ (as seen directly from the diagonal elements). The ground state, with energy $-1.737$, is $|\lambda_0\rangle = -0.188\,|0\rangle + 0.659\,|1\rangle + 0.728\,|2\rangle$, so $|0\rangle$ has the smallest overlap of the three basis states. The Fock state here with the largest amplitude ($|2\rangle$) has a single reference energy of $\langle 2 | H | 2\rangle=0.1$, which is higher than that of $\ket{0}$.

In Fig. \ref{fig:O3_linear}, we present an additional example - linear \ce{O3} in the triplet state, with each O$-$O bond set to $2$ {\AA}. We compare our search for an improved Fock state solution (via a MaxCut calculation \cite{ralli2025bridging} followed by a 1 minute brute force search) against results from PySCF using various SCF initialization strategies keeping the MO basis fixed from the optimized PySCF calculation. In many cases ($24$ out of $36$), we identify a lower-energy bitstring, further demonstrating that it is possible to obtain a lower energy single–Fock-state solution in the correct sector. Moreover, these results highlight that there is significant scope for developing new solver heuristics, both classical and quantum. In this example, the cc-pVDZ, cc-pVTZ, and cc-pVQZ basis sets correspond to search spaces of sizes $\binom{42}{13}\binom{42}{11} \approx 1.09\times 10^{20}$, $\binom{90}{13}\binom{90}{11} \approx 6.84\times 10^{28}$ and $ \binom{165}{13}\binom{165}{11} \approx 2.92 \times10^{35}$, respectively. The tabulated results are given in the supporting information \ref{sec:O3_data}, and the raw results and analysis can be found at \cite{gas_github}.

Stability analyzes can identify self-consistent field (SCF) solutions that correspond to local rather than global minima of the SCF energy landscape \cite{seeger1977self}. In such cases, the molecular orbital (MO) basis will be suboptimal with respect to the chosen single reference Fock state. Although stability methods can be used to obtain improved initial starting points for restarting the SCF procedure, they increase the computational cost and do not guarantee convergence to the global minimum. A well-known challenging example is the chromium dimer. We provide a \ce{Cr2} study in the supplemental information \ref{sec:Cr2res}. In contrast, within a fixed MO basis, GAS can identify improved reference bitstrings without requiring orbital reoptimization. This can help restarting the SCF optimization and can also be advantageous when using localized, natural, or other noncanonical orbitals, where preserving the chosen orbital representation is desirable.

Finally, in conventional SCF methods, the Hartree-Fock (HF) solution landscape is basis-set dependent: increasing the basis set does not merely improve the accuracy of an existing solution, but can introduce entirely new stationary points and instability directions, including solutions that preserve or break spin and/or spatial symmetries. Conversely, small basis sets such as STO-3G and 6-31G may artificially suppress such solutions because the required orbital degrees of freedom are unavailable. Fortunately, the GAS component of GAS-SCF does not suffer from this basis-set dependence of the solution landscape, as it amplifies any Fock state with the correct symmetries whose energy is lower than the threshold value $y$. This assumes that the desired symmetries are explicitly enforced within the GAS search, either directly through the construction of the search space or penalty functions in the cost function or via ancilla registers that identify states belonging to the target symmetry sectors.

\section{Utility and Advantage for GAS-SCF} \label{sec:Q_ad}
In quantum computing, advantage refers to problems that a  quantum computer can solve but a classical computer cannot. Quantum utility refers to a case where a classical calculation is feasible but a quantum computer can obtain the same result with lower costs in time, space or energy. 

The Hartree-Fock SCF problem is equivalent to NP-complete combinatorial optimization problems~\cite{whitfield2014np,ralli2025bridging}. In the worst case, it is widely believed that quantum computers cannot solve NP-complete problems in polynomial time. However, exact solutions in the worst case are not the only possibility. Approximate algorithms are evaluated by their {\em approximation ratio} - the fraction of the exact answer they achieve. Bounds on the approximation ratio achievable by classical algorithms are reasonable targets for quantum advantage. For example, the best approximation ratio achievable by semi-definite programming (SDP) for MAXCUT is $0.878$, meaning that SDP will achieve at least this approximation ratio for all instances of MAXCUT~\cite{goemans1994879}. A quantum algorithm that achieved an approximation ratio of $>0.878$ for all instances would achieve quantum advantage for MAXCUT.

The three main algorithms for quantum optimization of combinatorial optimization problems are adiabatic quantum computation (AQC)~\cite{farhi2001quantum}, the quantum approximate optimization algorithm (QAOA)~\cite{farhi2014quantum}, and decoded quantum interferometry (DQI)~\cite{jordan2025optimization}. There is recent evidence that they may be able to obtain improvements in the approximation ratio for some problems~\cite{jordan2025optimization,farhi2025lower}. These results mean that combinatorial optimization is once again a promising target quantum advantage.

In the context of SCF problems, there exist many molecular problems with convergence issues \cite{daniels2000converging, van2006starting, vaucher2017steering, lehtola2019assessment, greiner2025reusablelibrarysecondorderorbital, ralli2025bridging, schultz2005databases, Kudin_Scuseria_2007}. Using classical approaches to approximately solve these hard SCF instances can be used to yield the ``best'' classical solution. This can be used to seed the initial choice of $y$ in the GAS-SCF routine. GAS-SCF succeeds if the method outputs a bitstring that surpasses all classical solutions. A key strength of this approach is GAS-SCF marks and boosts the amplitudes of all lower-energy states over the best available classical solution. Such solutions are easy to verify, as they correspond to a single Fock states (single bitstrings). If no better solution is found, one can always fall back on the classical solution. The \ce{OH-} model above illustrates the existence of problem instances where it is possible for GAS-SCF to improve the classical solution. While this small model is well within classical tractability \cite{ralli2025bridging}, it does indicate that for larger problems that are classically hard to simulate, it may be possible to improve upon classical SCF solutions. The \ce{O2} and \ce{O3} results in this work further reinforce this point.

SCF calculations are routinely applied to large molecular systems. For example, \cite{palethorpe2024advanced} studies Ubiquitin in the 6-31G${}^{\ast}$ basis set, for which the Hilbert space dimension is $2^{2(10273 )}\approx 9.168 \times 10^{6184}$. While \cite{palethorpe2024advanced} demonstrates the feasibility of treating such a system using SCF on conventional computational hardware, the optimality of the resulting classical solutions is not guaranteed, and are unlikely to correspond to the global optimum and are more likely to reflect local minima.

At present, it is not clear whether a classical solver run for an equivalent wall‑clock time would offset any advantage of the GAS-SCF routine; addressing this question will require systematic benchmarking studies. Additionally, an alternating hybrid strategy that interleaves quantum and classical SCF iterations may offer further benefits. Ultimately, these possibilities can only be assessed through real execution.

SCF routines are the basis for virtually all post-Hartree-Fock single reference electronic structure methods, such as M{\o}ller–Plesset perturbation theory, coupled cluster, and configuration interaction methods. Therefore, any improvement in SCF results in knock-on benefits for the more accurate approaches that build on an SCF solution, by identifying a better molecular orbital basis in which to describe the system.  

Quantum algorithms for quantum chemistry, including quantum phase estimation \cite{kitaev1995quantummeasurementsabelianstabilizer, babbush2014adiabatic}, are similarly affected. Here, SCF methods define the MO basis, linearly independent vectors that span the Hilbert space, and therefore different choices can alter the support of the ground state. This leads to interrelated consequences: (1) initializing QPE near from a local minimum can be problematic, and (2) the choice of basis influences the support of each eigenstate and thus its overlap with the QPE input state. Improved SCF solutions by GAS-SCF or other quantum optimization techniques offer a means to mitigate both issues.

\section{Future directions} \label{sec:future}

In GAS-SCF, it is assumed that the problem structure is utilized as much as possible to establish a classical threshold ($y$), after which unstructured search is used to boost the amplitude of all lower energy Fock states. However, the number of GAS-SCF iterations explicitly depends on the size of the search space $N$, as shown in equation \ref{eq:Lrepeats}. Consequently, the required number of Grover iterations can become prohibitively large, posing a significant challenge for GAS-SCF.

This observation raises an interesting question: how can one effectively restrict the search space to a smaller, targeted region of Hilbert space (i.e., reducing $N$ through states other than $\ket{s}$ and $\ket{d}$ as used in this work)? A straightforward strategy is to impose restrictions via alternative ansatz circuits. For instance, applying powers of the Hamiltonian, analogous to Krylov subspace methods, would generate Fock states in a controlled and structured manner. While such approaches indeed reduce the size of the search space, they also make the problem more amenable to classical simulation. A more meaningful improvement would therefore involve constraining the search space to a region of Hilbert space that remains classically intractable, while not being exponentially large along with retaining efficient implementation on quantum hardware. Achieving this will likely require exploiting additional structure present in the electronic structure problem.

% A potential starting point in this direction is the unitary cluster Jastrow (uCJ) anstaz. Notably, the unitary coupled cluster (UCC) ansatz involves exponentiating the antihermitian cluster operator, which in turn generates support over the full Hilbert space (in the number sector defined by the reference state). This leads to an exponentially large search space, precisely the regime we seek to avoid. In contrast, the $k$-uCJ ansatz \cite{matsuzawa2020jastrow}, defined as $k$ sequential applications of the uCJ circuit, does not fully saturate this support. As a result, it provides a controlled and systematic means of restricting the accessible portion of Hilbert space, making it a particularly attractive starting point for reducing the effective search space in GAS-SCF.

Finally, another promising direction for GAS-SCF is to retain the exponential search space but improve the quadratic speedup to higher-order scaling (e.g., cubic or quartic). While the asymptotic runtime would remain exponential, such improvements could nevertheless provide a realizable quantum speedup for the combinatorial search problem. In this context, quantum speedups for branch-and-bound algorithms, such as the approach proposed by Montanaro in \cite{montanaro2020quantum}, provide a natural and compelling starting point for further exploration.

\section{Methodology} \label{sec:methodology}
The \texttt{PySCF} package \cite{sun2020recent} was used to obtain the molecular integrals needed to construct the Hamiltonian (Equation~\ref{eq:qubo_full}) for linear \ce{H3-} and \ce{LiH} in the STO-3G basis set \cite{hehre1969self} and \ce{OH-} in the 6-31G basis set \cite{hehre1972self} in a (6o, (3e,3e)) active space. The bond lengths studied were $1.0$ {\AA}, $1.5$ {\AA} and $3.0$ {\AA} respectively. A restricted Hartree-Fock (RHF) calculation was performed for each problem. All Hamiltonian operators were then written in the canonical basis, where their coefficients were multiplied by a constant then rounded to integer values in the corresponding QUBO problem. We  verified that this method preserved the ground eigenstate. We supply both the original and rounded Hamiltonian data in the online repository \cite{gas_github}. The first solution for the \ce{OH-} RHF calculation was intentionally chosen as it gives an example of SCF converging onto a local minimum.

These simulations were performed using the statevector method and no actual quantum computers were employed. The python package \texttt{symmer} \cite{symmerGithub} was used to build each chemical Hamiltonian and \texttt{Qiskit} \cite{qiskit2024} was used to simulate the quantum circuits. All the results in this paper can be reproduced by the code provided in the repository \cite{gas_github}. The quantum circuits to generate Dicke states were based on the circuits derived by B{\"a}rtschi and Eidenbenz \cite{bartschi2019deterministic}, using code modified from \cite{dicke_github}.

The results for triplet \ce{O2} at a bond length of 2.55 {\AA} were also obtained using \texttt{PySCF} and \texttt{symmer}  and can be found on the online repository \cite{gas_github}. An internal stability calculation for this system was performed to ensure that the restricted open-shell Hartree-Fock (ROHF) reference was not a saddle point in both basis sets. 

The ROHF and second-order ROHF (Newton method in PySCF) results for linear triplet \ce{O3} (O$-$O bond lengths each set to 2 {\AA}) were also computed using the cc-pVDZ, cc-pVTZ, and cc-pVQZ basis sets \cite{dunning1989gaussian} with different initialization strategies and are provided in the online repository \cite{gas_github}. The maximum number of SCF iterations was set to $500$ for each calculation. An additional single internal stability check was then performed. We report the internal stability results but do not restart the SCF if an internal instability was found. A summary of the energies and convergence data is provided in the Supporting Information (Section \ref{sec:O3_data}). The improved solutions were found by performing a MaxCut calculation (in the same MO basis) \cite{ralli2025bridging} followed by brute-force search in the appropriate number symmetry sector, for a 1-minute time limit. We confirmed that all the improved solutions had the correct $S^{2}=2$ value.

No point-group symmetries were imposed for any calculations (including for the PySCF results).

\section{Conclusion} \label{sec:conclusion}

In this paper, we defined the Grover Adaptive Search Self-Consistent Field (GAS-SCF) routine for SCF chemistry algorithms. We presented complete circuit constructions for two valid implementations: one including a symmetry-flagging step to ensure only solutions within the correct symmetry sector are marked and subsequently amplified, and another which initializes the circuit in a Dicke state and thus restricts us to valid solutions from the beginning. 

To investigate the performance of this algorithm, we conducted statevector simulations for several molecular problems. The \ce{OH-} example, in particular, highlights that the GAS-SCF routine can outperform classical simulation methods: our statevector-based quantum circuit simulation produces a result superior to that of a conventional chemistry package, which converges to a local minimum. Although the classical result could be improved due to the small system size, scaling to larger and more complex problems remains a significant challenge for classical approaches. This scalability limitation underscores the potential for quantum improvements for SCF problems. However, due to the substantial circuit complexity associated with implementing GAS-SCF, large-scale execution of the algorithm remains beyond the capabilities of current quantum hardware.

We then examined some of the criteria required for this algorithm to exhibit quantum advantage, motivating our discussion through a detailed study of triplet \ce{O2} and \ce{O3} (linear) - the largest problem considered was $330$ qubits. These systems, together with other molecular problems exhibiting SCF convergence challenges, form a promising class of candidate optimization problems. Improvements in approximation ratios for these problems would make them quantum advantage candidates for GAS-SCF, and for other quantum approximation algorithms such as AQC, QAOA and DQI. However, benchmarking against classical heuristics is essential before any claim of quantum advantage can be made. 

With respect to quantum speedup, it is important to note that classical SCF solvers rely on heuristic methods, most commonly iterative Fock matrix diagonalization or second-order orbital optimization, rather than exhaustive search. As a result, the quadratic speedup of GAS-SCF (defined relative to brute force) is not comparable to the performance of these classical heuristics. However, it is not an advantage in runtime that could be achieved but an improvement in quality of solution. Solution quality remains an open question for these classical heuristics, as seen here and in approaches such as MaxCut-SCF \cite{ralli2025bridging} and GAS-SCF establishes a useful baseline for future quantum improvements of SCF methods.

\section*{Data and Software Availability} 
The data underlying this study are available in the published article, the Supporting Information, and at \cite{ZenodoGas2026}. The scripts and code used to generate and analyze the results are also available at this link.

\section*{Acknowledgments} \label{sec:acknowledgements}
A.R. and T.W. are supported by QMatter, Inc. T.M.B. is supported by the Engineering and Physical Sciences Research Council (grant numbers EP/T517793/1, EP/W524335/1). P.V.C. is grateful for funding from the European Commission for VECMA (800925) and EPSRC for SEAVEA (EP/W007711/1). P.J.L. acknowledge support by the NSF STAQ project (PHY-1818914/232580) and by the NSF NQVL:QSTD:Pilot: Quantum Advantage-Class Trapped Ion system (QACTI) project NSF award number 2410675.

\section*{Supporting Information} 
The Supporting Information covers quantum arithmetic in the Fourier basis (Section \ref{sec:Fourier_add}), Two's complement encoding (Section \ref{sec:twosC}), GAS-SCF quantum circuits (Section \ref{sec:qcircuitsfull}), results for \ce{H3-}, \ce{O3}, and the Chromium dimer (Sections \ref{sec:H3_data}, \ref{sec:O3_data}, and \ref{sec:Cr2res}), an alternative quantum circuit compilation strategy (Section \ref{sec:FT_comp}), and the enforcement of point-group and $S^2$ symmetries in GAS-SCF (Section \ref{sec:mol_syms}).

% \clearpage
% \bibliographystyle{apsrev4-1}

% \bibliography{references}
% \bibliographystyle{apsrev4-2}
\bibliography{references}

@article{babbush2018low,
  title={Low-depth quantum simulation of materials},
  author={Babbush, Ryan and Wiebe, Nathan and McClean, Jarrod and McClain, James and Neven, Hartmut and Chan, Garnet Kin-Lic},
  journal={Physical Review X},
  volume={8},
  number={1},
  pages={011044},
  year={2018},
  publisher={APS},
  doi={https://doi.org/10.1103/PhysRevX.8.011044}
}

@article{babbush2018encoding,
  title={Encoding electronic spectra in quantum circuits with linear T complexity},
  author={Babbush, Ryan and Gidney, Craig and Berry, Dominic W and Wiebe, Nathan and McClean, Jarrod and Paler, Alexandru and Fowler, Austin and Neven, Hartmut},
  journal={Physical Review X},
  volume={8},
  number={4},
  pages={041015},
  year={2018},
  publisher={APS},
  doi={https://doi.org/10.1103/PhysRevX.8.041015}
}

@article{kivlichan2018quantum,
  title={Quantum simulation of electronic structure with linear depth and connectivity},
  author={Kivlichan, Ian D and McClean, Jarrod and Wiebe, Nathan and Gidney, Craig and Aspuru-Guzik, Al{\'a}n and Chan, Garnet Kin-Lic and Babbush, Ryan},
  journal={Physical review letters},
  volume={120},
  number={11},
  pages={110501},
  year={2018},
  publisher={APS},
  doi={https://doi.org/10.1103/PhysRevLett.120.110501}
}

@article{babbush2019quantum,
  title={Quantum simulation of chemistry with sublinear scaling in basis size},
  author={Babbush, Ryan and Berry, Dominic W and McClean, Jarrod R and Neven, Hartmut},
  journal={npj Quantum Information},
  volume={5},
  number={1},
  pages={92},
  year={2019},
  publisher={Nature Publishing Group UK London},
  doi={https://doi.org/10.1038/s41534-019-0199-y}
}

@article{lee2021even,
  title={Even more efficient quantum computations of chemistry through tensor hypercontraction},
  author={Lee, Joonho and Berry, Dominic W and Gidney, Craig and Huggins, William J and McClean, Jarrod R and Wiebe, Nathan and Babbush, Ryan},
  journal={PRX quantum},
  volume={2},
  number={3},
  pages={030305},
  year={2021},
  publisher={APS},
  doi={https://doi.org/10.1103/PRXQuantum.2.030305}
}

@article{su2021fault,
  title={Fault-tolerant quantum simulations of chemistry in first quantization},
  author={Su, Yuan and Berry, Dominic W and Wiebe, Nathan and Rubin, Nicholas and Babbush, Ryan},
  journal={PRX Quantum},
  volume={2},
  number={4},
  pages={040332},
  year={2021},
  publisher={APS},
  doi={https://doi.org/10.1103/PRXQuantum.2.040332}
}

@article{berry2025rapid,
  title={Rapid initial-state preparation for the quantum simulation of strongly correlated molecules},
  author={Berry, Dominic W and Tong, Yu and Khattar, Tanuj and White, Alec and Kim, Tae In and Low, Guang Hao and Boixo, Sergio and Ding, Zhiyan and Lin, Lin and Lee, Seunghoon and others},
  journal={PRX Quantum},
  volume={6},
  number={2},
  pages={020327},
  year={2025},
  publisher={APS},
  doi={https://doi.org/10.1103/PRXQuantum.6.020327}
}

@article{low2025fast,
  title={Fast quantum simulation of electronic structure by spectral amplification},
  author={Low, Guang Hao and King, Robbie and Berry, Dominic W and Han, Qiushi and DePrince III, A Eugene and White, Alec F and Babbush, Ryan and Somma, Rolando D and Rubin, Nicholas C},
  journal={Physical Review X},
  volume={15},
  number={4},
  pages={041016},
  year={2025},
  publisher={APS},
  doi={https://doi.org/10.1103/pb2g-j9cw}
}

@article{jordan2025optimization,
  title={Optimization by decoded quantum interferometry},
  author={Jordan, Stephen P and Shutty, Noah and Wootters, Mary and Zalcman, Adam and Schmidhuber, Alexander and King, Robbie and Isakov, Sergei V and Khattar, Tanuj and Babbush, Ryan},
  journal={Nature},
  volume={646},
  number={8086},
  pages={831--836},
  year={2025},
  publisher={Nature Publishing Group UK London},
  doi={https://doi.org/10.1038/s41586-025-09527-5}
}

@article{cain2026shor,
  title={Shor's algorithm is possible with as few as 10,000 reconfigurable atomic qubits},
  author={Cain, Madelyn and Xu, Qian and King, Robbie and Picard, Lewis RB and Levine, Harry and Endres, Manuel and Preskill, John and Huang, Hsin-Yuan and Bluvstein, Dolev},
  journal={arXiv preprint arXiv:2603.28627},
  year={2026},
  url={https://arxiv.org/abs/2603.28627}, 
}

@article{farhi2001quantum,
  title={A quantum adiabatic evolution algorithm applied to random instances of an NP-complete problem},
  author={Farhi, Edward and Goldstone, Jeffrey and Gutmann, Sam and Lapan, Joshua and Lundgren, Andrew and Preda, Daniel},
  journal={Science},
  volume={292},
  number={5516},
  pages={472--475},
  year={2001},
  publisher={American Association for the Advancement of Science}
}

@article{Hartree_1928, title={The Wave Mechanics of an Atom with a Non-Coulomb Central Field. Part I. Theory and Methods}, volume={24}, DOI={10.1017/S0305004100011919}, number={1}, journal={Mathematical Proceedings of the Cambridge Philosophical Society}, author={Hartree, D. R.}, year={1928}, pages={89–110}}

@article{santagati2024drug,
  title={Drug design on quantum computers},
  author={Santagati, Raffaele and Aspuru-Guzik, Alan and Babbush, Ryan and Degroote, Matthias and Gonz{\'a}lez, Leticia and Kyoseva, Elica and Moll, Nikolaj and Oppel, Markus and Parrish, Robert M and Rubin, Nicholas C and others},
  journal={Nature Physics},
  volume={20},
  number={4},
  pages={549--557},
  year={2024},
  publisher={Nature Publishing Group UK London},
 doi={https://doi.org/10.1038/s41567-024-02411-5}
}

@article{lee2023evaluating,
  title={Evaluating the evidence for exponential quantum advantage in ground-state quantum chemistry},
  author={Lee, Seunghoon and Lee, Joonho and Zhai, Huanchen and Tong, Yu and Dalzell, Alexander M and Kumar, Ashutosh and Helms, Phillip and Gray, Johnnie and Cui, Zhi-Hao and Liu, Wenyuan and others},
  journal={Nature communications},
  volume={14},
  number={1},
  pages={1952},
  year={2023},
  publisher={Nature Publishing Group UK London},
  doi={https://doi.org/10.1038/s41467-023-37587-6}
}

@article{babbush2023quantum,
  title={Quantum simulation of exact electron dynamics can be more efficient than classical mean-field methods},
  author={Babbush, Ryan and Huggins, William J and Berry, Dominic W and Ung, Shu Fay and Zhao, Andrew and Reichman, David R and Neven, Hartmut and Baczewski, Andrew D and Lee, Joonho},
  journal={Nature Communications},
  volume={14},
  number={1},
  pages={4058},
  year={2023},
  publisher={Nature Publishing Group UK London},
  doi={https://doi.org/10.1038/s41467-023-39024-0}
}

@article{chen2025framework,
  title={A framework for robust quantum speedups in practical correlated electronic structure and dynamics},
  author={Chen, Jielun and Chan, Garnet Kin},
  journal={arXiv preprint arXiv:2508.15765},
  year={2025},
  url={https://arxiv.org/abs/2508.15765}, 
}

@article{hoefler2023disentangling,
  title={Disentangling hype from practicality: On realistically achieving quantum advantage},
  author={Hoefler, Torsten and H{\"a}ner, Thomas and Troyer, Matthias},
  journal={Communications of the ACM},
  volume={66},
  number={5},
  pages={82--87},
  year={2023},
  publisher={ACM New York, NY, USA},
  doi={https://doi.org/10.1145/3571725}
}

@article{google2020hartree,
  title={Hartree-Fock on a superconducting qubit quantum computer},
  author={Google AI Quantum and Collaborators*† and Arute, Frank and Arya, Kunal and Babbush, Ryan and Bacon, Dave and Bardin, Joseph C and Barends, Rami and Boixo, Sergio and Broughton, Michael and Buckley, Bob B and others},
  journal={Science},
  volume={369},
  number={6507},
  pages={1084--1089},
  year={2020},
  publisher={American Association for the Advancement of Science},
  doi={https://doi.org/10.1126/science.abb9811}
}

@article{gilliam2021grover,
  title={Grover adaptive search for constrained polynomial binary optimization},
  author={Gilliam, Austin and Woerner, Stefan and Gonciulea, Constantin},
  journal={Quantum},
  volume={5},
  pages={428},
  year={2021},
  doi={https://doi.org/10.22331/q-2021-04-08-428}
}

@book{helgaker2013molecular,
  title={Molecular electronic-structure theory},
  author={Helgaker, Trygve and Jorgensen, Poul and Olsen, Jeppe},
  year={2013},
  publisher={John Wiley \& Sons},
  isbn={9781118531471}
}

@article{farhi2025lower,
  title={Lower bounding the MaxCut of high girth 3-regular graphs using the QAOA},
  author={Farhi, Edward and Gutmann, Sam and Ranard, Daniel and Villalonga, Benjamin},
  journal={arXiv preprint arXiv:2503.12789},
  year={2025}
}

@inproceedings{goemans1994879,
  title={. 879-approximation algorithms for max cut and max 2sat},
  author={Goemans, Michel X and Williamson, David P},
  booktitle={Proceedings of the twenty-sixth annual ACM symposium on Theory of computing},
  pages={422--431},
  year={1994},
  doi={https://doi.org/10.1145/195058.195216}
}

@article{vaucher2017steering,
  title={Steering orbital optimization out of local minima and saddle points toward lower energy},
  author={Vaucher, Alain C and Reiher, Markus},
  journal={Journal of Chemical Theory and Computation},
  volume={13},
  number={3},
  pages={1219--1228},
  year={2017},
  publisher={ACS Publications},
  doi={https://doi.org/10.1021/acs.jctc.7b00011}
}

@article{lehtola2019assessment,
  title={Assessment of initial guesses for self-consistent field calculations. Superposition of atomic potentials: Simple yet efficient},
  author={Lehtola, Susi},
  journal={Journal of chemical theory and computation},
  volume={15},
  number={3},
  pages={1593--1604},
  year={2019},
  publisher={ACS Publications},
  doi={https://doi.org/10.1021/acs.jctc.8b01089}
}

@article{seeger1977self,
  title={Self-consistent molecular orbital methods. XVIII. Constraints and stability in Hartree--Fock theory},
  author={Seeger, Rolf and Pople, John A},
  journal={The Journal of Chemical Physics},
  volume={66},
  number={7},
  pages={3045--3050},
  year={1977},
  publisher={AIP Publishing},
  doi={https://doi.org/10.1063/1.434318}
}

@article{van2006starting,
  title={Starting SCF calculations by superposition of atomic densities},
  author={Van Lenthe, JH and Zwaans, Renate and Van Dam, Huub JJ and Guest, MF},
  journal={Journal of computational chemistry},
  volume={27},
  number={8},
  pages={926--932},
  year={2006},
  publisher={Wiley Online Library},
  doi={https://doi.org/10.1002/jcc.20393}
}

@article{sun2020recent,
  title={Recent developments in the PySCF program package},
  author={Sun, Qiming and Zhang, Xing and Banerjee, Samragni and Bao, Peng and Barbry, Marc and Blunt, Nick S and Bogdanov, Nikolay A and Booth, George H and Chen, Jia and Cui, Zhi-Hao and others},
  journal={The Journal of chemical physics},
  volume={153},
  number={2},
  year={2020},
  publisher={AIP Publishing},
  doi={https://doi.org/10.1063/5.0006074}
}

@misc{qiskit2024,
      title={Quantum computing with {Q}iskit},
      author={Javadi-Abhari, Ali and Treinish, Matthew and Krsulich, Kevin and Wood, Christopher J. and Lishman, Jake and Gacon, Julien and Martiel, Simon and Nation, Paul D. and Bishop, Lev S. and Cross, Andrew W. and Johnson, Blake R. and Gambetta, Jay M.},
      year={2024},
      doi={10.48550/arXiv.2405.08810},
      eprint={2405.08810},
      archivePrefix={arXiv},
      primaryClass={quant-ph}
}

@article{whitfield2014np,
  title={On the NP-completeness of the Hartree-Fock method for translationally invariant systems},
  author={Whitfield, James Daniel and Zimbor{\'a}s, Zolt{\'a}n},
  journal={The Journal of chemical physics},
  volume={141},
  number={23},
  year={2014},
  publisher={AIP Publishing},
  doi={https://doi.org/10.1063/1.4903453}
}

@book{szabo2012modern,
  title     = "Modern Quantum Chemistry: Introduction to Advanced Electronic Structure Theory",
  author    = "Szabo, Attila and Ostlund, Neil S",
  publisher = "Dover Publications",
  series    = "Dover Books on Chemistry",
  year      =  1996,
  address   = "Mineola, NY",
  isbn ={0486691861}
}

@article{yen2019exact,
  title={Exact and approximate symmetry projectors for the electronic structure problem on a quantum computer},
  author={Yen, Tzu-Ching and Lang, Robert A and Izmaylov, Artur F},
  journal={The Journal of chemical physics},
  volume={151},
  number={16},
  year={2019},
  publisher={AIP Publishing},
  doi={https://doi.org/10.1063/1.5110682}
}

@misc{farhi2014quantum,
      title={A Quantum Approximate Optimization Algorithm}, 
      author={Edward Farhi and Jeffrey Goldstone and Sam Gutmann},
      year={2014},
      eprint={1411.4028},
      archivePrefix={arXiv},
      primaryClass={quant-ph},
      url={https://arxiv.org/abs/1411.4028}, 
}

@article{baritompa2005grover,
  title={Grover's quantum algorithm applied to global optimization},
  author={Baritompa, William P and Bulger, David W and Wood, Graham R},
  journal={SIAM Journal on Optimization},
  volume={15},
  number={4},
  pages={1170--1184},
  year={2005},
  publisher={SIAM},
  doi={https://doi.org/10.1137/040605072}
}

@article{bulger2003implementing,
  title={Implementing pure adaptive search with Grover's quantum algorithm},
  author={Bulger, David and Baritompa, William P and Wood, Graham R},
  journal={Journal of optimization theory and applications},
  volume={116},
  pages={517--529},
  year={2003},
  publisher={Springer},
  doi={https://doi.org/10.1023/A:1023061218864}
}

@article{cao2019quantum,
  title={Quantum chemistry in the age of quantum computing},
  author={Cao, Yudong and Romero, Jonathan and Olson, Jonathan P and Degroote, Matthias and Johnson, Peter D and Kieferov{\'a}, M{\'a}ria and Kivlichan, Ian D and Menke, Tim and Peropadre, Borja and Sawaya, Nicolas PD and others},
  journal={Chemical reviews},
  volume={119},
  number={19},
  pages={10856--10915},
  year={2019},
  publisher={ACS Publications},
  doi={10.1021/acs.chemrev.8b00803 }
}

@article{aspuru2005simulated,
  title={Simulated quantum computation of molecular energies},
  author={Aspuru-Guzik, Al{\'a}n and Dutoi, Anthony D and Love, Peter J and Head-Gordon, Martin},
  journal={Science},
  volume={309},
  number={5741},
  pages={1704--1707},
  year={2005},
  publisher={American Association for the Advancement of Science},
  doi={10.1126/science.1113479}
}

@article{hehre1972self,
  title={Self—consistent molecular orbital methods. XII. Further extensions of Gaussian—type basis sets for use in molecular orbital studies of organic molecules},
  author={Hehre, Warren J and Ditchfield, Robert and Pople, John A},
  journal={The Journal of Chemical Physics},
  volume={56},
  number={5},
  pages={2257--2261},
  year={1972},
  publisher={American Institute of Physics},
  doi={https://doi.org/10.1063/1.1677527}
}

@inproceedings{grover1996fast,
author = {Grover, Lov K.},
title = {A fast quantum mechanical algorithm for database search},
year = {1996},
isbn = {0897917855},
publisher = {Association for Computing Machinery},
address = {New York, NY, USA},
url = {https://doi.org/10.1145/237814.237866},
doi = {10.1145/237814.237866},
booktitle = {Proceedings of the Twenty-Eighth Annual ACM Symposium on Theory of Computing},
pages = {212–219},
numpages = {8},
location = {Philadelphia, Pennsylvania, USA},
series = {STOC '96}
}

@article{dunning1989gaussian,
  title={Gaussian basis sets for use in correlated molecular calculations. I. The atoms boron through neon and hydrogen},
  author={Dunning Jr, Thom H},
  journal={The Journal of chemical physics},
  volume={90},
  number={2},
  pages={1007--1023},
  year={1989},
  publisher={American Institute of Physics},
  doi={https://doi.org/10.1063/1.456153}
}

@article{peruzzo2014variational,
  title={A variational eigenvalue solver on a photonic quantum processor},
  author={Peruzzo, Alberto and McClean, Jarrod and Shadbolt, Peter and Yung, Man-Hong and Zhou, Xiao-Qi and Love, Peter J and Aspuru-Guzik, Al{\'a}n and O’brien, Jeremy L},
  journal={Nature communications},
  volume={5},
  number={1},
  pages={4213},
  year={2014},
  publisher={Nature Publishing Group UK London},
  doi={https://doi.org/10.1038/ncomms5213}
}

@misc{symmerGithub,
  author = {Ralli, Alexis and Weaving, Tim J},
  doi = {},
  title = {symmer},
  howpublished= {\url{https://github.com/qmatter-labs/symmer}},
  year = {2025}
}

@misc{dicke_github,
  author = {Juan, Andr{\'e}},
  doi = {},
  title = {{D}icke states preparation},
  howpublished= {\url{https://github.com/andre-juan/dicke_states_preparation/}},
  year = {2025}
}

@misc{gas_github,
  author = {Ralli, Alexis and Bickley, Tom},
  doi = {},
  title = {{{G}rover {A}daptive {S}earch}},
  howpublished= {\url{https://github.com/qmatter-labs/GAS}},
  year = {2025}
}

@misc{draper2000additionquantumcomputer,
      title={Addition on a Quantum Computer}, 
      author={Thomas G. Draper},
      year={2000},
      eprint={quant-ph/0008033},
      archivePrefix={arXiv},
      primaryClass={quant-ph},
      url={https://arxiv.org/abs/quant-ph/0008033}, 
}

@article{norimoto2023quantum,
  author={Norimoto, Masaya and Mori, Ryuhei and Ishikawa, Naoki},
  journal={IEEE Transactions on Communications}, 
  title={Quantum Algorithm for Higher-Order Unconstrained Binary Optimization and MIMO Maximum Likelihood Detection}, 
  year={2023},
  volume={71},
  number={4},
  pages={1926-1939},
  doi={10.1109/TCOMM.2023.3244924}}

@article{ralli2025bridging,
author = {Ralli, Alexis and Weaving, Tim and Coveney, Peter V. and Love, Peter J.},
title = {Bridging Quantum Chemistry and MaxCut: Classical Performance Guarantees and Quantum Algorithms for the Hartree–Fock Method},
journal = {Journal of Chemical Theory and Computation},
volume = {21},
number = {19},
pages = {9511-9524},
year = {2025},
doi = {10.1021/acs.jctc.5c00948},
    note ={PMID: 40985214},
}

@misc{durr1999quantumalgorithmfindingminimum,
      title={A Quantum Algorithm for Finding the Minimum}, 
      author={Christoph Durr and Peter Hoyer},
      year={1999},
      eprint={quant-ph/9607014},
      archivePrefix={arXiv},
      primaryClass={quant-ph},
      url={https://arxiv.org/abs/quant-ph/9607014}, 
}

@misc{greiner2025reusablelibrarysecondorderorbital,
      title={A Reusable Library for Second-Order Orbital Optimization Using the Trust Region Method}, 
      author={Jonas Greiner and Ida-Marie Høyvik and Susi Lehtola and Janus J. Eriksen},
      year={2025},
      eprint={2509.13931},
      archivePrefix={arXiv},
      primaryClass={physics.chem-ph},
      url={https://arxiv.org/abs/2509.13931}, 
}

@article{ominato2024grover,
  author={Ominato, Hiroaki and Ohyama, Takahiro and Yamaguchi, Koichiro},
  journal={IEEE Access}, 
  title={Grover Adaptive Search With Fewer Queries}, 
  year={2024},
  volume={12},
  number={},
  pages={74619-74632},
  doi={10.1109/ACCESS.2024.3403200}}

@InProceedings{bartschi2019deterministic,
author="B{\"a}rtschi, Andreas
and Eidenbenz, Stephan",
editor="G{\k{a}}sieniec, Leszek Antoni
and Jansson, Jesper
and Levcopoulos, Christos",
title="Deterministic Preparation of Dicke States",
booktitle="Fundamentals of Computation Theory",
year="2019",
publisher="Springer International Publishing",
address="Cham",
pages="126--139",
isbn="978-3-030-25027-0",
doi={https://doi.org/10.1007/978-3-030-25027-0_9},
url={https://doi.org/10.1007/978-3-030-25027-0_9}
}

@article{
brassard1997searching,
author = {Gilles Brassard },
title = {Searching a Quantum Phone Book},
journal = {Science},
volume = {275},
number = {5300},
pages = {627-628},
year = {1997},
doi = {10.1126/science.275.5300.627},
}

@inproceedings{faro2025scaling,
title = {Scaling Grover’s Search for Large Solution Spaces},
author = {Faro, Simone and Marino, Francesco Pio},
booktitle={Proceedings of the 34th International Symposium on High-Performance Parallel and Distributed Computing},
year = {2025},
isbn = {9798400718694},
publisher = {Association for Computing Machinery},
address = {New York, NY, USA},
doi = {10.1145/3731545.3744149},
url = {https://doi.org/10.1145/3731545.3744149},
articleno = {43},
numpages = {8},
location = {University of Notre Dame Conference Facilities, Notre Dame, IN, USA},
series = {HPDC '25}
}

@article{hehre1969self,
  title={Self-consistent molecular-orbital methods. I. Use of Gaussian expansions of Slater-type atomic orbitals},
  author={Hehre, Warren J and Stewart, Robert F and Pople, John A},
  journal={The Journal of Chemical Physics},
  volume={51},
  number={6},
  pages={2657--2664},
  year={1969},
  publisher={American Institute of Physics},
 doi={
https://doi.org/10.1063/1.1672392
}
}

@INPROCEEDINGS{bartschi2022short,
  author={Bärtschi, Andreas and Eidenbenz, Stephan},
  booktitle={2022 IEEE International Conference on Quantum Computing and Engineering (QCE)}, 
  title={Short-Depth Circuits for Dicke State Preparation}, 
  year={2022},
  volume={},
  number={},
  pages={87-96},
  doi={10.1109/QCE53715.2022.00027}}

@article{Gidney2018halvingcostof,
  doi = {10.22331/q-2018-06-18-74},
  url = {https://doi.org/10.22331/q-2018-06-18-74},
  title = {Halving the cost of quantum addition},
  author = {Gidney, Craig},
  journal = {{Quantum}},
  issn = {2521-327X},
  publisher = {{Verein zur F{\"{o}}rderung des Open Access Publizierens in den Quantenwissenschaften}},
  volume = {2},
  pages = {74},
  month = jun,
  year = {2018}
}

@article{ORTS2020102810,
title = {A review on reversible quantum adders},
journal = {Journal of Network and Computer Applications},
volume = {170},
pages = {102810},
year = {2020},
issn = {1084-8045},
doi = {https://doi.org/10.1016/j.jnca.2020.102810},
url = {https://www.sciencedirect.com/science/article/pii/S1084804520302812},
author = {F. Orts and G. Ortega and E.F. Combarro and E.M. Garzón},
}

@article{sun2015libcint,
author = {Sun, Qiming},
title = {Libcint: An efficient general integral library for Gaussian basis functions},
journal = {Journal of Computational Chemistry},
volume = {36},
number = {22},
pages = {1664-1671},
doi = {https://doi.org/10.1002/jcc.23981},
url = {https://onlinelibrary.wiley.com/doi/abs/10.1002/jcc.23981},
eprint = {https://onlinelibrary.wiley.com/doi/pdf/10.1002/jcc.23981},
year = {2015}
}

@misc{Brassard_2002,
   title={Quantum amplitude amplification and estimation},
   ISSN={0271-4132},
   url={http://dx.doi.org/10.1090/conm/305/05215},
   DOI={10.1090/conm/305/05215},
   journal={Quantum Computation and Information},
   publisher={American Mathematical Society},
   author={Brassard, Gilles and Høyer, Peter and Mosca, Michele and Tapp, Alain},
   year={2002},
   pages={53–74} }

@misc{kitaev1995quantummeasurementsabelianstabilizer,
      title={Quantum measurements and the Abelian Stabilizer Problem}, 
      author={A. Yu. Kitaev},
      year={1995},
      eprint={quant-ph/9511026},
      archivePrefix={arXiv},
      primaryClass={quant-ph},
      url={https://arxiv.org/abs/quant-ph/9511026}, 
}

@article{Chuang2014FixedPoint,
  title = {Fixed-Point Quantum Search with an Optimal Number of Queries},
  author = {Yoder, Theodore J. and Low, Guang Hao and Chuang, Isaac L.},
  journal = {Phys. Rev. Lett.},
  volume = {113},
  issue = {21},
  pages = {210501},
  numpages = {5},
  year = {2014},
  month = {Nov},
  publisher = {American Physical Society},
  doi = {10.1103/PhysRevLett.113.210501},
  url = {https://link.aps.org/doi/10.1103/PhysRevLett.113.210501}
}

@article{Grover2005Fixedpoint,
  title = {Fixed-Point Quantum Search},
  author = {Grover, Lov K.},
  journal = {Phys. Rev. Lett.},
  volume = {95},
  issue = {15},
  pages = {150501},
  numpages = {4},
  year = {2005},
  month = {Oct},
  publisher = {American Physical Society},
  doi = {10.1103/PhysRevLett.95.150501},
  url = {https://link.aps.org/doi/10.1103/PhysRevLett.95.150501}
}

@article{daniels2000converging,
author ="Daniels, Andrew D. and Scuseria, Gustavo E.",
title  ="Converging difficult SCF cases with conjugate gradient density matrix search",
journal  ="Phys. Chem. Chem. Phys.",
year  ="2000",
volume  ="2",
issue  ="10",
pages  ="2173-2176",
publisher  ="The Royal Society of Chemistry",
doi  ="10.1039/B000618L"
}

@article{babbush2021focus,
  title = {Focus beyond Quadratic Speedups for Error-Corrected Quantum Advantage},
  author = {Babbush, Ryan and McClean, Jarrod R. and Newman, Michael and Gidney, Craig and Boixo, Sergio and Neven, Hartmut},
  journal = {PRX Quantum},
  volume = {2},
  issue = {1},
  pages = {010103},
  numpages = {11},
  year = {2021},
  month = {Mar},
  publisher = {American Physical Society},
  doi = {10.1103/PRXQuantum.2.010103},
  url = {https://link.aps.org/doi/10.1103/PRXQuantum.2.010103}
}

@article{Campbell2019applyingquantum,
  doi = {10.22331/q-2019-07-18-167},
  url = {https://doi.org/10.22331/q-2019-07-18-167},
  title = {Applying quantum algorithms to constraint satisfaction problems},
  author = {Campbell, Earl and Khurana, Ankur and Montanaro, Ashley},
  journal = {{Quantum}},
  issn = {2521-327X},
  publisher = {{Verein zur F{\"{o}}rderung des Open Access Publizierens in den Quantenwissenschaften}},
  volume = {3},
  pages = {167},
  month = jul,
  year = {2019}
}

@article{sanders2020compilation,
  title = {Compilation of Fault-Tolerant Quantum Heuristics for Combinatorial Optimization},
  author = {Sanders, Yuval R. and Berry, Dominic W. and Costa, Pedro C.S. and Tessler, Louis W. and Wiebe, Nathan and Gidney, Craig and Neven, Hartmut and Babbush, Ryan},
  journal = {PRX Quantum},
  volume = {1},
  issue = {2},
  pages = {020312},
  numpages = {70},
  year = {2020},
  month = {Nov},
  publisher = {American Physical Society},
  doi = {10.1103/PRXQuantum.1.020312},
  url = {https://link.aps.org/doi/10.1103/PRXQuantum.1.020312}
}

@book{nielsen2010quantum,
  title={Quantum computation and quantum information},
  author={Nielsen, Michael A and Chuang, Isaac L},
  year={2010},
  publisher={Cambridge university press}
}

@article{boyer1998tight,
  title={Tight bounds on quantum searching},
  author={Boyer, Michel and Brassard, Gilles and H{\o}yer, Peter and Tapp, Alain},
  journal={Fortschritte der Physik: Progress of Physics},
  volume={46},
  number={4-5},
  pages={493--505},
  year={1998},
  publisher={Wiley Online Library},
 doi={https://doi.org/10.1002/(SICI)1521-3978(199806)46:4/5%3C493::AID-PROP493%3E3.0.CO;2-P}
}

@article{cremer2013configuration,
author = {Cremer, Dieter},
title = {From configuration interaction to coupled cluster theory: The quadratic configuration interaction approach},
journal = {WIREs Computational Molecular Science},
volume = {3},
number = {5},
pages = {482-503},
doi = {https://doi.org/10.1002/wcms.1131},
year = {2013}
}

@article{Kudin_Scuseria_2007,
title={Converging self-consistent field equations in quantum chemistry – recent achievements and remaining challenges}, volume={41}, 
DOI={10.1051/m2an:2007022}, 
number={2}, 
journal={ESAIM: Mathematical Modelling and Numerical Analysis}, author={Kudin, Konstantin N. and Scuseria, Gustavo E.}, year={2007}, 
pages={281–296}
}

@article{schultz2005databases,
author = {Schultz, Nathan E. and Zhao, Yan and Truhlar, Donald G.},
title = {Databases for Transition Element Bonding: Metal−Metal Bond Energies and Bond Lengths and Their Use To Test Hybrid, Hybrid Meta, and Meta Density Functionals and Generalized Gradient Approximations},
journal = {The Journal of Physical Chemistry A},
volume = {109},
number = {19},
pages = {4388-4403},
year = {2005},
doi = {10.1021/jp0504468},
}

@article{rubin2022compressing,
author = {Rubin, Nicholas C. and Lee, Joonho and Babbush, Ryan},
title = {Compressing Many-Body Fermion Operators under Unitary Constraints},
journal = {Journal of Chemical Theory and Computation},
volume = {18},
number = {3},
pages = {1480-1488},
year = {2022},
doi = {10.1021/acs.jctc.1c00912},
}

@article{slater1929theory,
  title = {The Theory of Complex Spectra},
  author = {Slater, J. C.},
  journal = {Phys. Rev.},
  volume = {34},
  issue = {10},
  pages = {1293-1322},
  numpages = {0},
  year = {1929},
  month = {Nov},
  publisher = {American Physical Society},
  doi = {10.1103/PhysRev.34.1293},
}

@article{condon1930theory,
  title = {The Theory of Complex Spectra},
  author = {Condon, E. U.},
  journal = {Phys. Rev.},
  volume = {36},
  issue = {7},
  pages = {1121--1133},
  numpages = {0},
  year = {1930},
  month = {Oct},
  publisher = {American Physical Society},
  doi = {10.1103/PhysRev.36.1121},
}

@article{szalay2012multiconfiguration,
  title={Multiconfiguration self-consistent field and multireference configuration interaction methods and applications},
  author={Szalay, Peter G and Muller, Thomas and Gidofalvi, Gergely and Lischka, Hans and Shepard, Ron},
  journal={Chemical reviews},
  volume={112},
  number={1},
  pages={108--181},
  year={2012},
  publisher={ACS Publications},
  doi = {10.1021/cr200137a},
}

@inproceedings{barca2020scaling,
author = {Barca, Giuseppe M. J. and Poole, David L. and Vallejo, Jorge L. Galvez and Alkan, Melisa and Bertoni, Colleen and Rendell, Alistair P. and Gordon, Mark S.},
title = {Scaling the hartree-fock matrix build on summit},
year = {2020},
isbn = {9781728199986},
publisher = {IEEE Press},
booktitle = {Proceedings of the International Conference for High Performance Computing, Networking, Storage and Analysis},
articleno = {81},
numpages = {14},
location = {Atlanta, Georgia},
series = {SC '20},
doi={https://doi.org/10.1109/SC41405.2020.00085 }
}

@article{palethorpe2024advanced,
author = {Palethorpe, Elise and Stocks, Ryan and Barca, Giuseppe M. J.},
title = {Advanced Techniques for High-Performance Fock Matrix Construction on GPU Clusters},
journal = {Journal of Chemical Theory and Computation},
volume = {20},
number = {23},
pages = {10424-10442},
year = {2024},
doi = {10.1021/acs.jctc.4c00994},
    note ={PMID: 39586097},
}

@article{THOULESS1960225,
title = {Stability conditions and nuclear rotations in the Hartree-Fock theory},
journal = {Nuclear Physics},
volume = {21},
pages = {225-232},
year = {1960},
issn = {0029-5582},
doi = {https://doi.org/10.1016/0029-5582(60)90048-1},
author = {D.J. Thouless},
}

@article{babbush2014adiabatic,
  title={Adiabatic quantum simulation of quantum chemistry},
  author={Babbush, Ryan and Love, Peter J and Aspuru-Guzik, Al{\'a}n},
  journal={Scientific reports},
  volume={4},
  number={1},
  pages={6603},
  year={2014},
  publisher={Nature Publishing Group UK London},
  doi={https://doi.org/10.1038/srep06603}
}

@article{almlof1982principles,
  title={Principles for a direct SCF approach to LICAO--MOab-initio calculations},
  author={Alml{\"o}f, Jan and F{\ae}gri Jr, Knut and Korsell, Knut},
  journal={Journal of Computational Chemistry},
  volume={3},
  number={3},
  pages={385--399},
  year={1982},
  publisher={Wiley Online Library},
  doi={https://doi.org/10.1002/jcc.540030314}
}

@article{Simon2025ladderoperatorblock,
  doi = {10.22331/q-2025-12-22-1953},
  url = {https://doi.org/10.22331/q-2025-12-22-1953},
  title = {Ladder {O}perator {B}lock-{E}ncoding},
  author = {Simon, William A. and Gustin, Carter M. and Serafin, Kamil and Ralli, Alexis and Goldstein, Gary R. and Love, Peter J.},
  journal = {{Quantum}},
  issn = {2521-327X},
  publisher = {{Verein zur F{\"{o}}rderung des Open Access Publizierens in den Quantenwissenschaften}},
  volume = {9},
  pages = {1953},
  month = dec,
  year = {2025}
}

@article{mazziotti2011large,
  title = {Large-Scale Semidefinite Programming for Many-Electron Quantum Mechanics},
  author = {Mazziotti, David A.},
  journal = {Phys. Rev. Lett.},
  volume = {106},
  issue = {8},
  pages = {083001},
  numpages = {4},
  year = {2011},
  month = {Feb},
  publisher = {American Physical Society},
  doi = {10.1103/PhysRevLett.106.083001},
  url = {https://link.aps.org/doi/10.1103/PhysRevLett.106.083001}
}

@incollection{coleman2007reduced,
  title={Reduced-density-matrix mechanics: with application to many-electron atoms and molecules},
  author={Coleman, A John},
  booktitle={Reduced-Density-Matrix Mechanics: with applications to many-electron atoms and molecules},
  year={2007},
  publisher={Wiley Online Library},
  doi={10.1002/0470106603},
  url = {https://www.doi.org/10.1002/0470106603},
}

@article{mazziotti2020dual,
  title = {Dual-cone variational calculation of the two-electron reduced density matrix},
  author = {Mazziotti, David A.},
  journal = {Phys. Rev. A},
  volume = {102},
  issue = {5},
  pages = {052819},
  numpages = {8},
  year = {2020},
  month = {Nov},
  publisher = {American Physical Society},
  doi = {10.1103/PhysRevA.102.052819},
  url = {https://link.aps.org/doi/10.1103/PhysRevA.102.052819}
}

@misc{zhai2026classicalsolutionfemocofactormodel,
      title={Classical solution of the FeMo-cofactor model to chemical accuracy and its implications}, 
      author={Huanchen Zhai and Chenghan Li and Xing Zhang and Zhendong Li and Seunghoon Lee and Garnet Kin-Lic Chan},
      year={2026},
      eprint={2601.04621},
      archivePrefix={arXiv},
      primaryClass={physics.chem-ph},
      url={https://arxiv.org/abs/2601.04621}, 
}

@article{montanaro2020quantum,
  title={Quantum speedup of branch-and-bound algorithms},
  author={Montanaro, Ashley},
  journal={Physical Review Research},
  volume={2},
  number={1},
  pages={013056},
  year={2020},
  publisher={APS},
  doi={ https://doi.org/10.1103/PhysRevResearch.2.013056}
}

@MISC{ZenodoGas2026,
  title     = "Grover Adaptive Search Self-Consistent Field [Computer software]. Zenodo",
  author    = "Ralli, Alexis",
  publisher = "Zenodo",
  year      =  2026,
  url = "https://doi.org/10.5281/zenodo.21397153",
  doi = "https://doi.org/10.5281/zenodo.21397153"
}

@article{dobrautz2019efficient,
  title={Efficient formulation of full configuration interaction quantum Monte Carlo in a spin eigenbasis via the graphical unitary group approach},
  author={Dobrautz, Werner and Smart, Simon D and Alavi, Ali},
  journal={The Journal of Chemical Physics},
  volume={151},
  number={9},
  year={2019},
  publisher={AIP Publishing},
  doi={https://doi.org/10.1063/1.5108908}
}

\onecolumngrid
\renewcommand{\thefigure}{S\arabic{figure}}
\renewcommand{\theequation}{S\arabic{equation}}
\renewcommand{\thetable}{S\arabic{table}}
\renewcommand{\thesection}{S.\Roman{section}}
\renewcommand{\thesubsection}{S.\arabic{section}\arabic{subsection}}
\renewcommand{\thesubsubsection}{S.\arabic{section}\arabic{subsection}.\arabic{subsubsection}}
\setcounter{section}{0}
\setcounter{figure}{0}
\setcounter{equation}{0}
\clearpage
\section*{Supporting Information}
This document provides the Supporting Information for the main paper. It includes background on quantum arithmetic (Section \ref{sec:Fourier_add}) and two’s complement representation (Section \ref{sec:twosC}), as well as a complete description of the quantum circuit components required to implement GAS-SCF introduced in the main text (Section \ref{sec:qcircuitsfull}). Section \ref{sec:O3_data} provides the raw results for the \ce{O3} result, including convergence information. Finally, Section \ref{sec:FT_comp} provides an alternate compilation for GAS-SCF.

\section{Fourier Basis and Addition} \label{sec:Fourier_add}

In this section, we summarize the work of Draper \cite{draper2000additionquantumcomputer}, who demonstrated how to perform arithmetic operations on a quantum computer using the Fourier basis.

The quantum Fourier transform (QFT) can be written as:

\begin{equation} \label{eq:QFT_eq}
\begin{aligned} 
    \ket{a} &= \ket{a_{1} a_{2} \dots a_{\omega}} \\   &\xleftrightharpoons[QFT]{QFT^{\dagger}} \\
    \frac{1}{\sqrt{2^{\omega}}} \sum_{k=0}^{2^{\omega}-1} e^{2 \pi i  a k/2^{\omega}} \ket{k} &= \frac{1}{\sqrt{2^{\omega}}} \sum_{k=0}^{2^{\omega}-1} e^{2 \pi i  (\sum_{l=1}^{\omega} k_{l}/2^{l})a}  \ket{k_{1} k_{2} \dots k_{\omega}} \\
&= \frac{1}{\sqrt{2^{\omega}}} \sum_{k=0}^{2^{\omega}-1} \prod_{l=1}^{\omega} e^{2 \pi i  ak_{l}/2^{l}}  \ket{k_{1} k_{2} \dots k_{\omega}} \\
    &= \frac{1}{\sqrt{2^{\omega}}} \bigotimes_{l=1}^{\omega}\bigg( \ket{0} + e^{2\pi i a /2^{l}}\ket{1} \bigg) \\
    &= \big( \ket{0} + e^{2\pi i a /2^{1}}\ket{1} \big) \otimes \big( \ket{0} + e^{2\pi i a /2^{2}}\ket{1} \big) \otimes \dots \otimes \big( \ket{0} + e^{2\pi i a /2^{\omega}}\ket{1} \big)
    \\
    &= \ket{\phi_{1}(a)} \otimes \ket{\phi_{2}(a)} \otimes \dots \otimes \ket{\phi_{\omega}(a)},
    \end{aligned}
\end{equation}
% https://quantum.cloud.ibm.com/learning/en/courses/utility-scale-quantum-computing/quantum-phase-estimation
where:

\begin{equation}
    \ket{\phi_{j}(a)} = \frac{1}{\sqrt{2}} \bigg( \ket{0} + e^{2\pi i a /2^{j}}\ket{1} \bigg).
\end{equation}
We observe that in the final two lines of Equation~\ref{eq:QFT_eq}, the output is represented as a tensor product of single-qubit states, which corresponds to a product state \footnote{This holds only for input states that are single computational basis states.}. Each single-qubit state can therefore be prepared using a single quantum phase gate $P(\theta)$:

% \begin{subequations}
% \begin{equation} \label{eq:HGate}
%     H = \frac{1}{\sqrt{2}}\begin{bmatrix}
% 1 & 1 \\
% 1 & -1 \\
% \end{bmatrix},
% \end{equation}    
% \begin{equation} \label{eq:PhaseGate}
%     P(\theta) = \begin{bmatrix}
% 1 & 0 \\
% 0 & e^{i\theta} \\
% \end{bmatrix},
% \end{equation}
% \end{subequations}

\begin{equation} \label{eq:PhaseGate}
    P(\theta) = \begin{bmatrix}
1 & 0 \\
0 & e^{i\theta} \\
\end{bmatrix},
\end{equation}
applied on qubit $j \in [1,\omega]$, in the $\ket{+} = H\ket{0} = \frac{1}{\sqrt{2}} (\ket{0} + \ket{1})$ state. The angle of rotation given $a$ is obtained via:
\begin{equation} \label{eq:theta_phase}
\theta =\frac{2\pi a}{2^{j}} =\frac{\pi a}{2^{j-1}}.
% \theta =\pi a\cdot 2^{j-\omega}.
\end{equation}
The gate is equivalent to a $R_{z}$ gate up to a phase factor: $P(\theta) = e^{i \theta/2} R_{z}(\theta)$.

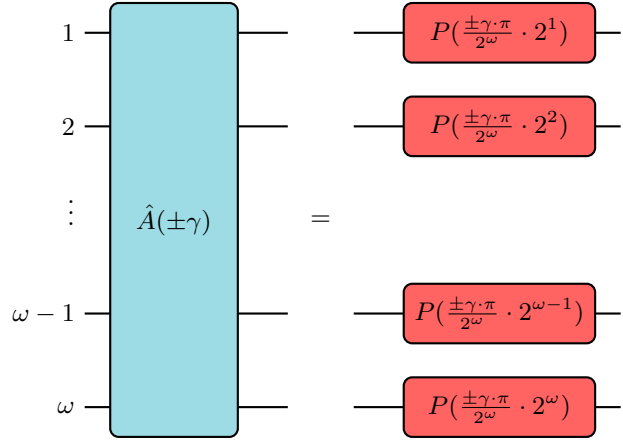
\begin{figure*}[t]
     \centering
     \begin{adjustbox}{width=0.5\textwidth}
\tikzset{every picture/.style={line width=0.75pt}} %set default line width to 0.75pt        

\definecolor{mygreen}{RGB}{34,139,33}
\definecolor{myblue}{RGB}{157,220,229}
\definecolor{myred}{RGB}{255,99,98}
\begin{tikzpicture}[scale=1.500000,x=1pt,y=1pt]
\filldraw[color=white] (0.000000, -11.000000) rectangle (126.000000, 99.000000);
% Drawing wires
% Line 17: q0 W  1
\draw[color=black] (0.000000,88.000000) -- (126.000000,88.000000);
\draw[color=black] (0.000000,88.000000) node[left] {$1$};
% Line 18: q1 W  2
\draw[color=black] (0.000000,66.000000) -- (126.000000,66.000000);
\draw[color=black] (0.000000,66.000000) node[left] {$2$};
% Line 19: ...b W
\draw[color=black] (0.000000,44.000000) node[anchor=mid east] {$\vdots$};
% Line 20: q2 W \omega-1
\draw[color=black] (0.000000,22.000000) -- (126.000000,22.000000);
\draw[color=black] (0.000000,22.000000) node[left] {$\omega-1$};
% Line 21: q3 W \omega
\draw[color=black] (0.000000,0.000000) -- (126.000000,0.000000);
\draw[color=black] (0.000000,0.000000) node[left] {$\omega$};
% Done with wires; drawing gates
% Line 23: q0 q1 q2 q3 G:state width=30 $\hat A(\pm \gamma)$
\draw[rounded corners=3pt] (21.000000,88.000000) -- (21.000000,0.000000);
\begin{scope}[rounded corners=3pt]
\begin{scope}
\draw[fill=myblue] (21.000000, 44.000000) +(-45.000000:21.213203pt and 72.124892pt) -- +(45.000000:21.213203pt and 72.124892pt) -- +(135.000000:21.213203pt and 72.124892pt) -- +(225.000000:21.213203pt and 72.124892pt) -- cycle;
\clip (21.000000, 44.000000) +(-45.000000:21.213203pt and 72.124892pt) -- +(45.000000:21.213203pt and 72.124892pt) -- +(135.000000:21.213203pt and 72.124892pt) -- +(225.000000:21.213203pt and 72.124892pt) -- cycle;
\draw (21.000000, 44.000000) node {$\hat A(\pm \gamma)$};
\end{scope}
\end{scope}
% Line 25: =
\draw[fill=white,color=white] (48.000000, -7.000000) rectangle (63.000000, 95.000000);
\draw (55.500000, 44.000000) node {$=$};
% Line 27: q0 G:env width=45 $P(\frac{\pm \gamma \cdot \pi}{2^{\omega}} \cdot 2^{1})$
\begin{scope}[rounded corners=3pt]
\begin{scope}
\draw[fill=myred] (97.500000, 88.000000) +(-45.000000:31.819805pt and 9.899495pt) -- +(45.000000:31.819805pt and 9.899495pt) -- +(135.000000:31.819805pt and 9.899495pt) -- +(225.000000:31.819805pt and 9.899495pt) -- cycle;
\clip (97.500000, 88.000000) +(-45.000000:31.819805pt and 9.899495pt) -- +(45.000000:31.819805pt and 9.899495pt) -- +(135.000000:31.819805pt and 9.899495pt) -- +(225.000000:31.819805pt and 9.899495pt) -- cycle;
\draw (97.500000, 88.000000) node {$P(\frac{\pm \gamma \cdot \pi}{2^{\omega}} \cdot 2^{1})$};
\end{scope}
\end{scope}
% Line 28: q1 G:env width=45 $P(\frac{\pm \gamma \cdot \pi}{2^{\omega}}\cdot 2^{2})$
\begin{scope}[rounded corners=3pt]
\begin{scope}
\draw[fill=myred] (97.500000, 66.000000) +(-45.000000:31.819805pt and 9.899495pt) -- +(45.000000:31.819805pt and 9.899495pt) -- +(135.000000:31.819805pt and 9.899495pt) -- +(225.000000:31.819805pt and 9.899495pt) -- cycle;
\clip (97.500000, 66.000000) +(-45.000000:31.819805pt and 9.899495pt) -- +(45.000000:31.819805pt and 9.899495pt) -- +(135.000000:31.819805pt and 9.899495pt) -- +(225.000000:31.819805pt and 9.899495pt) -- cycle;
\draw (97.500000, 66.000000) node {$P(\frac{\pm \gamma \cdot \pi}{2^{\omega}}\cdot 2^{2})$};
\end{scope}
\end{scope}
% Line 29: q2 G:env width=45 $P(\frac{\pm \gamma \cdot \pi}{2^{\omega}}\cdot 2^{\omega-1})$
\begin{scope}[rounded corners=3pt]
\begin{scope}
\draw[fill=myred] (97.500000, 22.000000) +(-45.000000:31.819805pt and 9.899495pt) -- +(45.000000:31.819805pt and 9.899495pt) -- +(135.000000:31.819805pt and 9.899495pt) -- +(225.000000:31.819805pt and 9.899495pt) -- cycle;
\clip (97.500000, 22.000000) +(-45.000000:31.819805pt and 9.899495pt) -- +(45.000000:31.819805pt and 9.899495pt) -- +(135.000000:31.819805pt and 9.899495pt) -- +(225.000000:31.819805pt and 9.899495pt) -- cycle;
\draw (97.500000, 22.000000) node {$P(\frac{\pm \gamma \cdot \pi}{2^{\omega}}\cdot 2^{\omega-1})$};
\end{scope}
\end{scope}
% Line 30: q3 G:env width=45 $P(\frac{\pm \gamma \cdot \pi}{2^{\omega}}\cdot 2^{\omega})$
\begin{scope}[rounded corners=3pt]
\begin{scope}
\draw[fill=myred] (97.500000, -0.000000) +(-45.000000:31.819805pt and 9.899495pt) -- +(45.000000:31.819805pt and 9.899495pt) -- +(135.000000:31.819805pt and 9.899495pt) -- +(225.000000:31.819805pt and 9.899495pt) -- cycle;
\clip (97.500000, -0.000000) +(-45.000000:31.819805pt and 9.899495pt) -- +(45.000000:31.819805pt and 9.899495pt) -- +(135.000000:31.819805pt and 9.899495pt) -- +(225.000000:31.819805pt and 9.899495pt) -- cycle;
\draw (97.500000, -0.000000) node {$P(\frac{\pm \gamma \cdot \pi}{2^{\omega}}\cdot 2^{\omega})$};
\end{scope}
\end{scope}
% Done with gates; drawing ending labels
\draw[color=black] (126.000000,44.000000) node[anchor=mid west] {$\vdots$};
% Done with ending labels; drawing cut lines and comments
% Done with comments
\end{tikzpicture}
\end{adjustbox}
        \caption{Quantum addition circuit operating on $\omega$-qubits in the Fourier basis, assuming each qubit is unentangled and its Bloch vector lies in the XY-plane. The single qubit phase gate $P$ is defined in Equation~\ref{eq:PhaseGate}.}
        \label{fig:adder}
\end{figure*}

Overall, given a computational basis state (the binary representation of an integer), we can express this state directly in the Fourier basis by applying single-qubit Hadamard and phase gates to each qubit. Next, we will see how this idea enables addition in the Fourier basis.

The quantum ``adder'' works due to the fact that rotations along the same direction (about the same axis) in SO(3) combine by simple addition of their angles. Therefore:

\begin{equation}
\begin{aligned}
\hat{A}(x) \hat{A}(m) H^{\otimes m}\ket{0} &= \hat{A}(x_{1},x_{2}, \dots, x_{\omega}) \hat{A}(m_{1},m_{2}, \dots, m_{\omega}) \ket{+} \\
&= \hat{A}(m_{1}+x_{1},m_{2}+x_{2}, \dots, m_{\omega} + x_{\omega} ) \ket{+}\\
&= \big( \ket{0} + e^{2\pi i (m+x) /2^{1}}\ket{1} \big) \otimes \big( \ket{0} + e^{2\pi i (m+x) /2^{2}}\ket{1} \big) \otimes \dots \otimes \big( \ket{0} + e^{2\pi i (m+x) /2^{l}}\ket{1} \big)\\
& \xrightarrow{QFT^{\dagger}} \ket{(m+x) \text{ mod } 2^{\omega}}.
    \end{aligned}
\end{equation}
Geometrically, Hadamard gates first map each qubit onto the XY-plane. Phase gates are then used to rotate the Bloch vector of each qubit around the Z-axis. Each rotation represent the addition of a different integer. The inverse QFT transform is then used to interfere these carefully constructed phases such that they produce the desired binary output. Figure~\ref{fig:adder} shows the quantum circuit required to implement $\hat{A}$.

The next Section shows how binary can be used to represent signed integers. 

\section{Two's compliment} \label{sec:twosC}

\begin{figure*}[b]
    \centering
    \begin{subfigure}[b]{0.45\textwidth}
\includegraphics[width=0.85\linewidth]{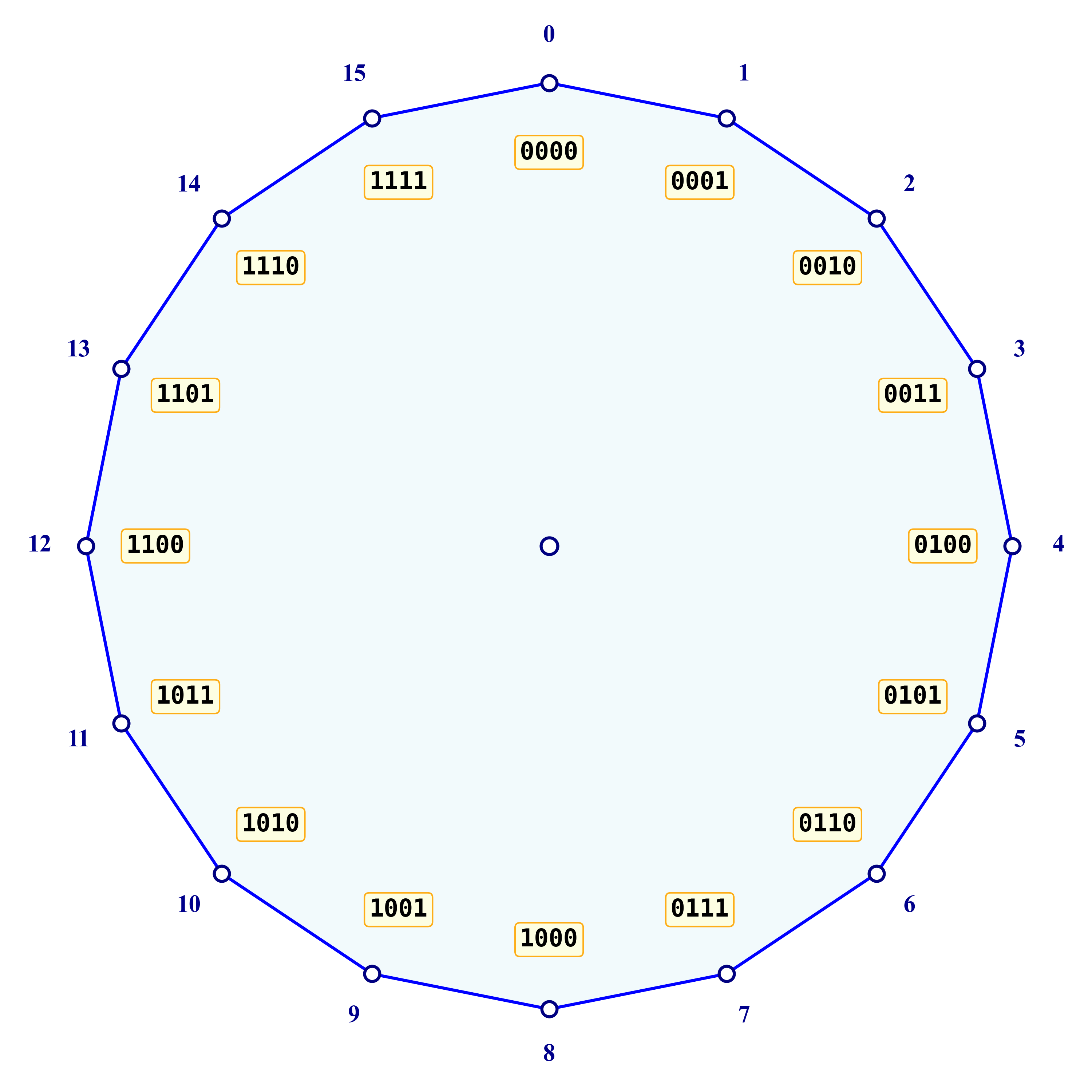}
\caption{Unsigned} \label{fig:unsigned_empty}
\end{subfigure}
    \hspace{0.05\textwidth}
    \begin{subfigure}[b]{0.45\textwidth}
\includegraphics[width=0.85\linewidth]{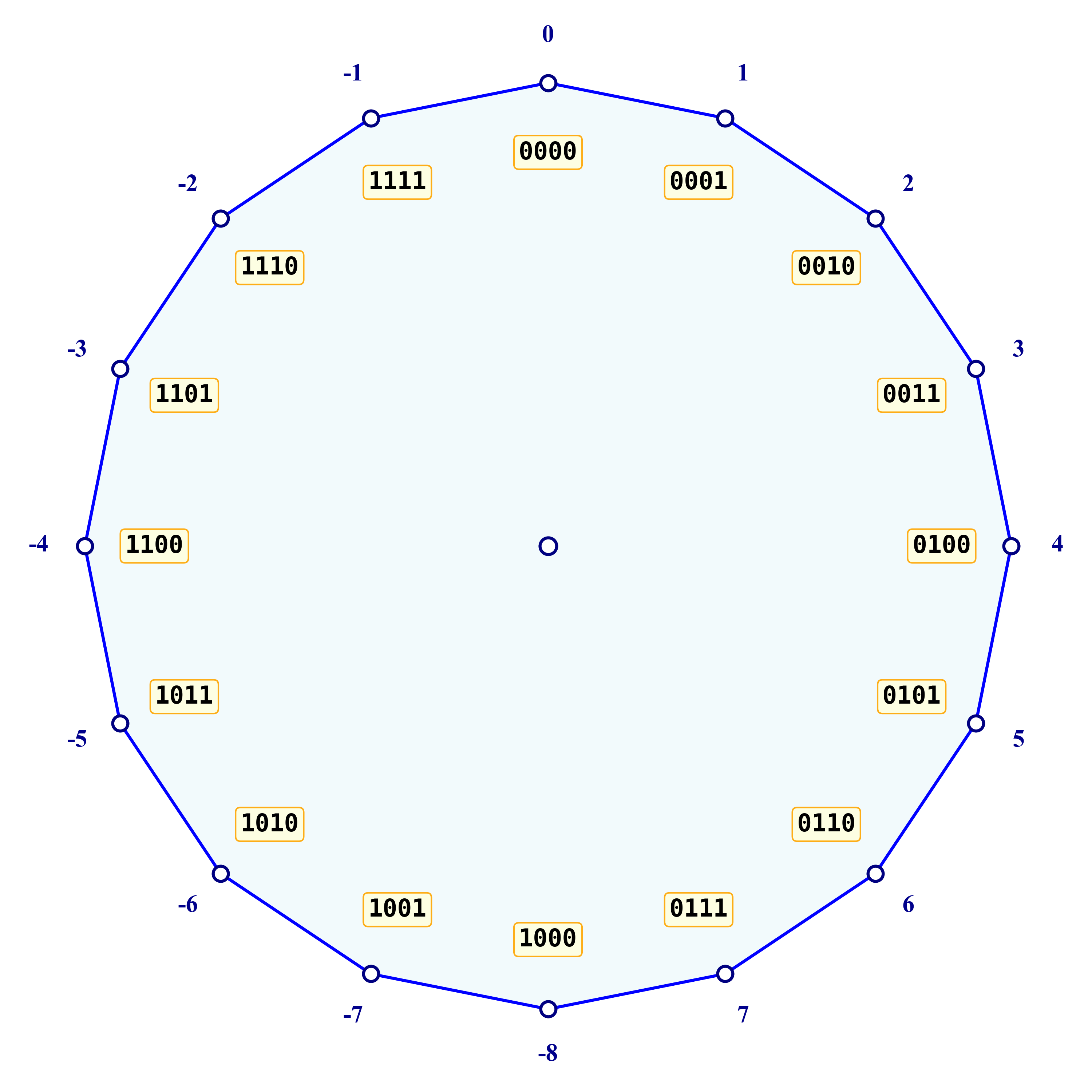}
\caption{Signed} \label{fig:signed_empty}
    \end{subfigure}
    \caption{Clock‑face illustration of four‑bit signed and unsigned binary integer representations.}
    \label{fig:empty_clock}
\end{figure*}

Two's complement is a method of representing signed integers (positive and negative). In this system, the most significant bit (the left‑most bit) determines the sign: $0$ for non‑negative numbers and $1$ for negative numbers. Figure \ref{fig:empty_clock} illustrates this using a 4‑bit clock. On the signed clock (Figure \ref{fig:signed_empty}), the value at the top is zero. Moving clockwise increases the value in positive steps, while moving counter‑clockwise decreases it in negative steps. In contrast, the unsigned clock (Figure \ref{fig:unsigned_empty}) contains only non‑negative values, encoded in standard binary. Both systems span the same total number of representable states $2^{n}$; for $n=4$ this is $16$. The signed clock has the integer range $[-2^{n-1}, 2^{n-1}) \cap \mathbb{Z}\equiv \{-8,-7,-6,-5,-4,-3,-2,-1,0,1,2,3,4,5,6,7 \}$ and the unsigned clock covers $[0, 2^{n}) \cap \mathbb{Z} \equiv \{ 0,1,2,3,4,5,6,7,8,9,10,11,12,13,14,15 \} $.

There is an important structural pattern visible in the signed clock (Figure \ref{fig:signed_empty}). If we ignore the most significant (leftmost) bit, the sequences
$0 \mapsto 7$ and $-8 \mapsto -1$ each use the standard binary encodings $000$ through $111$. The positive range follows the standard counting order, and the negative values follow the reverse order. If we could ``flip'' the negative region so that the magnitude of the number is represented in standard binary then we would have the entire number line would be represented contiguously in binary. Two’s complement achieves exactly this: it reinterprets the bit patterns so that the negative numbers wrap around and appear after the positive ones in standard binary order. The same structure arises for an arbitrary $n$-bit binary clock, not just the $4$‑bit example.

To determine the value of a number encoded in two’s complement, we follow the procedure summarized in Algorithm in Figure  \ref{alg:twocompalg}. If the most significant (left‑most) bit is $1$, the encoded number is negative. Its magnitude can be obtained by inverting all bits, adding one, and then converting the resulting bitstring from binary to its base‑10 value; the final result is the negation of this value. If the most significant bit is $0$, the bitstring is interpreted using the standard binary‑to‑decimal conversion. This process of invert and add one is precisely the ``flip'' referred to in the previous paragraph, which aligns the negative portion of the number line with the standard binary ordering for representing the magnitude of the negative number.

\begin{figure}[t]
  \centering
  \begin{minipage}{0.9\linewidth} % keeps it nicely boxed
    \hrule height 0.6pt
    \vspace{0.5em}
    \begin{algorithmic}[1] % [1] for line numbering
      \STATE \textbf{Input:} $[b_1, b_2, \dots, b_\tau]$ where $b_i \in \{0,1\}$
      \STATE {}
      \IF{$b_1 == 1$}
        \STATE $out \gets -\big(1 + \sum_{j=1}^{\tau} [\frac{b_j+1}{2} \cdot 2^{\tau-j}] \big)$
      \ELSE
        \STATE $out \gets \sum_{j=1}^{\tau} b_j \cdot 2^{\tau-j}$
      \ENDIF
      \STATE {}
      \STATE \textbf{Return:} $out \in [-2^{\tau-1}, 2^{\tau-1})$
    \end{algorithmic}
    \vspace{0.5em}
    % Bottom line
    \hrule height 0.6pt
  \end{minipage}
  \caption{Algorithm to convert a two’s complement $\tau$-bit binary number to signed decimal.}
  \label{alg:twocompalg}
\end{figure}

Take the binary number $1110$ as an example. Since the leftmost bit is $1$, it represents a negative number in two's complement form. First, invert all the bits:
$1110 \mapsto 0001$. Then, add 1 to the result:
$0001_{2}+1_{10}=1_{10}+1_{10}=2$. Finally, apply the negative sign: $-2$. So, in decimal, $1110_{2} = -2_{10}$. We could also work this out in the integer representation. Here $1110_{2}$ is $14_{10}$ in standard binary, but in two's compliment we get $-(2^{4}-14)=-(16-14)=-2$. 

In conventional modular arithmetic, subtraction (moving anti-clockwise) can be represented by addition (moving clockwise). In further detail, for $\text{mod} \: w$, we note that subtracting $x$ is the same as adding $(w-x) \: \text{mod} \: w$, where it is assumed $0\leq x\leq w$. As an example, take a look at Figure \ref{fig:unsigned_2_14}. If we wanted to subtract $4$ from $2$: $(2-4) \: \text{mod} \: 2^{4} = -2 \: \text{mod} \: 16 = 14$. This is the same thing as adding (moving clockwise): $(2^{4}-4) \: \text{mod} \: 2^{4} = (16-4) \: \text{mod} \: 16 = 12 \: \text{mod} \: 16 = 12$. Moving the green arrow at $2$ clockwise twelve units is the same thing as subtracting, moving anticlockwise, four units. 

In Two's Complement representation, the same thing is true. The number line has just been shifted from $[0, 2^{n})$ to $[-2^{n-1}, 2^{n-1})$. The movement of the clock acts in the same way. As an example, in Figure \ref{fig:signed_2_24} if we wanted to subtract four from two we see that in the same way we either move clockwise twelve units (addition) or moving anticlockwise (subtraction) four units. The only difference is now the encoding represents $-2$ rather than $14$ in standard binary.

Two’s complement preserves the usual behaviour of addition and subtraction on the integer number line; deviations arise only when an operation exceeds the representable range and overflow occurs. For instance, in the $4$‑bit system, subtracting $12$ from two would ordinarily yield $-10$; however, the result wraps to $6$ because $-10$ cannot be represented with only four bits in this encoding.

In two’s‑complement encoding with $\omega$-bits, any integer in the range $-2^{\omega -1}\leq a < 2^{\omega-1}$, including negative values, can also be represented using the phases defined in Equation \ref{eq:theta_phase}.

For Grover Adaptive Search the defining characteristic of this encoding is that the leftmost qubit serves as an indicator of whether the corresponding base-10 value is negative. In the  GAS quantum circuit, this qubit is specifically employed to control the marking oracle, thereby enabling the phase adjustment (marking) of lower-energy states for Grover’s search. As illustrated in Figure~\ref{fig:GAS-fig} and Figure~\ref{fig:GAS-Dicke} in the main text, the \emph{sign} qubit is positioned directly beneath the register of $m$ qubits.

In the next Section we discuss each component of the GAS quantum circuit for GAS-SCF.

% Appendix \ref{sec:Fourier_add} and Appendix \ref{sec:twosC} constitute all the components required for GAS if no additional constraints were necessary. However, in our approach, we must also enforce the number of spin-up and spin-down electrons, which introduces extra control operations. In the next Section we discuss each component of the GAS quantum circuit.

\begin{figure*}[h]
    \centering
    \begin{subfigure}[b]{0.45\textwidth}
\includegraphics[width=0.85\linewidth]{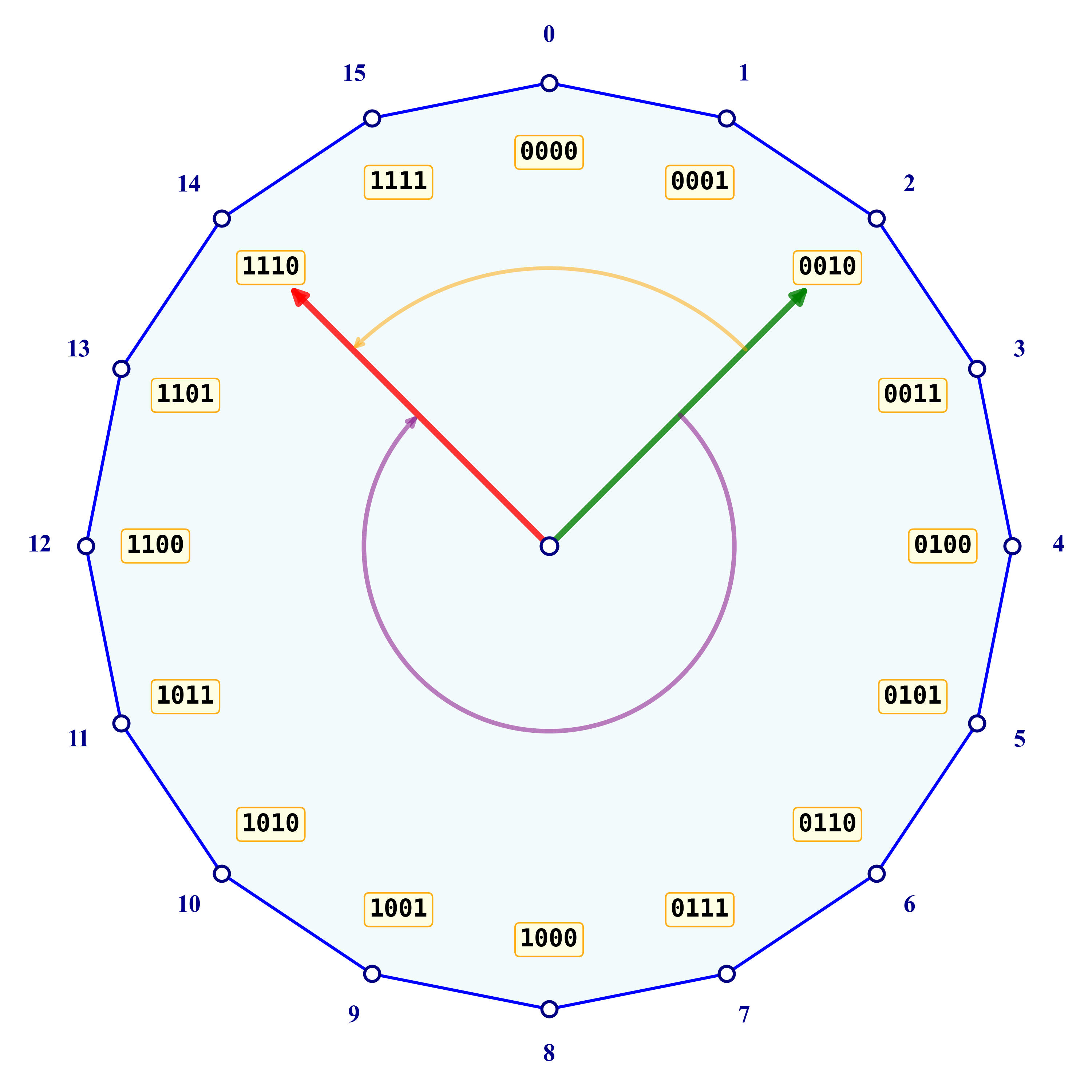}
\caption{Unsigned} \label{fig:unsigned_2_14}
\end{subfigure}
    \hspace{0.05\textwidth}
    \begin{subfigure}[b]{0.45\textwidth}
\includegraphics[width=0.85\linewidth]{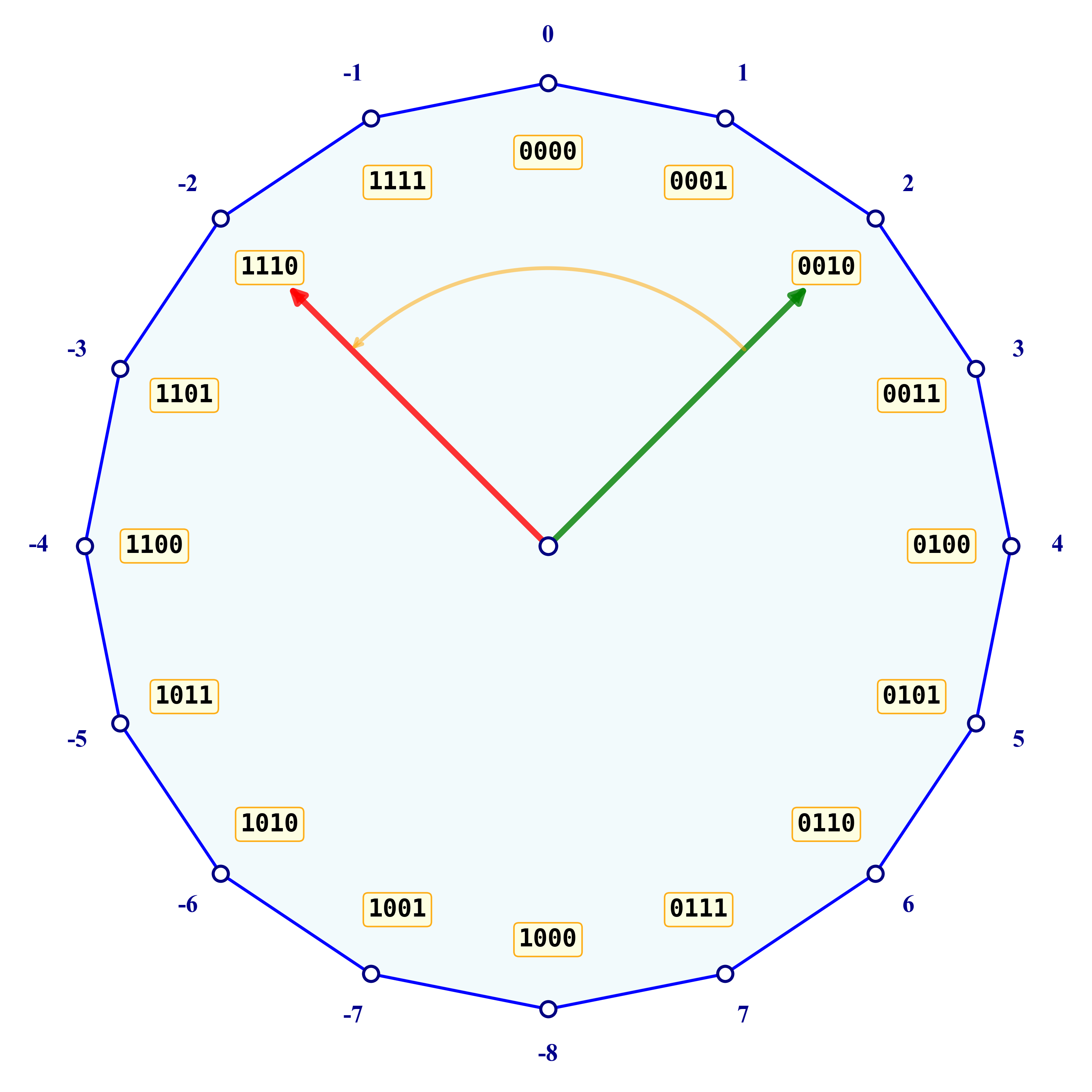}
\caption{Signed.} \label{fig:signed_2_24}
    \end{subfigure}
    \caption{Clock‑face illustration demonstrating how addition and subtraction coincide in modular arithmetic. In this case, adding $12$ produces the same result as subtracting $4$ (in $\text{mod} \: 16$). }
    \label{fig:clock_2_24}
\end{figure*}

\clearpage
\section{GAS-SCF Quantum Circuit Components \label{sec:qcircuitsfull}}

\begin{figure*}[t!]
    \centering
    \begin{subfigure}[t]{0.75\textwidth}
        \centering
\definecolor{mygreen}{RGB}{34,139,33}
\definecolor{myblue}{RGB}{157,220,229}
\definecolor{myred}{RGB}{255,99,98}
\begin{tikzpicture}[scale=1.500000,x=1pt,y=1pt]
\filldraw[color=white] (0.000000, -11.000000) rectangle (63.333333, 33.000000);
% Drawing wires
% Line 17: q0 W
\draw[color=black] (0.000000,22.000000) -- (63.333333,22.000000);
% Line 18: q2 W
\draw[color=black] (0.000000,0.000000) -- (63.333333,0.000000);
% Done with wires; drawing gates
% Line 20: q0 / n
\draw (6.000000, 15.000000) -- (15.333333, 29.000000);
\draw (13.000000, 25.500000) node[right] {$\scriptstyle{n}$};
% Line 21: q2 / m+1
\draw (6.000000, -7.000000) -- (15.333333, 7.000000);
\draw (13.000000, 3.500000) node[right] {$\scriptstyle{m+1}$};
% Line 24: q2 G:state width=30 $g(\vec{x})$ q0:shape=4
\draw[rounded corners=3pt] (42.333333,22.000000) -- (42.333333,0.000000);
\begin{scope}[rounded corners=3pt]
\begin{scope}
\draw[fill=myblue] (42.333333, -0.000000) +(-45.000000:21.213203pt and 9.899495pt) -- +(45.000000:21.213203pt and 9.899495pt) -- +(135.000000:21.213203pt and 9.899495pt) -- +(225.000000:21.213203pt and 9.899495pt) -- cycle;
\clip (42.333333, -0.000000) +(-45.000000:21.213203pt and 9.899495pt) -- +(45.000000:21.213203pt and 9.899495pt) -- +(135.000000:21.213203pt and 9.899495pt) -- +(225.000000:21.213203pt and 9.899495pt) -- cycle;
\draw (42.333333, -0.000000) node {$g(\vec{x})$};
\end{scope}
\end{scope}
\begin{scope}
\draw[fill=white] (42.333333, 22.000000) +(-45.000000:3.000000pt) -- +(45.000000:3.000000pt) -- +(135.000000:3.000000pt) -- +(225.000000:3.000000pt) -- cycle;
\clip (42.333333, 22.000000) +(-45.000000:3.000000pt) -- +(45.000000:3.000000pt) -- +(135.000000:3.000000pt) -- +(225.000000:3.000000pt) -- cycle;
\draw (39.333333, 22.000000) -- (45.333333, 22.000000);
\draw (42.333333, 19.000000) -- (42.333333, 25.000000);
\end{scope}
% Done with gates; drawing ending labels
% Done with ending labels; drawing cut lines and comments
% Done with comments
\end{tikzpicture}
        \caption{QUBO adder circuit}
    \end{subfigure}%
\vspace{1em} % optional space between subfigures
    \begin{subfigure}[t]{0.75\textwidth}
        \centering
        \begin{adjustbox}{width=1\linewidth}
       \providecommand{\ket}[1]{\left |#1\right\rangle}
\definecolor{mygreen}{RGB}{34,139,33}
\definecolor{myblue}{RGB}{157,220,229}
\definecolor{myred}{RGB}{255,99,98}
\begin{tikzpicture}[scale=1.500000,x=1pt,y=1pt]
\filldraw[color=white] (0.000000, -11.000000) rectangle (300.333333, 99.000000);
% Drawing wires
% Line 17: q0 W
\draw[color=black] (0.000000,88.000000) -- (300.333333,88.000000);
% Line 18: q1 W
\draw[color=black] (0.000000,66.000000) -- (300.333333,66.000000);
% Line 19: q2 W
\draw[color=black] (0.000000,44.000000) -- (300.333333,44.000000);
% Line 20: q3 W
\draw[color=black] (0.000000,22.000000) -- (300.333333,22.000000);
% Line 21: q4 W
\draw[color=black] (0.000000,0.000000) -- (300.333333,0.000000);
% Done with wires; drawing gates
% Line 23: q4 / m+1
\draw (6.000000, -7.000000) -- (15.333333, 7.000000);
\draw (13.000000, 3.500000) node[right] {$\scriptstyle{m+1}$};
% Line 25: q4 G:state width=30 $g(\vec{x})$ q0:shape=4 q1:shape=4 q2:shape=4 q3:shape=4
\draw[rounded corners=3pt] (42.333333,88.000000) -- (42.333333,0.000000);
\begin{scope}[rounded corners=3pt]
\begin{scope}
\draw[fill=myblue] (42.333333, -0.000000) +(-45.000000:21.213203pt and 9.899495pt) -- +(45.000000:21.213203pt and 9.899495pt) -- +(135.000000:21.213203pt and 9.899495pt) -- +(225.000000:21.213203pt and 9.899495pt) -- cycle;
\clip (42.333333, -0.000000) +(-45.000000:21.213203pt and 9.899495pt) -- +(45.000000:21.213203pt and 9.899495pt) -- +(135.000000:21.213203pt and 9.899495pt) -- +(225.000000:21.213203pt and 9.899495pt) -- cycle;
\draw (42.333333, -0.000000) node {$g(\vec{x})$};
\end{scope}
\end{scope}
\begin{scope}
\draw[fill=white] (42.333333, 88.000000) +(-45.000000:3.000000pt) -- +(45.000000:3.000000pt) -- +(135.000000:3.000000pt) -- +(225.000000:3.000000pt) -- cycle;
\clip (42.333333, 88.000000) +(-45.000000:3.000000pt) -- +(45.000000:3.000000pt) -- +(135.000000:3.000000pt) -- +(225.000000:3.000000pt) -- cycle;
\draw (39.333333, 88.000000) -- (45.333333, 88.000000);
\draw (42.333333, 85.000000) -- (42.333333, 91.000000);
\end{scope}
\begin{scope}
\draw[fill=white] (42.333333, 66.000000) +(-45.000000:3.000000pt) -- +(45.000000:3.000000pt) -- +(135.000000:3.000000pt) -- +(225.000000:3.000000pt) -- cycle;
\clip (42.333333, 66.000000) +(-45.000000:3.000000pt) -- +(45.000000:3.000000pt) -- +(135.000000:3.000000pt) -- +(225.000000:3.000000pt) -- cycle;
\draw (39.333333, 66.000000) -- (45.333333, 66.000000);
\draw (42.333333, 63.000000) -- (42.333333, 69.000000);
\end{scope}
\begin{scope}
\draw[fill=white] (42.333333, 44.000000) +(-45.000000:3.000000pt) -- +(45.000000:3.000000pt) -- +(135.000000:3.000000pt) -- +(225.000000:3.000000pt) -- cycle;
\clip (42.333333, 44.000000) +(-45.000000:3.000000pt) -- +(45.000000:3.000000pt) -- +(135.000000:3.000000pt) -- +(225.000000:3.000000pt) -- cycle;
\draw (39.333333, 44.000000) -- (45.333333, 44.000000);
\draw (42.333333, 41.000000) -- (42.333333, 47.000000);
\end{scope}
\begin{scope}
\draw[fill=white] (42.333333, 22.000000) +(-45.000000:3.000000pt) -- +(45.000000:3.000000pt) -- +(135.000000:3.000000pt) -- +(225.000000:3.000000pt) -- cycle;
\clip (42.333333, 22.000000) +(-45.000000:3.000000pt) -- +(45.000000:3.000000pt) -- +(135.000000:3.000000pt) -- +(225.000000:3.000000pt) -- cycle;
\draw (39.333333, 22.000000) -- (45.333333, 22.000000);
\draw (42.333333, 19.000000) -- (42.333333, 25.000000);
\end{scope}
% Line 26: =
\draw[fill=white,color=white] (69.333333, -7.000000) rectangle (84.333333, 95.000000);
\draw (76.833333, 44.000000) node {$=$};
% Line 28: q4 G:state width=30 $\hat{A}(a)$ q0
\draw[rounded corners=3pt] (111.333333,88.000000) -- (111.333333,0.000000);
\begin{scope}[rounded corners=3pt]
\begin{scope}
\draw[fill=myblue] (111.333333, -0.000000) +(-45.000000:21.213203pt and 9.899495pt) -- +(45.000000:21.213203pt and 9.899495pt) -- +(135.000000:21.213203pt and 9.899495pt) -- +(225.000000:21.213203pt and 9.899495pt) -- cycle;
\clip (111.333333, -0.000000) +(-45.000000:21.213203pt and 9.899495pt) -- +(45.000000:21.213203pt and 9.899495pt) -- +(135.000000:21.213203pt and 9.899495pt) -- +(225.000000:21.213203pt and 9.899495pt) -- cycle;
\draw (111.333333, -0.000000) node {$\hat{A}(a)$};
\end{scope}
\end{scope}
\filldraw (111.333333, 88.000000) circle(1.500000pt);
% Line 29: q4 G:state width=30 $\hat{A}(b)$ q1 q2
\draw[rounded corners=3pt] (153.333333,66.000000) -- (153.333333,0.000000);
\begin{scope}[rounded corners=3pt]
\begin{scope}
\draw[fill=myblue] (153.333333, -0.000000) +(-45.000000:21.213203pt and 9.899495pt) -- +(45.000000:21.213203pt and 9.899495pt) -- +(135.000000:21.213203pt and 9.899495pt) -- +(225.000000:21.213203pt and 9.899495pt) -- cycle;
\clip (153.333333, -0.000000) +(-45.000000:21.213203pt and 9.899495pt) -- +(45.000000:21.213203pt and 9.899495pt) -- +(135.000000:21.213203pt and 9.899495pt) -- +(225.000000:21.213203pt and 9.899495pt) -- cycle;
\draw (153.333333, -0.000000) node {$\hat{A}(b)$};
\end{scope}
\end{scope}
\filldraw (153.333333, 66.000000) circle(1.500000pt);
\filldraw (153.333333, 44.000000) circle(1.500000pt);
% Line 30: q4 G:state width=30 $\hat{A}(c)$ q0 q2
\draw[rounded corners=3pt] (195.333333,88.000000) -- (195.333333,0.000000);
\begin{scope}[rounded corners=3pt]
\begin{scope}
\draw[fill=myblue] (195.333333, -0.000000) +(-45.000000:21.213203pt and 9.899495pt) -- +(45.000000:21.213203pt and 9.899495pt) -- +(135.000000:21.213203pt and 9.899495pt) -- +(225.000000:21.213203pt and 9.899495pt) -- cycle;
\clip (195.333333, -0.000000) +(-45.000000:21.213203pt and 9.899495pt) -- +(45.000000:21.213203pt and 9.899495pt) -- +(135.000000:21.213203pt and 9.899495pt) -- +(225.000000:21.213203pt and 9.899495pt) -- cycle;
\draw (195.333333, -0.000000) node {$\hat{A}(c)$};
\end{scope}
\end{scope}
\filldraw (195.333333, 88.000000) circle(1.500000pt);
\filldraw (195.333333, 44.000000) circle(1.500000pt);
% Line 31: q4 G:state width=30 $\hat{A}(d)$ q3
\draw[rounded corners=3pt] (237.333333,22.000000) -- (237.333333,0.000000);
\begin{scope}[rounded corners=3pt]
\begin{scope}
\draw[fill=myblue] (237.333333, -0.000000) +(-45.000000:21.213203pt and 9.899495pt) -- +(45.000000:21.213203pt and 9.899495pt) -- +(135.000000:21.213203pt and 9.899495pt) -- +(225.000000:21.213203pt and 9.899495pt) -- cycle;
\clip (237.333333, -0.000000) +(-45.000000:21.213203pt and 9.899495pt) -- +(45.000000:21.213203pt and 9.899495pt) -- +(135.000000:21.213203pt and 9.899495pt) -- +(225.000000:21.213203pt and 9.899495pt) -- cycle;
\draw (237.333333, -0.000000) node {$\hat{A}(d)$};
\end{scope}
\end{scope}
\filldraw (237.333333, 22.000000) circle(1.500000pt);
% Line 32: q4 G:state width=30 $\hat{A}(e)$ q2 q3
\draw[rounded corners=3pt] (279.333333,44.000000) -- (279.333333,0.000000);
\begin{scope}[rounded corners=3pt]
\begin{scope}
\draw[fill=myblue] (279.333333, -0.000000) +(-45.000000:21.213203pt and 9.899495pt) -- +(45.000000:21.213203pt and 9.899495pt) -- +(135.000000:21.213203pt and 9.899495pt) -- +(225.000000:21.213203pt and 9.899495pt) -- cycle;
\clip (279.333333, -0.000000) +(-45.000000:21.213203pt and 9.899495pt) -- +(45.000000:21.213203pt and 9.899495pt) -- +(135.000000:21.213203pt and 9.899495pt) -- +(225.000000:21.213203pt and 9.899495pt) -- cycle;
\draw (279.333333, -0.000000) node {$\hat{A}(e)$};
\end{scope}
\end{scope}
\filldraw (279.333333, 44.000000) circle(1.500000pt);
\filldraw (279.333333, 22.000000) circle(1.500000pt);
% Done with gates; drawing ending labels
% Done with ending labels; drawing cut lines and comments
% Done with comments
\end{tikzpicture}
\end{adjustbox}
        \caption{Example: $g(\vec{x}) = a x_{0} + b x_{1} x_{2} + c x_{0} x_{2} + d x_{3} + e x_{2} x_{3}$}
    \end{subfigure}
    \caption{Quantum circuits to implement addition in a polynomial Boolean function. (a) Shows the generic circuit and (b) Provides a simple example where each coefficient is a positive/negative integer: $\{ a,b,c,d,e\} \in \mathbb{Z}$. We assume here the $m+1$ qubits is enough to store the maximum value of the Boolean function $|g(\vec{x})|$. Each $\hat{A}$ gate is defined in Figure~\ref{fig:adder}. Note that each qubit in the $m+1$ register is assumed to have been rotated into the XY-plane. This circuit produces the sum in the Fourier basis, which can be converted back to the computational (decimal) basis using the inverse QFT.\label{fig:QUBO-adder}} 
\end{figure*}
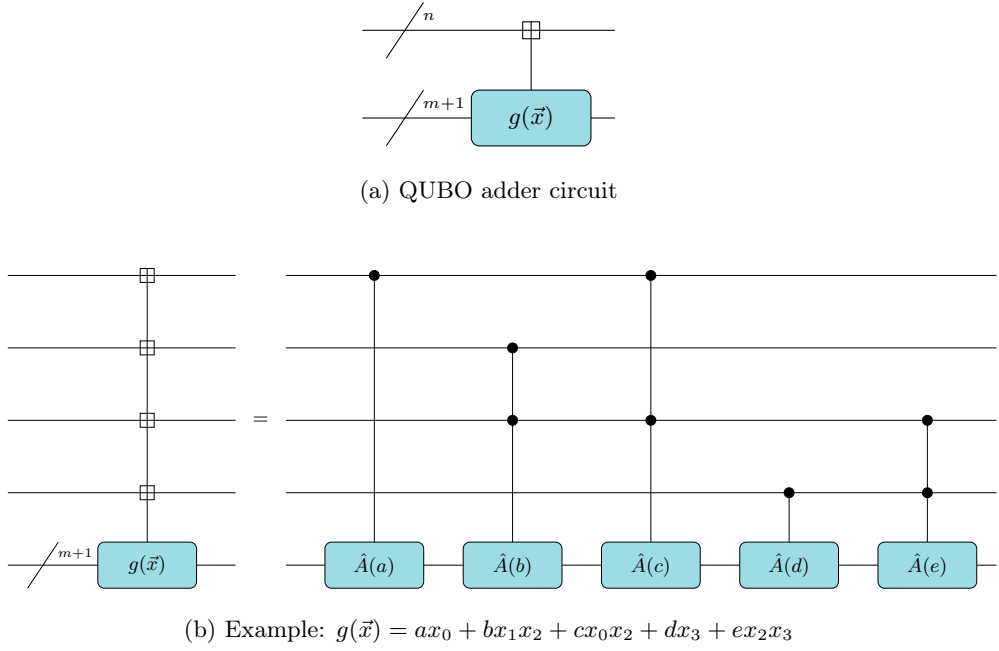
 
In this section we detail all the non-standard circuit components required to implement GAS-SCF.

Gilliam, Woerner, and Gonciulea \cite{gilliam2021grover} showed how to implement a pseudo-Boolean function via a quantum circuit. We summarize their approach in Figure~\ref{fig:QUBO-adder}. However, for SCF simulations we also require solutions to be in the correct symmetry sector. We can write the number operator as a Boolean function composed of linear terms:

\begin{subequations} \label{eq:number_qubos}
\begin{equation}
  \hat{N}_{\alpha}(\vec{x}) = \sum_{i \in \mathcal{A}} x_{i} ,
\end{equation}    
\begin{equation}
  \hat{N}_{\beta}(\vec{x}) = \sum_{i \in \mathcal{B}} x_{i},
\end{equation}
\end{subequations}
where $\mathcal{A}$ and $\mathcal{B}$ contain the spin indices for the spin-up and spin-down electrons respectively. Both these functions can be implemented according to the circuits in Figure~\ref{fig:QUBO-adder}. Here each polynomial term is linear and thus at most each term will require a single control. 

To ensure that the correct number of particles is ensured, we need to flag all bitstrings that have the correct Hamming weight. Figure~\ref{fig:QUBO-proj} details the general quantum circuit required for $\gamma$ electrons.

Note that the control sequence is used to find all states with a particular occupation number (Hamming weight). For example, on the system register $\ket{0111110}$, $\ket{1011101}$ and $\ket{1101011}$ all have $N(x)=5$ electrons. Running $N(x)$ (Figure~\ref{fig:QUBO-proj}) applied to these input states, results in $\ket{101}$ being stored in the $N_{\gamma(x)}$ register (post $QFT^{\dagger}$). We see that Figure~\ref{fig:QUBO-proj-b} would therefore flag these states. In other words, this circuit should not be mistaken for selecting the state $\ket{101}$ in the system register.

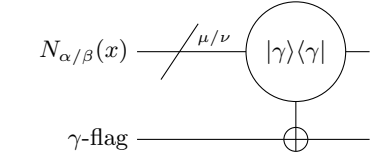
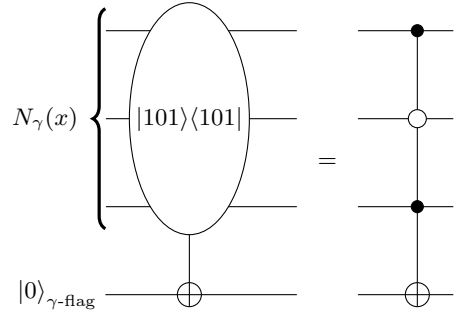
\begin{figure*}[b]
    \centering
    \begin{subfigure}[t]{0.85\textwidth}
        \centering
\definecolor{mygreen}{RGB}{34,139,33}
\definecolor{myblue}{RGB}{157,220,229}
\definecolor{myred}{RGB}{255,99,98}
\begin{tikzpicture}[scale=1.500000,x=1pt,y=1pt]
\filldraw[color=white] (0.000000, -11.000000) rectangle (58.333333, 33.000000);
% Drawing wires
% Line 17: q0 W N_{\alpha / \beta}(x)
\draw[color=black] (0.000000,22.000000) -- (58.333333,22.000000);
\draw[color=black] (0.000000,22.000000) node[left] {$N_{\alpha / \beta}(x)$};
% Line 18: q1 W \gamma\text{-flag}
\draw[color=black] (0.000000,0.000000) -- (58.333333,0.000000);
\draw[color=black] (0.000000,0.000000) node[left] {$\gamma\text{-flag}$};
% Done with wires; drawing gates
% Line 20: q0 / {\mu / \nu}
\draw (6.000000, 15.000000) -- (15.333333, 29.000000);
\draw (13.000000, 25.500000) node[right] {$\scriptstyle{{\mu / \nu}}$};
% Line 22: q0 P:width=25 $| \gamma \rangle \langle \gamma |$ +q1
\draw (39.833333,22.000000) -- (39.833333,0.000000);
\begin{scope}
\draw[fill=white] (39.833333, 22.000000) circle(12.500000pt);
\clip (39.833333, 22.000000) circle(12.500000pt);
\draw (39.833333, 22.000000) node {$| \gamma \rangle \langle \gamma |$};
\end{scope}
\begin{scope}
\draw[fill=white] (39.833333, 0.000000) circle(3.000000pt);
\clip (39.833333, 0.000000) circle(3.000000pt);
\draw (36.833333, 0.000000) -- (42.833333, 0.000000);
\draw (39.833333, -3.000000) -- (39.833333, 3.000000);
\end{scope}
% Done with gates; drawing ending labels
% Done with ending labels; drawing cut lines and comments
% Done with comments
\end{tikzpicture}
        \caption{General number flagging circuit}
    \end{subfigure}%
\vspace{1em} % optional space between subfigures
    \begin{subfigure}[t]{0.35\textwidth}
        \centering
        \begin{adjustbox}{width=1\linewidth}
%! \usetikzlibrary{decorations.pathreplacing,decorations.pathmorphing}
\definecolor{mygreen}{RGB}{34,139,33}
\definecolor{myblue}{RGB}{157,220,229}
\definecolor{myred}{RGB}{255,99,98}
\begin{tikzpicture}[scale=1.500000,x=1pt,y=1pt]
\filldraw[color=white] (0.000000, -11.000000) rectangle (87.000000, 77.000000);
% Drawing wires
% Line 20: q0 q1 q2 W N_{\gamma}(x)<
\draw[color=black] (0.000000,66.000000) -- (87.000000,66.000000);
%   Deferring wire label at (0.000000,66.000000)
% Line 20: q0 q1 q2 W N_{\gamma}(x)<
\draw[color=black] (0.000000,44.000000) -- (87.000000,44.000000);
%   Deferring wire label at (0.000000,44.000000)
% Line 20: q0 q1 q2 W N_{\gamma}(x)<
\draw[color=black] (0.000000,22.000000) -- (87.000000,22.000000);
\filldraw[color=white,fill=white] (0.000000,16.500000) rectangle (-4.666667,71.500000);
\draw[decorate,decoration={brace,amplitude = 4.666667pt},very thick] (0.000000,16.500000) -- (0.000000,71.500000);
\draw[color=black] (-4.666667,44.000000) node[left] {$N_{\gamma}(x)$};
% Line 22: q3 W \ket{0}_{\gamma\text{-flag}}
\draw[color=black] (0.000000,0.000000) -- (87.000000,0.000000);
\draw[color=black] (0.000000,0.000000) node[left] {$\ket{0}_{\gamma\text{-flag}}$};
% Done with wires; drawing gates
% Line 26: q0 q1 q2  P:width=30 $| 101 \rangle \langle 101 |$ +q3
\draw (21.000000,66.000000) -- (21.000000,0.000000);
\begin{scope}
\draw[fill=white] (21.000000, 44.000000) ellipse(15.000000pt and 29.000000pt);
\clip (21.000000, 44.000000) ellipse(15.000000pt and 29.000000pt);
\draw (21.000000, 44.000000) node {$| 101 \rangle \langle 101 |$};
\end{scope}
\begin{scope}
\draw[fill=white] (21.000000, 0.000000) circle(3.000000pt);
\clip (21.000000, 0.000000) circle(3.000000pt);
\draw (18.000000, 0.000000) -- (24.000000, 0.000000);
\draw (21.000000, -3.000000) -- (21.000000, 3.000000);
\end{scope}
% Line 27: =
\draw[fill=white,color=white] (48.000000, -7.000000) rectangle (63.000000, 73.000000);
\draw (55.500000, 33.000000) node {$=$};
% Line 28: q0 -q1 q2 +q3
\draw (78.000000,66.000000) -- (78.000000,0.000000);
\filldraw (78.000000, 66.000000) circle(1.500000pt);
\draw[fill=white] (78.000000, 44.000000) circle(2.250000pt);
\filldraw (78.000000, 22.000000) circle(1.500000pt);
\begin{scope}
\draw[fill=white] (78.000000, 0.000000) circle(3.000000pt);
\clip (78.000000, 0.000000) circle(3.000000pt);
\draw (75.000000, 0.000000) -- (81.000000, 0.000000);
\draw (78.000000, -3.000000) -- (78.000000, 3.000000);
\end{scope}
% Done with gates; drawing ending labels
% Done with ending labels; drawing cut lines and comments
% Done with comments
\end{tikzpicture}

\end{adjustbox}
        \caption{Example circuit for flagging $101_{2}=5_{10}$ electron states \label{fig:QUBO-proj-b}}
    \end{subfigure}
    \caption{Quantum circuits to flag  equality constraint needed for ensuring correct number of electrons. Note it is assumed that the $N_{\gamma\in \{\alpha, \beta \}}(x)$ register has stored the number of electrons according to Equation~\ref{eq:number_qubos} implemented by a circuit in Figure~\ref{fig:QUBO-adder} followed by an inverse QFT so that the occupation numbers are represented in the computational basis rather than Fourier basis. We show the full structure in Figure~\ref{fig:GAS-fig}. (a) Shows the generic circuit for $\gamma$ electrons. (b) Provides a simple example for $\gamma=5_{10}=101_{2}$.  \label{fig:QUBO-proj}} 
\end{figure*}

Next, a quantum circuit is required for marking quantum states that are $f(\vec{x})-y<0$ in the $m+1$ qubit register (and potentially states with the correct symmetries). In GAS, single ancilla qubits in the $\ket{1}$ state are used to herald when states obey the required constraints. To mark these states therefore only requires controlling off these ancilla qubits. Figure~\ref{fig:QUBO-marker} shows how this unitary can be compiled as a quantum circuit. These templates stem from the following circuit identity:

\begin{equation} \label{eq:negII}
 \begin{aligned}
	CNOT(0,1)  \cdot  \big[I_{0} \otimes R_{z}(-2 \pi)_{1} \big]  \cdot   CNOT(0,1) &= CNOT(0,1)  \cdot  \big[ I_{0}\otimes -I_{1} \big]  \cdot   CNOT(0,1) \\
	&=  \bigg(-\ket{0}\bra{0} \otimes   I - \ket{1}\bra{1} \otimes   X  \bigg)    \cdot    \bigg(\ket{0}\bra{0} \otimes   I + \ket{1}\bra{1} \otimes   X \bigg)  \\ 
	&=  -\ket{0}\bra{0} \otimes   I - \ket{1}\bra{1} \otimes   I    \\ 
	&=  \frac{1}{2} \bigg( -(I+Z) \otimes  I - (I-Z ) \otimes  I \bigg) \\
	&=  \frac{1}{2} \bigg( -II-ZI - II +ZI  \bigg) \\
	&=  \frac{2(-II)}{2} = -II .
\end{aligned}
\end{equation}
We could have also leveraged the fact that the identity matrix commutes with all operators to move the $CNOT$ gates and cancel them out. The cascade of $CNOT$ gates enforces a global $-(I^{\otimes \omega})$ operation across all qubits. In contrast, if the qubits were not fully entangled, applying a single $R_{z}(- 2\pi)$ rotation would introduce a relative phase rather than a uniform global phase.

% We can see that actually using the fact that the identity matrix commutes to commute the CNOT gate to cancel each other out. The cascade of CNOT gate ensures that the $-I$ is applied across the qubits, as if the system qubits were not all entangled then a relative phase would be generated.

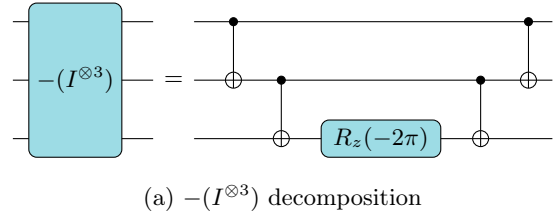
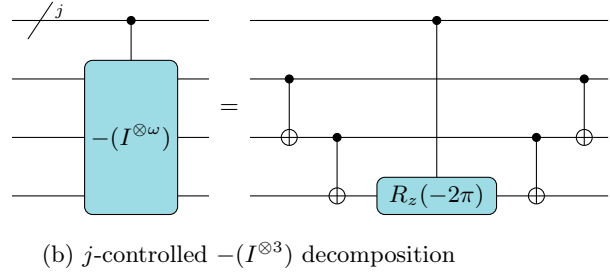
\begin{figure*}[t!]
    \centering
    \begin{subfigure}[t]{0.75\textwidth}
        \centering
\definecolor{mygreen}{RGB}{34,139,33}
\definecolor{myblue}{RGB}{157,220,229}
\definecolor{myred}{RGB}{255,99,98}
\begin{tikzpicture}[scale=1.000000,x=1pt,y=1pt]
\filldraw[color=white] (0.000000, -11.000000) rectangle (203.000000, 55.000000);
% Drawing wires
% Line 15: q0 W
\draw[color=black] (0.000000,44.000000) -- (203.000000,44.000000);
% Line 16: q1 W
\draw[color=black] (0.000000,22.000000) -- (203.000000,22.000000);
% Line 17: q2 W
\draw[color=black] (0.000000,0.000000) -- (203.000000,0.000000);
% Done with wires; drawing gates
% Line 21: q0 q1 q2 G:state width=35 $-(I^{\otimes 3})$
\draw[rounded corners=3pt] (23.500000,44.000000) -- (23.500000,0.000000);
\begin{scope}[rounded corners=3pt]
\begin{scope}
\draw[fill=myblue] (23.500000, 22.000000) +(-45.000000:24.748737pt and 41.012193pt) -- +(45.000000:24.748737pt and 41.012193pt) -- +(135.000000:24.748737pt and 41.012193pt) -- +(225.000000:24.748737pt and 41.012193pt) -- cycle;
\clip (23.500000, 22.000000) +(-45.000000:24.748737pt and 41.012193pt) -- +(45.000000:24.748737pt and 41.012193pt) -- +(135.000000:24.748737pt and 41.012193pt) -- +(225.000000:24.748737pt and 41.012193pt) -- cycle;
\draw (23.500000, 22.000000) node {$-(I^{\otimes 3})$};
\end{scope}
\end{scope}
% Line 34: =
\draw[fill=white,color=white] (53.000000, -7.000000) rectangle (68.000000, 51.000000);
\draw (60.500000, 22.000000) node {$=$};
% Line 35: q1 C q0
\draw (83.000000,44.000000) -- (83.000000,22.000000);
\begin{scope}
\draw[fill=white] (83.000000, 22.000000) circle(3.000000pt);
\clip (83.000000, 22.000000) circle(3.000000pt);
\draw (80.000000, 22.000000) -- (86.000000, 22.000000);
\draw (83.000000, 19.000000) -- (83.000000, 25.000000);
\end{scope}
\filldraw (83.000000, 44.000000) circle(1.500000pt);
% Line 36: q2 C q1
\draw (101.000000,22.000000) -- (101.000000,0.000000);
\begin{scope}
\draw[fill=white] (101.000000, 0.000000) circle(3.000000pt);
\clip (101.000000, 0.000000) circle(3.000000pt);
\draw (98.000000, 0.000000) -- (104.000000, 0.000000);
\draw (101.000000, -3.000000) -- (101.000000, 3.000000);
\end{scope}
\filldraw (101.000000, 22.000000) circle(1.500000pt);
% Line 38: q2 G:state width=45 $R_{z}(- 2 \pi)$
\begin{scope}[rounded corners=3pt]
\begin{scope}
\draw[fill=myblue] (138.500000, -0.000000) +(-45.000000:31.819805pt and 9.899495pt) -- +(45.000000:31.819805pt and 9.899495pt) -- +(135.000000:31.819805pt and 9.899495pt) -- +(225.000000:31.819805pt and 9.899495pt) -- cycle;
\clip (138.500000, -0.000000) +(-45.000000:31.819805pt and 9.899495pt) -- +(45.000000:31.819805pt and 9.899495pt) -- +(135.000000:31.819805pt and 9.899495pt) -- +(225.000000:31.819805pt and 9.899495pt) -- cycle;
\draw (138.500000, -0.000000) node {$R_{z}(- 2 \pi)$};
\end{scope}
\end{scope}
% Line 42: q2 C q1
\draw (176.000000,22.000000) -- (176.000000,0.000000);
\begin{scope}
\draw[fill=white] (176.000000, 0.000000) circle(3.000000pt);
\clip (176.000000, 0.000000) circle(3.000000pt);
\draw (173.000000, 0.000000) -- (179.000000, 0.000000);
\draw (176.000000, -3.000000) -- (176.000000, 3.000000);
\end{scope}
\filldraw (176.000000, 22.000000) circle(1.500000pt);
% Line 43: q1 C q0
\draw (194.000000,44.000000) -- (194.000000,22.000000);
\begin{scope}
\draw[fill=white] (194.000000, 22.000000) circle(3.000000pt);
\clip (194.000000, 22.000000) circle(3.000000pt);
\draw (191.000000, 22.000000) -- (197.000000, 22.000000);
\draw (194.000000, 19.000000) -- (194.000000, 25.000000);
\end{scope}
\filldraw (194.000000, 44.000000) circle(1.500000pt);
% Done with gates; drawing ending labels
% Done with ending labels; drawing cut lines and comments
% Done with comments
\end{tikzpicture}
        \caption{$-(I^{\otimes 3})$ decomposition}
    \end{subfigure}%
\vspace{1em} % optional space between subfigures
    \begin{subfigure}[t]{0.35\textwidth}
        \centering
\definecolor{myblue}{RGB}{157,220,229}
\definecolor{myred}{RGB}{255,99,98}
\begin{tikzpicture}[scale=1.000000,x=1pt,y=1pt]
\filldraw[color=white] (0.000000, -11.000000) rectangle (224.333333, 77.000000);
% Drawing wires
% Line 15: q0 W
\draw[color=black] (0.000000,66.000000) -- (224.333333,66.000000);
% Line 16: q1 W
\draw[color=black] (0.000000,44.000000) -- (224.333333,44.000000);
% Line 17: q2 W
\draw[color=black] (0.000000,22.000000) -- (224.333333,22.000000);
% Line 18: q3 W
\draw[color=black] (0.000000,0.000000) -- (224.333333,0.000000);
% Done with wires; drawing gates
% Line 20: q0 / j
\draw (6.000000, 59.000000) -- (15.333333, 73.000000);
\draw (13.000000, 69.500000) node[right] {$\scriptstyle{j}$};
% Line 22: q1 q2 q3 G:state width=35 $-(I^{\otimes \omega})$ q0
\draw[rounded corners=3pt] (44.833333,66.000000) -- (44.833333,0.000000);
\begin{scope}[rounded corners=3pt]
\begin{scope}
\draw[fill=myblue] (44.833333, 22.000000) +(-45.000000:24.748737pt and 41.012193pt) -- +(45.000000:24.748737pt and 41.012193pt) -- +(135.000000:24.748737pt and 41.012193pt) -- +(225.000000:24.748737pt and 41.012193pt) -- cycle;
\clip (44.833333, 22.000000) +(-45.000000:24.748737pt and 41.012193pt) -- +(45.000000:24.748737pt and 41.012193pt) -- +(135.000000:24.748737pt and 41.012193pt) -- +(225.000000:24.748737pt and 41.012193pt) -- cycle;
\draw (44.833333, 22.000000) node {$-(I^{\otimes \omega})$};
\end{scope}
\end{scope}
\filldraw (44.833333, 66.000000) circle(1.500000pt);
% Line 25: =
\draw[fill=white,color=white] (74.333333, -7.000000) rectangle (89.333333, 73.000000);
\draw (81.833333, 33.000000) node {$=$};
% Line 26: q2 C q1
\draw (104.333333,44.000000) -- (104.333333,22.000000);
\begin{scope}
\draw[fill=white] (104.333333, 22.000000) circle(3.000000pt);
\clip (104.333333, 22.000000) circle(3.000000pt);
\draw (101.333333, 22.000000) -- (107.333333, 22.000000);
\draw (104.333333, 19.000000) -- (104.333333, 25.000000);
\end{scope}
\filldraw (104.333333, 44.000000) circle(1.500000pt);
% Line 27: q3 C q2
\draw (122.333333,22.000000) -- (122.333333,0.000000);
\begin{scope}
\draw[fill=white] (122.333333, 0.000000) circle(3.000000pt);
\clip (122.333333, 0.000000) circle(3.000000pt);
\draw (119.333333, 0.000000) -- (125.333333, 0.000000);
\draw (122.333333, -3.000000) -- (122.333333, 3.000000);
\end{scope}
\filldraw (122.333333, 22.000000) circle(1.500000pt);
% Line 28: q3 G:state width=45 $R_{z}(- 2 \pi)$ q0
\draw[rounded corners=3pt] (159.833333,66.000000) -- (159.833333,0.000000);
\begin{scope}[rounded corners=3pt]
\begin{scope}
\draw[fill=myblue] (159.833333, -0.000000) +(-45.000000:31.819805pt and 9.899495pt) -- +(45.000000:31.819805pt and 9.899495pt) -- +(135.000000:31.819805pt and 9.899495pt) -- +(225.000000:31.819805pt and 9.899495pt) -- cycle;
\clip (159.833333, -0.000000) +(-45.000000:31.819805pt and 9.899495pt) -- +(45.000000:31.819805pt and 9.899495pt) -- +(135.000000:31.819805pt and 9.899495pt) -- +(225.000000:31.819805pt and 9.899495pt) -- cycle;
\draw (159.833333, -0.000000) node {$R_{z}(- 2 \pi)$};
\end{scope}
\end{scope}
\filldraw (159.833333, 66.000000) circle(1.500000pt);
% Line 29: q3 C q2
\draw (197.333333,22.000000) -- (197.333333,0.000000);
\begin{scope}
\draw[fill=white] (197.333333, 0.000000) circle(3.000000pt);
\clip (197.333333, 0.000000) circle(3.000000pt);
\draw (194.333333, 0.000000) -- (200.333333, 0.000000);
\draw (197.333333, -3.000000) -- (197.333333, 3.000000);
\end{scope}
\filldraw (197.333333, 22.000000) circle(1.500000pt);
% Line 30: q2 C q1
\draw (215.333333,44.000000) -- (215.333333,22.000000);
\begin{scope}
\draw[fill=white] (215.333333, 22.000000) circle(3.000000pt);
\clip (215.333333, 22.000000) circle(3.000000pt);
\draw (212.333333, 22.000000) -- (218.333333, 22.000000);
\draw (215.333333, 19.000000) -- (215.333333, 25.000000);
\end{scope}
\filldraw (215.333333, 44.000000) circle(1.500000pt);
% Done with gates; drawing ending labels
% Done with ending labels; drawing cut lines and comments
% Done with comments
\end{tikzpicture}

    \caption{$j$-controlled $-(I^{\otimes 3})$ decomposition}
    \end{subfigure}
    \caption{Quantum circuits to implement $-I$ operation on $\omega$-qubits. (a) Quantum circuit template to implement $-(I^{\otimes w})$ by recursively applying  Equation~\ref{eq:negII}. (b) Efficient implementation of the multi-control $-(I^{\otimes 3})$ operator. \label{fig:QUBO-marker}} 
\end{figure*}

Finally, Figure~\ref{fig:reflection} shows how the reflection step can be compiled efficiently as a quantum circuit. 

\begin{figure*}[b]
     \centering
     \begin{adjustbox}{width=0.95\textwidth}
\tikzset{every picture/.style={line width=0.75pt}} %set default line width to 0.75pt        

\definecolor{mygreen}{RGB}{34,139,33}
\definecolor{myblue}{RGB}{157,220,229}
\definecolor{myred}{RGB}{255,99,98}
\begin{tikzpicture}[scale=1.000000,x=1pt,y=1pt]
\filldraw[color=white] (0.000000, -11.000000) rectangle (658.000000, 99.000000);
% Drawing wires
% Line 14: q0 W 1
\draw[color=black] (0.000000,88.000000) -- (658.000000,88.000000);
\draw[color=black] (0.000000,88.000000) node[left] {$1$};
% Line 15: q1 W 2
\draw[color=black] (0.000000,66.000000) -- (658.000000,66.000000);
\draw[color=black] (0.000000,66.000000) node[left] {$2$};
% Line 16: ...b W
\draw[color=black] (0.000000,44.000000) node[anchor=mid east] {$\vdots$};
% Line 17: q2 W n-1
\draw[color=black] (0.000000,22.000000) -- (658.000000,22.000000);
\draw[color=black] (0.000000,22.000000) node[left] {$n-1$};
% Line 18: q3 W n
\draw[color=black] (0.000000,0.000000) -- (658.000000,0.000000);
\draw[color=black] (0.000000,0.000000) node[left] {$n$};
% Done with wires; drawing gates
% Line 22: q0 q1 q2 q3 G:env width=55 $2|u\rangle \langle u | - I$
\draw[rounded corners=3pt] (33.500000,88.000000) -- (33.500000,0.000000);
\begin{scope}[rounded corners=3pt]
\begin{scope}
\draw[fill=myred] (33.500000, 44.000000) +(-45.000000:38.890873pt and 72.124892pt) -- +(45.000000:38.890873pt and 72.124892pt) -- +(135.000000:38.890873pt and 72.124892pt) -- +(225.000000:38.890873pt and 72.124892pt) -- cycle;
\clip (33.500000, 44.000000) +(-45.000000:38.890873pt and 72.124892pt) -- +(45.000000:38.890873pt and 72.124892pt) -- +(135.000000:38.890873pt and 72.124892pt) -- +(225.000000:38.890873pt and 72.124892pt) -- cycle;
\draw (33.500000, 44.000000) node {$2|u\rangle \langle u | - I$};
\end{scope}
\end{scope}
% Line 24: =
\draw[fill=white,color=white] (73.000000, -7.000000) rectangle (88.000000, 95.000000);
\draw (80.500000, 44.000000) node {$=$};
% Line 26: q0 q1 q2 q3 G:state width=25 $U^{\dagger}$
\draw[rounded corners=3pt] (112.500000,88.000000) -- (112.500000,0.000000);
\begin{scope}[rounded corners=3pt]
\begin{scope}
\draw[fill=myblue] (112.500000, 44.000000) +(-45.000000:17.677670pt and 72.124892pt) -- +(45.000000:17.677670pt and 72.124892pt) -- +(135.000000:17.677670pt and 72.124892pt) -- +(225.000000:17.677670pt and 72.124892pt) -- cycle;
\clip (112.500000, 44.000000) +(-45.000000:17.677670pt and 72.124892pt) -- +(45.000000:17.677670pt and 72.124892pt) -- +(135.000000:17.677670pt and 72.124892pt) -- +(225.000000:17.677670pt and 72.124892pt) -- cycle;
\draw (112.500000, 44.000000) node {$U^{\dagger}$};
\end{scope}
\end{scope}
% Line 28: q0 q1 q2 q3 G:env width=55 $2|0\rangle \langle 0 | - I$
\draw[rounded corners=3pt] (164.500000,88.000000) -- (164.500000,0.000000);
\begin{scope}[rounded corners=3pt]
\begin{scope}
\draw[fill=myred] (164.500000, 44.000000) +(-45.000000:38.890873pt and 72.124892pt) -- +(45.000000:38.890873pt and 72.124892pt) -- +(135.000000:38.890873pt and 72.124892pt) -- +(225.000000:38.890873pt and 72.124892pt) -- cycle;
\clip (164.500000, 44.000000) +(-45.000000:38.890873pt and 72.124892pt) -- +(45.000000:38.890873pt and 72.124892pt) -- +(135.000000:38.890873pt and 72.124892pt) -- +(225.000000:38.890873pt and 72.124892pt) -- cycle;
\draw (164.500000, 44.000000) node {$2|0\rangle \langle 0 | - I$};
\end{scope}
\end{scope}
% Line 30: q0 q1 q2 q3 G:state width=25 $U$
\draw[rounded corners=3pt] (216.500000,88.000000) -- (216.500000,0.000000);
\begin{scope}[rounded corners=3pt]
\begin{scope}
\draw[fill=myblue] (216.500000, 44.000000) +(-45.000000:17.677670pt and 72.124892pt) -- +(45.000000:17.677670pt and 72.124892pt) -- +(135.000000:17.677670pt and 72.124892pt) -- +(225.000000:17.677670pt and 72.124892pt) -- cycle;
\clip (216.500000, 44.000000) +(-45.000000:17.677670pt and 72.124892pt) -- +(45.000000:17.677670pt and 72.124892pt) -- +(135.000000:17.677670pt and 72.124892pt) -- +(225.000000:17.677670pt and 72.124892pt) -- cycle;
\draw (216.500000, 44.000000) node {$U$};
\end{scope}
\end{scope}
% Line 32: =
\draw[fill=white,color=white] (241.000000, -7.000000) rectangle (256.000000, 95.000000);
\draw (248.500000, 44.000000) node {$=$};
% Line 34: q0 q1 q2 q3 G:state width=25 $U^{\dagger}$
\draw[rounded corners=3pt] (280.500000,88.000000) -- (280.500000,0.000000);
\begin{scope}[rounded corners=3pt]
\begin{scope}
\draw[fill=myblue] (280.500000, 44.000000) +(-45.000000:17.677670pt and 72.124892pt) -- +(45.000000:17.677670pt and 72.124892pt) -- +(135.000000:17.677670pt and 72.124892pt) -- +(225.000000:17.677670pt and 72.124892pt) -- cycle;
\clip (280.500000, 44.000000) +(-45.000000:17.677670pt and 72.124892pt) -- +(45.000000:17.677670pt and 72.124892pt) -- +(135.000000:17.677670pt and 72.124892pt) -- +(225.000000:17.677670pt and 72.124892pt) -- cycle;
\draw (280.500000, 44.000000) node {$U^{\dagger}$};
\end{scope}
\end{scope}
% Line 36: q0 G:env $X$
\begin{scope}[rounded corners=3pt]
\begin{scope}
\draw[fill=myred] (312.000000, 88.000000) +(-45.000000:9.899495pt and 9.899495pt) -- +(45.000000:9.899495pt and 9.899495pt) -- +(135.000000:9.899495pt and 9.899495pt) -- +(225.000000:9.899495pt and 9.899495pt) -- cycle;
\clip (312.000000, 88.000000) +(-45.000000:9.899495pt and 9.899495pt) -- +(45.000000:9.899495pt and 9.899495pt) -- +(135.000000:9.899495pt and 9.899495pt) -- +(225.000000:9.899495pt and 9.899495pt) -- cycle;
\draw (312.000000, 88.000000) node {$X$};
\end{scope}
\end{scope}
% Line 37: q1 G:env $X$
\begin{scope}[rounded corners=3pt]
\begin{scope}
\draw[fill=myred] (312.000000, 66.000000) +(-45.000000:9.899495pt and 9.899495pt) -- +(45.000000:9.899495pt and 9.899495pt) -- +(135.000000:9.899495pt and 9.899495pt) -- +(225.000000:9.899495pt and 9.899495pt) -- cycle;
\clip (312.000000, 66.000000) +(-45.000000:9.899495pt and 9.899495pt) -- +(45.000000:9.899495pt and 9.899495pt) -- +(135.000000:9.899495pt and 9.899495pt) -- +(225.000000:9.899495pt and 9.899495pt) -- cycle;
\draw (312.000000, 66.000000) node {$X$};
\end{scope}
\end{scope}
% Line 38: q2 G:env $X$
\begin{scope}[rounded corners=3pt]
\begin{scope}
\draw[fill=myred] (312.000000, 22.000000) +(-45.000000:9.899495pt and 9.899495pt) -- +(45.000000:9.899495pt and 9.899495pt) -- +(135.000000:9.899495pt and 9.899495pt) -- +(225.000000:9.899495pt and 9.899495pt) -- cycle;
\clip (312.000000, 22.000000) +(-45.000000:9.899495pt and 9.899495pt) -- +(45.000000:9.899495pt and 9.899495pt) -- +(135.000000:9.899495pt and 9.899495pt) -- +(225.000000:9.899495pt and 9.899495pt) -- cycle;
\draw (312.000000, 22.000000) node {$X$};
\end{scope}
\end{scope}
% Line 39: q3 G:env $X$
\begin{scope}[rounded corners=3pt]
\begin{scope}
\draw[fill=myred] (312.000000, -0.000000) +(-45.000000:9.899495pt and 9.899495pt) -- +(45.000000:9.899495pt and 9.899495pt) -- +(135.000000:9.899495pt and 9.899495pt) -- +(225.000000:9.899495pt and 9.899495pt) -- cycle;
\clip (312.000000, -0.000000) +(-45.000000:9.899495pt and 9.899495pt) -- +(45.000000:9.899495pt and 9.899495pt) -- +(135.000000:9.899495pt and 9.899495pt) -- +(225.000000:9.899495pt and 9.899495pt) -- cycle;
\draw (312.000000, -0.000000) node {$X$};
\end{scope}
\end{scope}
% Line 42: q3 G:env width=20 $Z$ q0 q1 q2
\draw[rounded corners=3pt] (341.000000,88.000000) -- (341.000000,0.000000);
\begin{scope}[rounded corners=3pt]
\begin{scope}
\draw[fill=myred] (341.000000, -0.000000) +(-45.000000:14.142136pt and 9.899495pt) -- +(45.000000:14.142136pt and 9.899495pt) -- +(135.000000:14.142136pt and 9.899495pt) -- +(225.000000:14.142136pt and 9.899495pt) -- cycle;
\clip (341.000000, -0.000000) +(-45.000000:14.142136pt and 9.899495pt) -- +(45.000000:14.142136pt and 9.899495pt) -- +(135.000000:14.142136pt and 9.899495pt) -- +(225.000000:14.142136pt and 9.899495pt) -- cycle;
\draw (341.000000, -0.000000) node {$Z$};
\end{scope}
\end{scope}
\filldraw (341.000000, 88.000000) circle(1.500000pt);
\filldraw (341.000000, 66.000000) circle(1.500000pt);
\filldraw (341.000000, 22.000000) circle(1.500000pt);
% Line 44: q0 LABEL
% Line 45: q1 LABEL
% Line 46: q2 LABEL
% Line 47: q3 G:env $Z$
\begin{scope}[rounded corners=3pt]
\begin{scope}
\draw[fill=myred] (370.500000, -0.000000) +(-45.000000:9.899495pt and 9.899495pt) -- +(45.000000:9.899495pt and 9.899495pt) -- +(135.000000:9.899495pt and 9.899495pt) -- +(225.000000:9.899495pt and 9.899495pt) -- cycle;
\clip (370.500000, -0.000000) +(-45.000000:9.899495pt and 9.899495pt) -- +(45.000000:9.899495pt and 9.899495pt) -- +(135.000000:9.899495pt and 9.899495pt) -- +(225.000000:9.899495pt and 9.899495pt) -- cycle;
\draw (370.500000, -0.000000) node {$Z$};
\end{scope}
\end{scope}
% Line 49: q0 G:env $X$
\begin{scope}[rounded corners=3pt]
\begin{scope}
\draw[fill=myred] (397.000000, 88.000000) +(-45.000000:9.899495pt and 9.899495pt) -- +(45.000000:9.899495pt and 9.899495pt) -- +(135.000000:9.899495pt and 9.899495pt) -- +(225.000000:9.899495pt and 9.899495pt) -- cycle;
\clip (397.000000, 88.000000) +(-45.000000:9.899495pt and 9.899495pt) -- +(45.000000:9.899495pt and 9.899495pt) -- +(135.000000:9.899495pt and 9.899495pt) -- +(225.000000:9.899495pt and 9.899495pt) -- cycle;
\draw (397.000000, 88.000000) node {$X$};
\end{scope}
\end{scope}
% Line 50: q1 G:env $X$
\begin{scope}[rounded corners=3pt]
\begin{scope}
\draw[fill=myred] (397.000000, 66.000000) +(-45.000000:9.899495pt and 9.899495pt) -- +(45.000000:9.899495pt and 9.899495pt) -- +(135.000000:9.899495pt and 9.899495pt) -- +(225.000000:9.899495pt and 9.899495pt) -- cycle;
\clip (397.000000, 66.000000) +(-45.000000:9.899495pt and 9.899495pt) -- +(45.000000:9.899495pt and 9.899495pt) -- +(135.000000:9.899495pt and 9.899495pt) -- +(225.000000:9.899495pt and 9.899495pt) -- cycle;
\draw (397.000000, 66.000000) node {$X$};
\end{scope}
\end{scope}
% Line 51: q2 G:env $X$
\begin{scope}[rounded corners=3pt]
\begin{scope}
\draw[fill=myred] (397.000000, 22.000000) +(-45.000000:9.899495pt and 9.899495pt) -- +(45.000000:9.899495pt and 9.899495pt) -- +(135.000000:9.899495pt and 9.899495pt) -- +(225.000000:9.899495pt and 9.899495pt) -- cycle;
\clip (397.000000, 22.000000) +(-45.000000:9.899495pt and 9.899495pt) -- +(45.000000:9.899495pt and 9.899495pt) -- +(135.000000:9.899495pt and 9.899495pt) -- +(225.000000:9.899495pt and 9.899495pt) -- cycle;
\draw (397.000000, 22.000000) node {$X$};
\end{scope}
\end{scope}
% Line 54: q3 G:env $X$
\begin{scope}[rounded corners=3pt]
\begin{scope}
\draw[fill=myred] (397.000000, -0.000000) +(-45.000000:9.899495pt and 9.899495pt) -- +(45.000000:9.899495pt and 9.899495pt) -- +(135.000000:9.899495pt and 9.899495pt) -- +(225.000000:9.899495pt and 9.899495pt) -- cycle;
\clip (397.000000, -0.000000) +(-45.000000:9.899495pt and 9.899495pt) -- +(45.000000:9.899495pt and 9.899495pt) -- +(135.000000:9.899495pt and 9.899495pt) -- +(225.000000:9.899495pt and 9.899495pt) -- cycle;
\draw (397.000000, -0.000000) node {$X$};
\end{scope}
\end{scope}
% Line 55: q3 G:env $Z$
\begin{scope}[rounded corners=3pt]
\begin{scope}
\draw[fill=myred] (423.000000, -0.000000) +(-45.000000:9.899495pt and 9.899495pt) -- +(45.000000:9.899495pt and 9.899495pt) -- +(135.000000:9.899495pt and 9.899495pt) -- +(225.000000:9.899495pt and 9.899495pt) -- cycle;
\clip (423.000000, -0.000000) +(-45.000000:9.899495pt and 9.899495pt) -- +(45.000000:9.899495pt and 9.899495pt) -- +(135.000000:9.899495pt and 9.899495pt) -- +(225.000000:9.899495pt and 9.899495pt) -- cycle;
\draw (423.000000, -0.000000) node {$Z$};
\end{scope}
\end{scope}
% Line 57: q0 q1 q2 q3 G:state width=25 $U$
\draw[rounded corners=3pt] (454.500000,88.000000) -- (454.500000,0.000000);
\begin{scope}[rounded corners=3pt]
\begin{scope}
\draw[fill=myblue] (454.500000, 44.000000) +(-45.000000:17.677670pt and 72.124892pt) -- +(45.000000:17.677670pt and 72.124892pt) -- +(135.000000:17.677670pt and 72.124892pt) -- +(225.000000:17.677670pt and 72.124892pt) -- cycle;
\clip (454.500000, 44.000000) +(-45.000000:17.677670pt and 72.124892pt) -- +(45.000000:17.677670pt and 72.124892pt) -- +(135.000000:17.677670pt and 72.124892pt) -- +(225.000000:17.677670pt and 72.124892pt) -- cycle;
\draw (454.500000, 44.000000) node {$U$};
\end{scope}
\end{scope}
% Line 61: =
\draw[fill=white,color=white] (479.000000, -7.000000) rectangle (494.000000, 95.000000);
\draw (486.500000, 44.000000) node {$\propto$};
% Line 64: q0 q1 q2 q3 G:state width=25 $U^{\dagger}$
\draw[rounded corners=3pt] (518.500000,88.000000) -- (518.500000,0.000000);
\begin{scope}[rounded corners=3pt]
\begin{scope}
\draw[fill=myblue] (518.500000, 44.000000) +(-45.000000:17.677670pt and 72.124892pt) -- +(45.000000:17.677670pt and 72.124892pt) -- +(135.000000:17.677670pt and 72.124892pt) -- +(225.000000:17.677670pt and 72.124892pt) -- cycle;
\clip (518.500000, 44.000000) +(-45.000000:17.677670pt and 72.124892pt) -- +(45.000000:17.677670pt and 72.124892pt) -- +(135.000000:17.677670pt and 72.124892pt) -- +(225.000000:17.677670pt and 72.124892pt) -- cycle;
\draw (518.500000, 44.000000) node {$U^{\dagger}$};
\end{scope}
\end{scope}
% Line 66: q0 G:env $X$
\begin{scope}[rounded corners=3pt]
\begin{scope}
\draw[fill=myred] (550.000000, 88.000000) +(-45.000000:9.899495pt and 9.899495pt) -- +(45.000000:9.899495pt and 9.899495pt) -- +(135.000000:9.899495pt and 9.899495pt) -- +(225.000000:9.899495pt and 9.899495pt) -- cycle;
\clip (550.000000, 88.000000) +(-45.000000:9.899495pt and 9.899495pt) -- +(45.000000:9.899495pt and 9.899495pt) -- +(135.000000:9.899495pt and 9.899495pt) -- +(225.000000:9.899495pt and 9.899495pt) -- cycle;
\draw (550.000000, 88.000000) node {$X$};
\end{scope}
\end{scope}
% Line 67: q1 G:env $X$
\begin{scope}[rounded corners=3pt]
\begin{scope}
\draw[fill=myred] (550.000000, 66.000000) +(-45.000000:9.899495pt and 9.899495pt) -- +(45.000000:9.899495pt and 9.899495pt) -- +(135.000000:9.899495pt and 9.899495pt) -- +(225.000000:9.899495pt and 9.899495pt) -- cycle;
\clip (550.000000, 66.000000) +(-45.000000:9.899495pt and 9.899495pt) -- +(45.000000:9.899495pt and 9.899495pt) -- +(135.000000:9.899495pt and 9.899495pt) -- +(225.000000:9.899495pt and 9.899495pt) -- cycle;
\draw (550.000000, 66.000000) node {$X$};
\end{scope}
\end{scope}
% Line 68: q2 G:env $X$
\begin{scope}[rounded corners=3pt]
\begin{scope}
\draw[fill=myred] (550.000000, 22.000000) +(-45.000000:9.899495pt and 9.899495pt) -- +(45.000000:9.899495pt and 9.899495pt) -- +(135.000000:9.899495pt and 9.899495pt) -- +(225.000000:9.899495pt and 9.899495pt) -- cycle;
\clip (550.000000, 22.000000) +(-45.000000:9.899495pt and 9.899495pt) -- +(45.000000:9.899495pt and 9.899495pt) -- +(135.000000:9.899495pt and 9.899495pt) -- +(225.000000:9.899495pt and 9.899495pt) -- cycle;
\draw (550.000000, 22.000000) node {$X$};
\end{scope}
\end{scope}
% Line 69: q3 G:env $X$
\begin{scope}[rounded corners=3pt]
\begin{scope}
\draw[fill=myred] (550.000000, -0.000000) +(-45.000000:9.899495pt and 9.899495pt) -- +(45.000000:9.899495pt and 9.899495pt) -- +(135.000000:9.899495pt and 9.899495pt) -- +(225.000000:9.899495pt and 9.899495pt) -- cycle;
\clip (550.000000, -0.000000) +(-45.000000:9.899495pt and 9.899495pt) -- +(45.000000:9.899495pt and 9.899495pt) -- +(135.000000:9.899495pt and 9.899495pt) -- +(225.000000:9.899495pt and 9.899495pt) -- cycle;
\draw (550.000000, -0.000000) node {$X$};
\end{scope}
\end{scope}
% Line 71: q3 G:env width=20 $Z$ q0 q1 q2
\draw[rounded corners=3pt] (579.000000,88.000000) -- (579.000000,0.000000);
\begin{scope}[rounded corners=3pt]
\begin{scope}
\draw[fill=myred] (579.000000, -0.000000) +(-45.000000:14.142136pt and 9.899495pt) -- +(45.000000:14.142136pt and 9.899495pt) -- +(135.000000:14.142136pt and 9.899495pt) -- +(225.000000:14.142136pt and 9.899495pt) -- cycle;
\clip (579.000000, -0.000000) +(-45.000000:14.142136pt and 9.899495pt) -- +(45.000000:14.142136pt and 9.899495pt) -- +(135.000000:14.142136pt and 9.899495pt) -- +(225.000000:14.142136pt and 9.899495pt) -- cycle;
\draw (579.000000, -0.000000) node {$Z$};
\end{scope}
\end{scope}
\filldraw (579.000000, 88.000000) circle(1.500000pt);
\filldraw (579.000000, 66.000000) circle(1.500000pt);
\filldraw (579.000000, 22.000000) circle(1.500000pt);
% Line 73: q0 G:env $X$
\begin{scope}[rounded corners=3pt]
\begin{scope}
\draw[fill=myred] (608.000000, 88.000000) +(-45.000000:9.899495pt and 9.899495pt) -- +(45.000000:9.899495pt and 9.899495pt) -- +(135.000000:9.899495pt and 9.899495pt) -- +(225.000000:9.899495pt and 9.899495pt) -- cycle;
\clip (608.000000, 88.000000) +(-45.000000:9.899495pt and 9.899495pt) -- +(45.000000:9.899495pt and 9.899495pt) -- +(135.000000:9.899495pt and 9.899495pt) -- +(225.000000:9.899495pt and 9.899495pt) -- cycle;
\draw (608.000000, 88.000000) node {$X$};
\end{scope}
\end{scope}
% Line 74: q1 G:env $X$
\begin{scope}[rounded corners=3pt]
\begin{scope}
\draw[fill=myred] (608.000000, 66.000000) +(-45.000000:9.899495pt and 9.899495pt) -- +(45.000000:9.899495pt and 9.899495pt) -- +(135.000000:9.899495pt and 9.899495pt) -- +(225.000000:9.899495pt and 9.899495pt) -- cycle;
\clip (608.000000, 66.000000) +(-45.000000:9.899495pt and 9.899495pt) -- +(45.000000:9.899495pt and 9.899495pt) -- +(135.000000:9.899495pt and 9.899495pt) -- +(225.000000:9.899495pt and 9.899495pt) -- cycle;
\draw (608.000000, 66.000000) node {$X$};
\end{scope}
\end{scope}
% Line 75: q2 G:env $X$
\begin{scope}[rounded corners=3pt]
\begin{scope}
\draw[fill=myred] (608.000000, 22.000000) +(-45.000000:9.899495pt and 9.899495pt) -- +(45.000000:9.899495pt and 9.899495pt) -- +(135.000000:9.899495pt and 9.899495pt) -- +(225.000000:9.899495pt and 9.899495pt) -- cycle;
\clip (608.000000, 22.000000) +(-45.000000:9.899495pt and 9.899495pt) -- +(45.000000:9.899495pt and 9.899495pt) -- +(135.000000:9.899495pt and 9.899495pt) -- +(225.000000:9.899495pt and 9.899495pt) -- cycle;
\draw (608.000000, 22.000000) node {$X$};
\end{scope}
\end{scope}
% Line 76: q3 G:env $X$
\begin{scope}[rounded corners=3pt]
\begin{scope}
\draw[fill=myred] (608.000000, -0.000000) +(-45.000000:9.899495pt and 9.899495pt) -- +(45.000000:9.899495pt and 9.899495pt) -- +(135.000000:9.899495pt and 9.899495pt) -- +(225.000000:9.899495pt and 9.899495pt) -- cycle;
\clip (608.000000, -0.000000) +(-45.000000:9.899495pt and 9.899495pt) -- +(45.000000:9.899495pt and 9.899495pt) -- +(135.000000:9.899495pt and 9.899495pt) -- +(225.000000:9.899495pt and 9.899495pt) -- cycle;
\draw (608.000000, -0.000000) node {$X$};
\end{scope}
\end{scope}
% Line 79: q0 q1 q2 q3 G:state width=25 $U$
\draw[rounded corners=3pt] (639.500000,88.000000) -- (639.500000,0.000000);
\begin{scope}[rounded corners=3pt]
\begin{scope}
\draw[fill=myblue] (639.500000, 44.000000) +(-45.000000:17.677670pt and 72.124892pt) -- +(45.000000:17.677670pt and 72.124892pt) -- +(135.000000:17.677670pt and 72.124892pt) -- +(225.000000:17.677670pt and 72.124892pt) -- cycle;
\clip (639.500000, 44.000000) +(-45.000000:17.677670pt and 72.124892pt) -- +(45.000000:17.677670pt and 72.124892pt) -- +(135.000000:17.677670pt and 72.124892pt) -- +(225.000000:17.677670pt and 72.124892pt) -- cycle;
\draw (639.500000, 44.000000) node {$U$};
\end{scope}
\end{scope}
% Done with gates; drawing ending labels
\draw[color=black] (658.000000,44.000000) node[anchor=mid west] {$\vdots$};
% Done with ending labels; drawing cut lines and comments
% Done with comments
\end{tikzpicture}

\end{adjustbox}
        \caption{Quantum circuit to implement reflection around $\ket{u} = U\ket{0}$. In the context of this work, the unitary $U$ will either prepare: $U\ket{0}=\ket{s} = H^{\otimes n}\ket{0}$ or $U\ket{0}=\ket{d}$ (Dicke state). Note final right hand circuit is the same up to a global phase of $-1$. This circuit implements $(2|u\rangle\langle u|-I)\ket{\psi} = U(2 |0\rangle\langle0|-I)U^{\dagger}\ket{\psi}$ for an arbitrary input state $\ket{\psi}$.}
        \label{fig:reflection}
\end{figure*}
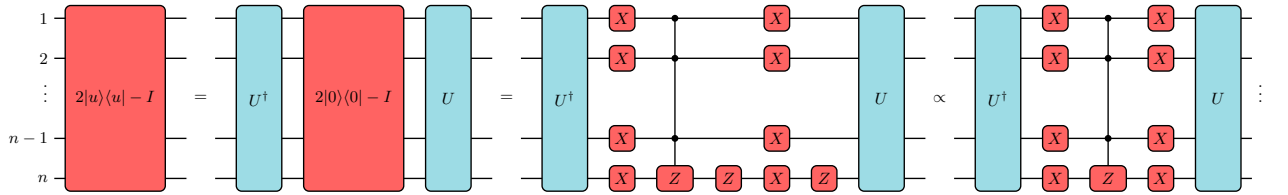

\clearpage
\section{\ce{H3-} results \label{sec:H3_data}}

This section introduces the toy \ce{H3-} model. We reiterate that this system is not classically hard; rather, it is used as a testbed for investigating the properties and behaviour of the GAS-SCF algorithm.

The QUBO corresponding to the determination of the lowest-energy bitstring for \ce{H3-} in the STO-3G basis, expressed in the canonical molecular orbital basis, is given by:

\begin{equation}
\begin{aligned}
    H_{QUBO} = &4 x_{0} x_{1} + 4 x_{0} x_{2} + 4 x_{0} x_{3} + 4 x_{0} x_{4} + 4 x_{0} x_{5} - 14 x_{0} + 4 x_{1} x_{2} + \\
    & 4 x_{1} x_{3} + 4 x_{1} x_{4} + 4 x_{1} x_{5} - 14 x_{1} + 4 x_{2} x_{3} + 4 x_{2} x_{4} + 4 x_{2} x_{5} - \\
    & 10 x_{2} + 4 x_{3} x_{4} + 4 x_{3} x_{5} - 10 x_{3} + 8 x_{4} x_{5} - 6 x_{4} - 6 x_{5}.
\end{aligned}
\end{equation}

The full eigenspectrum for this problem, in the correct particle number sector, can be enumerated as follows:

\begin{table}[h]
\centering
\begin{tabular}{ccc}
\hline
Fock state     & $\langle H \rangle$ & $\langle S^{2} \rangle$ \\ \hline
$\ket{111100}$ & $-24$                        & $0$                    \\
$\ket{111001}$ & $-20$                        & $1$                    \\
$\ket{101101}$ & $-16$                        & $1$                    \\
$\ket{110110}$ & $-20$                        & $1$                    \\
$\ket{110011}$ & $-12$                        & $0$                    \\
$\ket{100111}$ & $-8 $                        & $1$                    \\
$\ket{011110}$ & $-16$                        & $1$                    \\
$\ket{011011}$ & $-8 $                        & $1$                    \\
$\ket{001111}$ & $-4 $                        & $0$                    \\ 
\hline
\end{tabular}
\end{table}

\noindent where we have utilized a $\alpha,\beta,\alpha,\beta...$ spin ordering. 

We observe that there are nine possible states in the Dicke circuit realization of GAS-SCF. To avoid overbalancing, the number of marked states must not exceed half of the search space. In the present case, this limits the number of marked Fock states to at most four. We find that this condition is satisfied only for $y \leq-12$. The thresholding results for different values of $y$ are reported in Table~\ref{tab:GAS_H3_res}. We further note that, if the additional constraint $\langle S^{2}\rangle=0$ were imposed, overbalancing could not occur, since at most three states would be marked. In contrast, imposing $\langle S^{2}\rangle=1$ does not guarantee the absence of overbalancing, as six states could be marked which is still more than half the search space.

In practical GAS-SCF calculations, a conventional SCF solution can be used to inform the selection of $y$, thereby initializing the procedure in the vicinity of an optimized solution. Should overbalancing still occur, the amplitudes of higher-energy Fock states are increased, and thus their probability of being sampled too. Consequently, the frequent observation of energetically unfavorable Fock states (above $y$) provides a diagnostic signal that the threshold used is causing overbalancing. 

\begin{table}[]
\centering
\begin{tabular}{lc|c|c|c|c}
\hline
Iteration & $y=-4$                                                                                                                                                                                                                                                                                             & $y=-8$                                                                                                                                                                                                                                                                                             & $y=-12$                                                                                                                                                                                                                                                                                               & $y=-16$                                                                                                                                                                                                                                                                                               & $y=-20$                                                                                                                                                                                                                                                                                               \\ \hline
1         & \begin{tabular}[c]{@{}l@{}}$0.852 \ket{001111} + $ \\ $0.185 \ket{011110} + $\\ $0.185 \ket{011011} + $ \\ $0.185 \ket{100111} + $ \\ $0.185 \ket{110110} + $ \\ $0.185 \ket{111001} + $ \\ $0.185 \ket{101101} + $ \\ $0.185 \ket{110011} + $ \\ $ \mathbf{0.185 \ket{111100}}  $\end{tabular}  & \begin{tabular}[c]{@{}l@{}}$0.556 \ket{011011} + $ \\ $0.556 \ket{001111} + $ \\ $0.556 \ket{011110} + $ \\ $ \mathbf{0.111 \ket{111100}}+ $ \\ $0.111 \ket{110011} + $ \\ $0.111 \ket{111001} + $ \\ $0.111 \ket{101101} + $ \\ $0.111 \ket{100111} + $ \\ $0.111 \ket{110110}   $\end{tabular} & \begin{tabular}[c]{@{}l@{}}$0.407 \ket{100111} + $ \\ $0.407 \ket{011011} + $ \\ $0.407 \ket{011110} + $ \\ $0.407 \ket{001111} + $ \\ $ \mathbf{0.259 \ket{111100}}+ $ \\ $0.259 \ket{110011} + $ \\ $0.259 \ket{111001} + $ \\ $0.259 \ket{110110} + $ \\ $0.259 \ket{101101}   $\end{tabular} & \begin{tabular}[c]{@{}l@{}}$0.556 \ket{111001} + $ \\ $0.556 \ket{110011} + $ \\ $ \mathbf{0.556 \ket{111100}}+ $ \\ $0.111 \ket{110110} + $ \\ $0.111 \ket{001111} + $ \\ $0.111 \ket{100111} + $ \\ $0.111 \ket{101101} + $ \\ $0.111 \ket{011011} + $ \\ $0.111 \ket{011110}   $\end{tabular} & \begin{tabular}[c]{@{}l@{}}$ \mathbf{0.852 \ket{111100}}+ $ \\ $0.185 \ket{001111} + $ \\ $0.185 \ket{100111} + $ \\ $0.185 \ket{011011} + $ \\ $0.185 \ket{011110} + $ \\ $0.185 \ket{101101} + $ \\ $0.185 \ket{110011} + $ \\ $0.185 \ket{111001} + $ \\ $0.185 \ket{110110}   $\end{tabular} \\ \hline
2         & \begin{tabular}[c]{@{}l@{}}$0.992 \ket{001111} + $ \\ $0.045 \ket{110110} + $ \\ $0.045 \ket{111001} + $ \\ $0.045 \ket{110011} + $ \\ $0.045 \ket{101101} + $ \\ $0.045 \ket{100111} + $ \\ $0.045 \ket{011011} + $ \\ $ \mathbf{0.045 \ket{111100}}+ $ \\ $0.045 \ket{011110}   $\end{tabular} & \begin{tabular}[c]{@{}l@{}}$0.407 \ket{100111} + $ \\ $0.407 \ket{110110} + $ \\ $0.407 \ket{101101} + $ \\ $0.407 \ket{111001} + $ \\ $0.407 \ket{110011} + $ \\ $ \mathbf{0.407 \ket{111100}}+ $ \\ $0.037 \ket{011011} + $ \\ $0.037 \ket{011110} + $ \\ $0.037 \ket{001111}   $\end{tabular} & \begin{tabular}[c]{@{}l@{}}$0.391 \ket{110110} + $ \\ $0.391 \ket{101101} + $ \\ $0.391 \ket{111001} + $ \\ $0.391 \ket{110011} + $ \\ $ \mathbf{0.391 \ket{111100}}+ $ \\ $0.243 \ket{001111} + $ \\ $0.243 \ket{011011} + $ \\ $0.243 \ket{011110} + $ \\ $0.243 \ket{100111}   $\end{tabular} & \begin{tabular}[c]{@{}l@{}}$0.407 \ket{011110} + $ \\ $0.407 \ket{011011} + $ \\ $0.407 \ket{100111} + $ \\ $0.407 \ket{001111} + $ \\ $0.407 \ket{101101} + $ \\ $0.407 \ket{110110} + $ \\ $ \mathbf{0.037 \ket{111100}}+ $ \\ $0.037 \ket{111001} + $ \\ $0.037 \ket{110011}   $\end{tabular} & \begin{tabular}[c]{@{}l@{}}$ \mathbf{0.992 \ket{111100}}+ $ \\ $0.045 \ket{100111} + $ \\ $0.045 \ket{001111} + $ \\ $0.045 \ket{011011} + $ \\ $0.045 \ket{111001} + $ \\ $0.045 \ket{011110} + $ \\ $0.045 \ket{110110} + $ \\ $0.045 \ket{101101} + $ \\ $0.045 \ket{110011}   $\end{tabular} \\ \hline
3         & \begin{tabular}[c]{@{}l@{}}$0.691 \ket{001111} + $ \\ $0.256 \ket{011011} + $ \\ $0.256 \ket{100111} + $ \\ $0.256 \ket{111001} + $ \\ $0.256 \ket{110110} + $ \\ $0.256 \ket{101101} + $ \\ $0.256 \ket{110011} + $ \\ $0.256 \ket{011110} + $ \\ $ \mathbf{0.256 \ket{111100}}  $\end{tabular} & \begin{tabular}[c]{@{}l@{}}$0.531 \ket{011110} + $ \\ $0.531 \ket{011011} + $ \\ $0.531 \ket{001111} + $ \\ $0.16 \ket{100111} + $ \\ $0.16 \ket{110110} + $ \\ $ \mathbf{0.16 \ket{111100}}+ $ \\ $0.16 \ket{111001} + $ \\ $0.16 \ket{110011} + $ \\ $0.16 \ket{101101}   $\end{tabular}       & \begin{tabular}[c]{@{}l@{}}$0.461 \ket{001111} + $ \\ $0.461 \ket{011011} + $ \\ $0.461 \ket{011110} + $ \\ $0.461 \ket{100111} + $ \\ $0.172 \ket{110110} + $ \\ $0.172 \ket{111001} + $ \\ $0.172 \ket{101101} + $ \\ $0.172 \ket{110011} + $ \\ $ \mathbf{0.172 \ket{111100}}  $\end{tabular} & \begin{tabular}[c]{@{}l@{}}$0.531 \ket{110011} + $ \\ $0.531 \ket{111001} + $ \\ $ \mathbf{0.531 \ket{111100}}+ $ \\ $0.16 \ket{001111} + $ \\ $0.16 \ket{101101} + $ \\ $0.16 \ket{110110} + $ \\ $0.16 \ket{011011} + $ \\ $0.16 \ket{011110} + $ \\ $0.16 \ket{100111}   $\end{tabular}       & \begin{tabular}[c]{@{}l@{}}$ \mathbf{0.691 \ket{111100}}+ $ \\ $0.256 \ket{011011} + $ \\ $0.256 \ket{011110} + $ \\ $0.256 \ket{001111} + $ \\ $0.256 \ket{100111} + $ \\ $0.256 \ket{110110} + $ \\ $0.256 \ket{110011} + $ \\ $0.256 \ket{111001} + $ \\ $0.256 \ket{101101}   $\end{tabular} \\ \hline
4         & \begin{tabular}[c]{@{}l@{}}$0.352 \ket{011110} + $ \\ $0.352 \ket{011011} + $ \\ $0.352 \ket{100111} + $ \\ $0.352 \ket{110110} + $ \\ $0.352 \ket{101101} + $ \\ $0.352 \ket{111001} + $ \\ $0.352 \ket{110011} + $ \\ $ \mathbf{0.352 \ket{111100}}+ $ \\ $0.083 \ket{001111}   $\end{tabular} & \begin{tabular}[c]{@{}l@{}}$0.391 \ket{001111} + $ \\ $0.391 \ket{011110} + $ \\ $0.391 \ket{011011} + $ \\ $0.3 \ket{111001} + $ \\ $0.3 \ket{110110} + $ \\ $ \mathbf{0.3 \ket{111100}}+ $ \\ $0.3 \ket{110011} + $ \\ $0.3 \ket{100111} + $ \\ $0.3 \ket{101101}   $\end{tabular}             & \begin{tabular}[c]{@{}l@{}}$0.429 \ket{101101} + $ \\ $0.429 \ket{110011} + $ \\ $0.429 \ket{111001} + $ \\ $0.429 \ket{110110} + $ \\ $ \mathbf{0.429 \ket{111100}}+ $ \\ $0.14 \ket{011110} + $ \\ $0.14 \ket{001111} + $ \\ $0.14 \ket{011011} + $ \\ $0.14 \ket{100111}   $\end{tabular}     & \begin{tabular}[c]{@{}l@{}}$0.391 \ket{110011} + $ \\ $0.391 \ket{111001} + $ \\ $ \mathbf{0.391 \ket{111100}}+ $ \\ $0.3 \ket{001111} + $ \\ $0.3 \ket{100111} + $ \\ $0.3 \ket{110110} + $ \\ $0.3 \ket{011110} + $ \\ $0.3 \ket{101101} + $ \\ $0.3 \ket{011011}   $\end{tabular}             & \begin{tabular}[c]{@{}l@{}}$0.352 \ket{001111} + $ \\ $0.352 \ket{100111} + $ \\ $0.352 \ket{011011} + $ \\ $0.352 \ket{011110} + $ \\ $0.352 \ket{111001} + $ \\ $0.352 \ket{101101} + $ \\ $0.352 \ket{110110} + $ \\ $0.352 \ket{110011} + $ \\ $ \mathbf{0.083 \ket{111100}}  $\end{tabular} \\ \hline
\end{tabular}
\caption{Statevector obtained after applying different numbers of Grover iterations during a GAS calculation for various threshold values $y$. Note that the problem is overbalanced for all $y>-12$. The lowest energy Fock state is highlighted in bold.  \label{tab:GAS_H3_res}}
\end{table}

\clearpage
\section{Linear (triplet) \ce{O3} results \label{sec:O3_data}}
Tables \ref{tab:O3_ROHF} and \ref{tab:O3_SO_ROHF} present the raw results corresponding to Figure\ref{fig:O3_linear}. Tables \ref{tab:O3_ROHF_conv} and \ref{tab:O3_SO_ROHF_conv} provide the associated convergence data. All data is also available online at \cite{gas_github}. All reported Fock-state solutions were found in the correct particle-number and spin sector, corresponding to a triplet state with $S^{2}=2$.

\begin{table}[ht]
\centering
\begin{subtable}{0.65\textwidth}
\centering
\caption{Energies (Ha)}
\label{tab:O3_ROHF}
\begin{tabular}{llll}
\hline
 & cc-pVDZ & cc-pVTZ & cc-pVQZ \\
\hline
1e & (-224.00169, -223.88841) & (-224.05179, -224.05179) & (-224.21012, -224.21012) \\
atom & (-223.92771, -223.85721) & (-223.97658, -223.91233) & (-223.99071, -223.92923) \\
huckel & (-224.13916, -224.13916) & (-224.19363, -224.19363) & (-224.21012, -224.21012) \\
minao & (-223.92771, -223.85721) & (-223.97658, -223.91233) & (-223.99071, -223.92923) \\
sap & (-224.00169, -223.88841) & (-224.04739, -223.94295) & (-224.06144, -223.99572) \\
vsap & (-223.92771, -223.85721) & (-223.97658, -223.91233) & (-223.99071, -223.92923) \\
\hline
\end{tabular}

\end{subtable}
\hfill
\begin{subtable}{0.65\textwidth}
\centering
\caption{Convergence PySCF}
\label{tab:O3_ROHF_conv}
\begin{tabular}{llll}
\hline
 & cc-pVDZ & cc-pVTZ & cc-pVQZ \\
\hline
1e & (True, False) & (True, False) & (True, True) \\
atom & (True, False) & (True, False) & (True, False) \\
huckel & (True, True) & (True, True) & (True, True) \\
minao & (True, False) & (True, False) & (True, False) \\
sap & (True, False) & (True, False) & (True, False) \\
vsap & (True, False) & (True, False) & (True, False) \\
\hline
\end{tabular}
\end{subtable}
\caption{Top: Total energies (Ha) obtained for \ce{O3}, reported as ordered pairs $(E_{\ket{x}}, E_{\mathrm{PySCF}})$, where the left value corresponds to the alternative Fock-state solution and the right value corresponds to the PySCF ROHF solution. Bottom: Convergence information for the PySCF ROHF calculations of \ce{O3}. Each entry is reported as a Boolean pair $(\text{SCF converged}, \text{internally stable})$, indicating whether the SCF procedure converged and whether the resulting solution passed the internal stability analysis.}
\end{table}

\begin{table}[ht]
\centering
\begin{subtable}{0.65\textwidth}
\centering
\caption{Energies (Ha)}
\label{tab:O3_SO_ROHF}
\begin{tabular}{llll}
\hline
 & cc-pVDZ & cc-pVTZ & cc-pVQZ \\
\hline
1e & (-224.06688, -223.82783) & (-224.19363, -224.19363) & (-224.0023, -222.21971) \\
atom & (-223.97525, -223.82995) & (-224.11787, -223.88555) & (-224.13202, -223.90338) \\
huckel & (-224.13916, -224.13916) & (-224.19363, -224.19363) & (-224.21012, -224.21012) \\
minao & (-223.97525, -223.82995) & (-224.11787, -223.88555) & (-224.13202, -223.90338) \\
sap & (-224.07051, -224.07051) & (-224.19107, -224.19107) & (-224.20713, -224.20713) \\
vsap & (-223.97525, -223.82995) & (-224.11787, -223.88555) & (-224.13202, -223.90338) \\
\hline
\end{tabular}
\end{subtable}
\hfill
\begin{subtable}{0.65\textwidth}
\centering
\caption{Convergence PySCF}
\label{tab:O3_SO_ROHF_conv}
\begin{tabular}{llll}
\hline
 & cc-pVDZ & cc-pVTZ & cc-pVQZ \\
\hline
1e & (True, False) & (True, True) & (True, False) \\
atom & (True, False) & (True, False) & (True, False) \\
huckel & (True, True) & (True, True) & (True, True) \\
minao & (True, False) & (True, False) & (True, False) \\
sap & (True, False) & (True, True) & (True, True) \\
vsap & (True, False) & (True, False) & (True, False) \\
\hline
\end{tabular}
\end{subtable}
\caption{Top: Total energies (Ha) obtained for \ce{O3}, reported as ordered pairs $(E_{\ket{x}}, E_{\mathrm{PySCF}})$, where the left value corresponds to the alternative Fock-state solution and the right value corresponds to the PySCF SO-ROHF solution. Bottom: Convergence information for the PySCF SO-ROHF calculations of \ce{O3}. Each entry is reported as a Boolean pair $(\text{SCF converged}, \text{internally stable})$, indicating whether the SCF procedure converged and whether the resulting solution passed the internal stability analysis.}
\end{table}

\clearpage
\section{Fault tolerant compilation}\label{sec:FT_comp}

Simon \textit{et al.} \cite[Appendix D]{Simon2025ladderoperatorblock} describe how to perform quantum arithmetic for known classical integer values. These results can be applied directly to obtain the $T$-gate cost for GAS-SCF (note other adder constructions are also possible). Figures \ref{fig:single_adder} and \ref{fig:double_adder} summarize the adder circuits required.
The implementation of the singly controlled adder circuit requires $4(m - 1)$ $T$ gates and $(m - 1)$ clean ancilla qubits \cite{Simon2025ladderoperatorblock}. Each doubly controlled adder circuit (Figure \ref{fig:double_adder}) incurs an additional cost of $4$ $T$ gates and one extra ancilla, bringing the total to $4m$ $T$ gates with $m$ clean ancilla \cite{Gidney2018halvingcostof, Simon2025ladderoperatorblock}.
Importantly, this approach maintains the cost function values in two’s complement representation (without using the Fourier basis), and therefore does not require an inverse quantum Fourier transform. Table \ref{tab:circuit_cost2} provides a summary of the overall gate costs for a single iteration. In this realization $\mathcal{O}(m)$ extra ancilla qubits are required for the adder circuits controlled from the QUBO register onto the $m+1$ register and $\mathcal{O}(n)$ additional ancilla qubits for the adder circuits controlled from the QUBO register storing the number of electrons in the $\nu$ and $\mu$ registers. This $\mathcal{O}(n)$ extra ancilla is not required for the Dicke realization, as the search space is already in the correct number sector.

\begin{figure*}[h]
     \centering
     \begin{adjustbox}{width=0.55\textwidth}
\providecommand{\ket}[1]{|#1\rangle}
\providecommand{\bra}[1]{\langle #1|}
\definecolor{mygreen}{RGB}{34,139,33}
\definecolor{myblue}{RGB}{157,220,229}
\definecolor{myred}{RGB}{255,99,98}
\begin{tikzpicture}[scale=1.500000,x=1pt,y=1pt]
\filldraw[color=white] (0.000000, -11.000000) rectangle (170.666667, 55.000000);
% Drawing wires
% Line 17: q0 W
\draw[color=black] (0.000000,44.000000) -- (170.666667,44.000000);
% Line 18: q1 W
\draw[color=black] (98.833333,22.000000) -- (170.666667,22.000000);
% Line 19: q2 W
\draw[color=black] (0.000000,0.000000) -- (170.666667,0.000000);
% Done with wires; drawing gates
% Line 21: q2 / m
\draw (6.000000, -7.000000) -- (15.333333, 7.000000);
\draw (13.000000, 3.500000) node[right] {$\scriptstyle{m}$};
% Line 23: q2 G:state width=25 $\text{Add}(x)$ q0
\draw[rounded corners=3pt] (39.833333,44.000000) -- (39.833333,0.000000);
\begin{scope}[rounded corners=3pt]
\begin{scope}
\draw[fill=myblue] (39.833333, -0.000000) +(-45.000000:17.677670pt and 9.899495pt) -- +(45.000000:17.677670pt and 9.899495pt) -- +(135.000000:17.677670pt and 9.899495pt) -- +(225.000000:17.677670pt and 9.899495pt) -- cycle;
\clip (39.833333, -0.000000) +(-45.000000:17.677670pt and 9.899495pt) -- +(45.000000:17.677670pt and 9.899495pt) -- +(135.000000:17.677670pt and 9.899495pt) -- +(225.000000:17.677670pt and 9.899495pt) -- cycle;
\draw (39.833333, -0.000000) node {$\text{Add}(x)$};
\end{scope}
\end{scope}
\filldraw (39.833333, 44.000000) circle(1.500000pt);
% Line 25: =
\draw[fill=white,color=white] (64.333333, -7.000000) rectangle (79.333333, 51.000000);
\draw (71.833333, 22.000000) node {$=$};
% Line 27: q1 START
\draw[color=black] (106.333333,22.000000) node[fill=white,left,minimum height=22.000000pt,minimum width=15.000000pt,inner sep=0pt] {\phantom{$\ket{0^{\otimes w}}$}};
\draw[color=black] (106.333333,22.000000) node[left] {$\ket{0^{\otimes w}}$};
% Line 30: q2 / m
\draw (94.166667, -7.000000) -- (103.500000, 7.000000);
\draw (101.166667, 3.500000) node[right] {$\scriptstyle{m}$};
% Line 29: q1 / w
\draw (118.333333, 15.000000) -- (127.666667, 29.000000);
\draw (125.333333, 25.500000) node[right] {$\scriptstyle{w}$};
% Line 32: q1 q2 G:state width=25 $\text{Add}(x)$ q0
\draw[rounded corners=3pt] (152.166667,44.000000) -- (152.166667,0.000000);
\begin{scope}[rounded corners=3pt]
\begin{scope}
\draw[fill=myblue] (152.166667, 11.000000) +(-45.000000:17.677670pt and 25.455844pt) -- +(45.000000:17.677670pt and 25.455844pt) -- +(135.000000:17.677670pt and 25.455844pt) -- +(225.000000:17.677670pt and 25.455844pt) -- cycle;
\clip (152.166667, 11.000000) +(-45.000000:17.677670pt and 25.455844pt) -- +(45.000000:17.677670pt and 25.455844pt) -- +(135.000000:17.677670pt and 25.455844pt) -- +(225.000000:17.677670pt and 25.455844pt) -- cycle;
\draw (152.166667, 11.000000) node {$\text{Add}(x)$};
\end{scope}
\end{scope}
\filldraw (152.166667, 44.000000) circle(1.500000pt);
% Done with gates; drawing ending labels
\draw[color=black] (170.666667,22.000000) node[right] {$\ket{0^{\otimes w}}$};
% Done with ending labels; drawing cut lines and comments
% Done with comments
\end{tikzpicture}
\end{adjustbox}
        \caption{General singly controlled adder circuit. The number of clean ancilla required is $w = m - 1$. See \cite[Appendix D]{Simon2025ladderoperatorblock} for how to compile the adder circuit fully.} 
        \label{fig:single_adder}
\end{figure*}
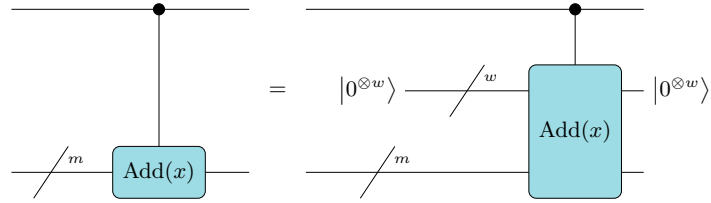

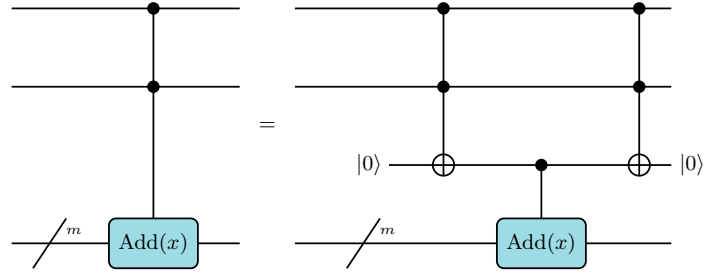
\begin{figure*}[h]
     \centering
     \begin{adjustbox}{width=0.55\textwidth}
\tikzset{every picture/.style={line width=0.75pt}} %set default line width to 0.75pt        
\providecommand{\ket}[1]{ |#1\rangle}
\providecommand{\bra}[1]{\langle #1|}
\definecolor{mygreen}{RGB}{34,139,33}
\definecolor{myblue}{RGB}{157,220,229}
\definecolor{myred}{RGB}{255,99,98}
\begin{tikzpicture}[scale=1.500000,x=1pt,y=1pt]
\filldraw[color=white] (0.000000, -11.000000) rectangle (185.333333, 77.000000);
% Drawing wires
% Line 17: q0 W
\draw[color=black] (0.000000,66.000000) -- (185.333333,66.000000);
% Line 18: q1 W
\draw[color=black] (0.000000,44.000000) -- (185.333333,44.000000);
% Line 19: q2 W
\draw[color=black] (98.833333,22.000000) -- (185.333333,22.000000);
% Line 20: q3 W
\draw[color=black] (0.000000,0.000000) -- (185.333333,0.000000);
% Done with wires; drawing gates
% Line 22: q3 / m
\draw (6.000000, -7.000000) -- (15.333333, 7.000000);
\draw (13.000000, 3.500000) node[right] {$\scriptstyle{m}$};
% Line 25: q3 G:state width=25 $\text{Add}(x)$ q0 q1
\draw[rounded corners=3pt] (39.833333,66.000000) -- (39.833333,0.000000);
\begin{scope}[rounded corners=3pt]
\begin{scope}
\draw[fill=myblue] (39.833333, -0.000000) +(-45.000000:17.677670pt and 9.899495pt) -- +(45.000000:17.677670pt and 9.899495pt) -- +(135.000000:17.677670pt and 9.899495pt) -- +(225.000000:17.677670pt and 9.899495pt) -- cycle;
\clip (39.833333, -0.000000) +(-45.000000:17.677670pt and 9.899495pt) -- +(45.000000:17.677670pt and 9.899495pt) -- +(135.000000:17.677670pt and 9.899495pt) -- +(225.000000:17.677670pt and 9.899495pt) -- cycle;
\draw (39.833333, -0.000000) node {$\text{Add}(x)$};
\end{scope}
\end{scope}
\filldraw (39.833333, 66.000000) circle(1.500000pt);
\filldraw (39.833333, 44.000000) circle(1.500000pt);
% Line 27: =
\draw[fill=white,color=white] (64.333333, -7.000000) rectangle (79.333333, 73.000000);
\draw (71.833333, 33.000000) node {$=$};
% Line 29: q2 START
\draw[color=black] (106.333333,22.000000) node[fill=white,left,minimum height=22.000000pt,minimum width=15.000000pt,inner sep=0pt] {\phantom{$\ket{0}$}};
\draw[color=black] (106.333333,22.000000) node[left] {$\ket{0}$};
% Line 31: q3 / m
\draw (94.166667, -7.000000) -- (103.500000, 7.000000);
\draw (101.166667, 3.500000) node[right] {$\scriptstyle{m}$};
% Line 33: q0 q1 +q2
\draw (121.333333,66.000000) -- (121.333333,22.000000);
\filldraw (121.333333, 66.000000) circle(1.500000pt);
\filldraw (121.333333, 44.000000) circle(1.500000pt);
\begin{scope}
\draw[fill=white] (121.333333, 22.000000) circle(3.000000pt);
\clip (121.333333, 22.000000) circle(3.000000pt);
\draw (118.333333, 22.000000) -- (124.333333, 22.000000);
\draw (121.333333, 19.000000) -- (121.333333, 25.000000);
\end{scope}
% Line 34: q3 G:state width=25 $\text{Add}(x)$ q2
\draw[rounded corners=3pt] (148.833333,22.000000) -- (148.833333,0.000000);
\begin{scope}[rounded corners=3pt]
\begin{scope}
\draw[fill=myblue] (148.833333, -0.000000) +(-45.000000:17.677670pt and 9.899495pt) -- +(45.000000:17.677670pt and 9.899495pt) -- +(135.000000:17.677670pt and 9.899495pt) -- +(225.000000:17.677670pt and 9.899495pt) -- cycle;
\clip (148.833333, -0.000000) +(-45.000000:17.677670pt and 9.899495pt) -- +(45.000000:17.677670pt and 9.899495pt) -- +(135.000000:17.677670pt and 9.899495pt) -- +(225.000000:17.677670pt and 9.899495pt) -- cycle;
\draw (148.833333, -0.000000) node {$\text{Add}(x)$};
\end{scope}
\end{scope}
\filldraw (148.833333, 22.000000) circle(1.500000pt);
% Line 35: q0 q1 +q2
\draw (176.333333,66.000000) -- (176.333333,22.000000);
\filldraw (176.333333, 66.000000) circle(1.500000pt);
\filldraw (176.333333, 44.000000) circle(1.500000pt);
\begin{scope}
\draw[fill=white] (176.333333, 22.000000) circle(3.000000pt);
\clip (176.333333, 22.000000) circle(3.000000pt);
\draw (173.333333, 22.000000) -- (179.333333, 22.000000);
\draw (176.333333, 19.000000) -- (176.333333, 25.000000);
\end{scope}
% Done with gates; drawing ending labels
\draw[color=black] (185.333333,22.000000) node[right] {$\ket{0}$};
% Done with ending labels; drawing cut lines and comments
% Done with comments
\end{tikzpicture}
\end{adjustbox}
        \caption{General doubly controlled adder circuit. This construction uses a single additional ancilla qubit to store the parity of the control qubits, after which a singly controlled adder (Figure \ref{fig:single_adder}) is applied. Gidney \cite{Gidney2018halvingcostof} demonstrates that this can be implemented efficiently using four $T$ gates together with a conditioned controlled-$Z$ operation.} 
        \label{fig:double_adder}
\end{figure*}

\begin{table*}[t]
    \centering
    \begin{adjustbox}{width=\textwidth}
    \begin{tabular}{c@{\hskip 20pt}c@{\hskip 20pt}c@{\hskip 20pt}c@{\hskip 20pt}c c}
        \hline \hline
            & \textbf{Gate Type} & \textbf{Gate Count} & \textbf{Asymptotic Gate count} & \textbf{Asymptotic T count} & \textbf{Note} \\ 
        \hline
        (1) 
            & Singly-controlled Adder & At most $2n$& $\mathcal{O}(n)$ & $\mathcal{O}(n m)$ & $f(\vec{x})$ linear terms \\
            & Doubly-controlled Adder & At most $2\binom{n}{2}$& $\mathcal{O}(n^{2})$ & $\mathcal{O}(n^{2}m)$ & $f(\vec{x})$ quadratic terms \\
            & $(n-1)$-controlled $Z$ & $1$& $\mathcal{O}(1)$ & $\mathcal{O}(n)$ & For $2|0^{\otimes n}\rangle \langle 0^{\otimes n}|-I^{\otimes n}$  reflection \\
            & $X$ & $2n$& $\mathcal{O}(n)$ & $0$ & For $2|0^{\otimes n}\rangle \langle 0^{\otimes n}|-I^{\otimes n}$ reflection  \\
            & $CNOT$ & $2(n-1)$& $\mathcal{O}(n)$ & $0$ & For marking operation \\
        \hline
        (2) 
            & Singly-controlled Adder & $2n$& $\mathcal{O}(n)$ & $\mathcal{O}(n m)$ & $\hat{N}_{\alpha / \beta}(x)$ linear terms \\
            & $\mu$-controlled $X$ & $2$& $\mathcal{O}(1)$ & $\mathcal{O}(n)$ & To mark correct Hamming weight states ($\alpha$ electrons)\\
            & $\nu$-controlled $X$ & $2$& $\mathcal{O}(1)$ & $\mathcal{O}(n)$ & To mark correct Hamming weight states ($\beta$ electrons)\\
            & Toffoli & $2$& $\mathcal{O}(1)$ & $\mathcal{O}(1)$ & Check for correct number of  ${\alpha}$ \& ${\beta}$ electrons \\
        \hline
        (3) & $H$ & $2n$& $\mathcal{O}(n)$ & $0$ & To reflect around $\ket{s}$ on $n$-qubits \\
            & Doubly-controlled $R_{Z}(-2\pi)$ & $1$ & $\mathcal{O}(1)$ & $\mathcal{O}(1)$ & For marking operation \\
        \hline
        (4) & Dicke state construction. See \textit{e.g.} \cite{bartschi2019deterministic, bartschi2022short} & 
        $2 \big( w(n/2,n_{\alpha}) + w(n/2,n_{\beta}) \big)$\footnote{$w(n,k)$  used as a placeholder for quantum circuit cost to generate $\binom{n}{k}$ Dicke state.} & $\mathcal{O}(kn)\leq \mathcal{O}(n^{2})$\footnote{Scaling for the Dicke circuit implementation we used is at worst quadratic when $k = \mathcal{O}(n)$ \cite{bartschi2019deterministic}; other approaches may differ.} & \cite{bartschi2019deterministic} & To reflect around $\ket{d}$ on $n$-qubits \\
            & Singly-controlled $R_{Z}(-2\pi)$ & $1$& $\mathcal{O}(1)$ & $0$ & For marking operation \\
        \hline \hline
    \end{tabular}
    \end{adjustbox}
    \caption{Gate requirements for one repetition  of GAS-SCF that is repeated $L$ times. Adder circuits for this implementation are given in Figures \ref{fig:single_adder} and \ref{fig:double_adder}. The table is broken into four sections as follows: (1), gates required for both $\ket{s}$ and $\ket{d}$ algorithms; (2), extra gates required to enforce occupation numbers (for $\ket{s}$ version only); (3), extra gates required for reflection (for $\ket{s}$ version only); (4), extra gates required for reflection (for $\ket{d}$ version only). Here $\alpha$ and $\beta$ denote the number of spin-up and spin-down electrons respectively. Note no $QFT^{\dagger}$ is required. In this realization $\mathcal{O}(m+n)$ extra ancilla qubits are required for the adder circuits for $\ket{s}$ and $\mathcal{O}(m)$ for $\ket{d}$. }
    \label{tab:circuit_cost2}
\end{table*}

\clearpage
\section{Chromium dimer study} \label{sec:Cr2res}

In Figure \ref{fig:Cr2_linear}, we present an additional example for a hard SCF problem: triplet \ce{Cr2}, with a bond length of $3$ {\AA}. We compare our search for an improved Fock state solution against results from PySCF using various SCF initialization strategies keeping the MO basis fixed from the optimized PySCF calculation. The same computational setup as the \ce{O3} results was used for these results. Just as the \ce{O3} results, we find many situations where within a fixed MO basis there exist lower energy single–Fock-state solutions in the correct sector. In this example, the cc-pVDZ, cc-pVTZ, and cc-pVQZ basis sets correspond to search spaces of sizes $\binom{86}{25}\binom{86}{23} \approx 1.45 \times 10^{42}$, $\binom{136}{25}\binom{136}{23} \approx 8.49 \times 10 ^{52}$ and $ \binom{208}{25}\binom{208}{23} \approx 2.90 \times 10^{62}$ respectively. All reported Fock-state solutions were found in the correct particle-number and spin sector, corresponding to a triplet state with $S^{2}=2$.

\begin{figure*}[ht]
    \centering
\includegraphics[width=0.95\linewidth]{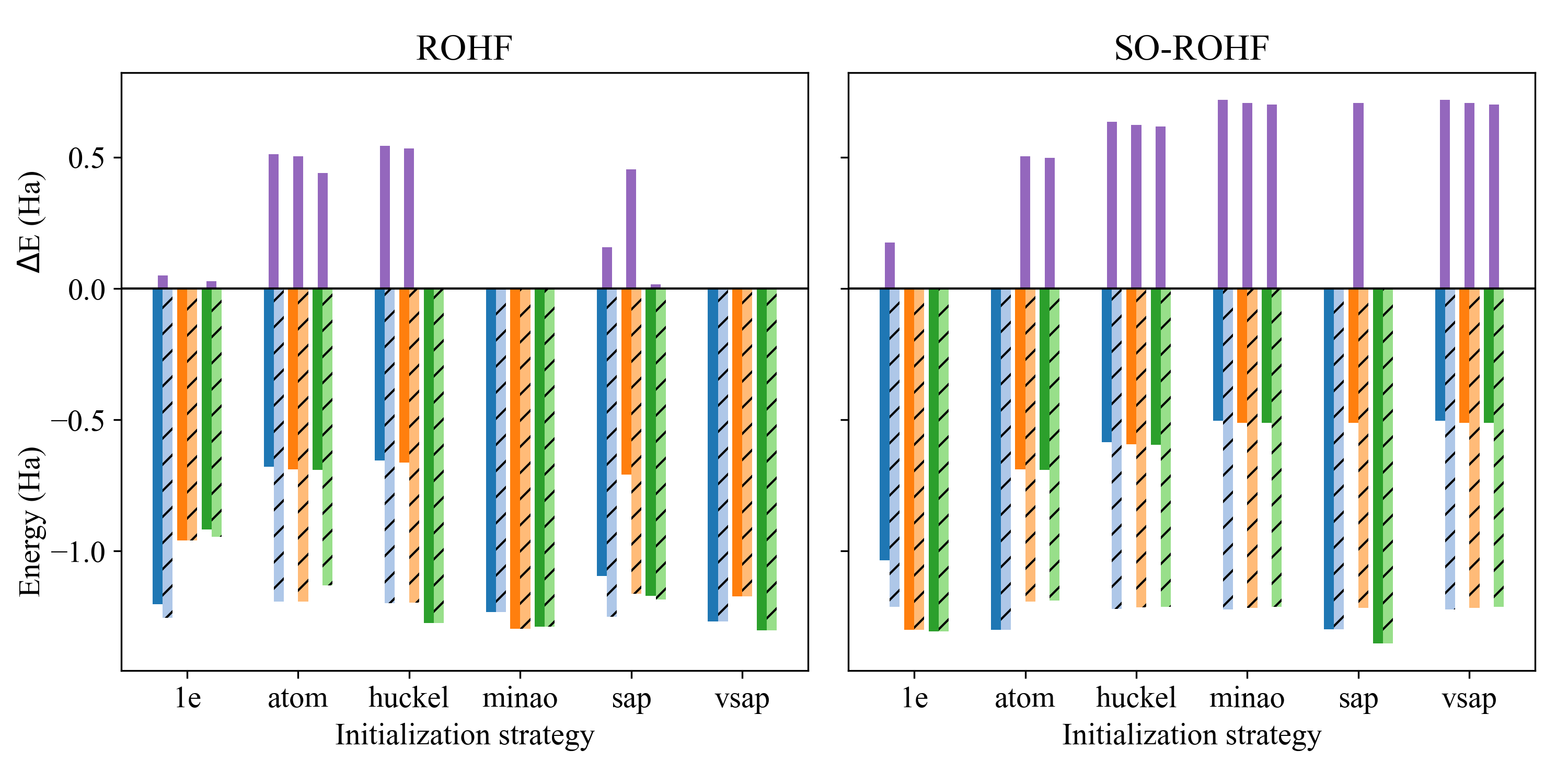}
    \caption{Comparison of SCF initialization strategies for triplet \ce{Cr2}, showing the PySCF solution and alternative single Fock-state solutions when found. The left panels show ROHF results, and the right panels show second-order ROHF (SO-ROHF) results. In the lower panels, results are grouped into cc-pVDZ, cc-pVTZ, and cc-pVQZ basis sets (from left to right); within each basis-set pair, the left bar corresponds to the PySCF solution and the right bar (hatched) to the alternative Fock-state solution $\ket{x}$. The upper panels show the corresponding energy difference, $\Delta E = E_{PySCF} - E_{\ket{x}} $, where positive values indicate that $\ket{x}$ is lower in energy than the PySCF solution. The lower panels show the total energies obtained from each calculation, shifted by $2085$ Ha for clarity. The initialization strategies presented are superposition of atomic densities (MINAO, atom) \cite{almlof1982principles, van2006starting}, the core-Hamiltonian guess (1e) \cite{lehtola2019assessment}, the parameter-free H\"uckel guess \cite{lehtola2019assessment}, and superposition of atomic potentials (VSAP, SAP) \cite{lehtola2019assessment}. Additional details, including the improved Fock-state solutions $\ket{x}$, are available in the online repository \cite{gas_github}. The cc-pVDZ, cc-pVTZ, and cc-pVQZ calculations correspond to systems with $172$, $272$, and $416$ qubits, respectively, excluding ancillary qubits.}
    \label{fig:Cr2_linear}
\end{figure*}

Tables \ref{tab:Cr2_ROHF},  \ref{tab:Cr2_SO_ROHF}, \ref{tab:Cr2_ROHF_conv} and \ref{tab:Cr2_SO_ROHF_conv} provide the associated numerical data. All data is also available online at \cite{gas_github}.

\begin{table}[ht]
\centering
\begin{subtable}{0.65\textwidth}
\centering
\caption{Energies (Ha)}
\label{tab:Cr2_ROHF}
\begin{tabular}{llll}
\hline
 & cc-pVDZ & cc-pVTZ & cc-pVQZ \\
\hline
1e & (-2086.2534, -2086.2028) & (-2085.95949, -2085.95949) & (-2085.94593, -2085.91855) \\
atom & (-2086.19189, -2085.67975) & (-2086.19199, -2085.68866) & (-2086.13023, -2085.69088) \\
huckel & (-2086.19826, -2085.65421) & (-2086.19716, -2085.66309) & (-2086.27389, -2086.27389) \\
minao & (-2086.23305, -2086.23305) & (-2086.29658, -2086.29658) & (-2086.28813, -2086.28813) \\
sap & (-2086.25095, -2086.09438) & (-2086.16261, -2085.70808) & (-2086.18511, -2086.16996) \\
vsap & (-2086.26777, -2086.26777) & (-2086.17176, -2086.17176) & (-2086.3019, -2086.3019) \\
\hline
\end{tabular}

\end{subtable}
\hfill
\begin{subtable}{0.65\textwidth}
\centering
\caption{Convergence PySCF}
\label{tab:Cr2_ROHF_conv}
\begin{tabular}{llll}
\hline
 & cc-pVDZ & cc-pVTZ & cc-pVQZ \\
\hline
1e & (False, False) & (True, False) & (True, False) \\
atom & (True, False) & (True, False) & (True, False) \\
huckel & (False, False) & (True, False) & (True, False) \\
minao & (False, False) & (False, False) & (False, False) \\
sap & (True, False) & (True, False) & (False, False) \\
vsap & (True, False) & (False, False) & (False, False) \\
\hline
\end{tabular}
\end{subtable}
\caption{Top: Total energies (Ha) obtained for \ce{Cr2}, reported as ordered pairs $(E_{\ket{x}}, E_{\mathrm{PySCF}})$, where the left value corresponds to the alternative Fock-state solution and the right value corresponds to the PySCF ROHF solution. Bottom: Convergence information for the PySCF ROHF calculations of \ce{Cr2}. Each entry is reported as a Boolean pair $(\text{SCF converged}, \text{internally stable})$, indicating whether the SCF procedure converged and whether the resulting solution passed the internal stability analysis.}
\end{table}

\begin{table}[ht]
\centering
\begin{subtable}{0.65\textwidth}
\centering
\caption{Energies (Ha)}
\label{tab:Cr2_SO_ROHF}
\begin{tabular}{llll}
\hline
 & cc-pVDZ & cc-pVTZ & cc-pVQZ \\
\hline
1e & (-2086.21154, -2086.03559) & (-2086.30069, -2086.30069) & (-2086.30649, -2086.30649) \\
atom & (-2086.30056, -2086.30056) & (-2086.19176, -2085.68866) & (-2086.18847, -2085.69088) \\
huckel & (-2086.21947, -2085.58477) & (-2086.21533, -2085.59226) & (-2086.2117, -2085.59449) \\
minao & (-2086.22183, -2085.5031) & (-2086.21705, -2085.51069) & (-2086.21247, -2085.51239) \\
sap & (-2086.29771, -2086.29771) & (-2086.21705, -2085.51069) & (-2086.35204, -2086.35204) \\
vsap & (-2086.22183, -2085.5031) & (-2086.21705, -2085.51069) & (-2086.21247, -2085.51239) \\
\hline
\end{tabular}
\end{subtable}
\hfill
\begin{subtable}{0.65\textwidth}
\centering
\caption{Convergence PySCF}
\label{tab:Cr2_SO_ROHF_conv}
\begin{tabular}{llll}
\hline
 & cc-pVDZ & cc-pVTZ & cc-pVQZ \\
\hline
1e & (True, False) & (False, False) & (False, False) \\
atom & (True, False) & (True, False) & (True, False) \\
huckel & (True, False) & (True, False) & (True, False) \\
minao & (True, False) & (True, False) & (True, False) \\
sap & (False, False) & (True, False) & (True, True) \\
vsap & (True, False) & (True, False) & (True, False) \\
\hline
\end{tabular}
\end{subtable}
\caption{Top: Total energies (Ha) obtained for \ce{Cr2}, reported as ordered pairs $(E_{\ket{x}}, E_{\mathrm{PySCF}})$, where the left value corresponds to the alternative Fock-state solution and the right value corresponds to the PySCF SO-ROHF solution. Bottom: Convergence information for the PySCF SO-ROHF calculations of \ce{Cr2}. Each entry is reported as a Boolean pair $(\text{SCF converged}, \text{internally stable})$, indicating whether the SCF procedure converged and whether the resulting solution passed the internal stability analysis.}
\end{table}

\newpage
\section{Molecular Symmetries} \label{sec:mol_syms}

In the following subsections, we show the total spin squared $S^{2}$ and molecular point group symmetries can be enforced within the GAS-SCF algorithm.

\subsection{Total spin squared}
The $S^{2}$ operator can be written as  \cite[Section 2.5.1]{szabo2012modern}\cite{yen2019exact}:

\begin{equation} \label{eq:S2_op}
    \begin{aligned}
S^{2} = S_{z} + S_{z}^{2} + S_{-} S_{+}.
    \end{aligned}
\end{equation}

where:

\begin{subequations}\label{eq:S2_terms}
\begin{equation} 
    \begin{aligned}
S_{+} = \sum_{j} \alpha_{j \alpha}^{\dagger} \alpha_{j \beta}
    \end{aligned}
\end{equation}
%%%%
\begin{equation}
S_{-} = (S_{+})^{\dagger}
\end{equation}
%%%%
\begin{equation}
S_{z} = \frac{1}{2}\sum_{j} \alpha_{j \alpha}^{\dagger} \alpha_{j \alpha} - \alpha_{j \beta}^{\dagger} \alpha_{j \beta}
\end{equation}
\end{subequations}
%%%%
\begin{equation}
\begin{aligned}
S_{-}S_{+} &= \sum_{j} \sum_{k} \alpha_{j \beta}^{\dagger} \alpha_{j \alpha} \alpha_{k \alpha}^{\dagger} \alpha_{k \beta} \\
% &= -\sum_{j} \sum_{k} \alpha_{j \beta}^{\dagger} \alpha_{j \alpha} \alpha_{k \beta}\alpha_{k \alpha}^{\dagger}  \\
% &= \sum_{j} \sum_{k} \alpha_{j \beta}^{\dagger} \alpha_{k \beta} \alpha_{j \alpha} \alpha_{k \alpha}^{\dagger}  \\
&= \sum_{j} \sum_{k} \alpha_{j \beta}^{\dagger} \alpha_{k \beta} (\delta_{kj} -\alpha_{k \alpha}^{\dagger} \alpha_{j \alpha})   \\
\end{aligned}
\end{equation}
here each sum over $j$ goes over the spatial orbital indices and $\alpha / \beta$ denotes the spin-orbital. However, as discussed in \cite{ralli2025bridging} we only need to consider single Fock states in single reference SCF methods and so only the diagonal components of $S^{2}$ are relevant.  We can write the diagonal form as:

\begin{equation} \label{eq:S2_op3}
    \begin{aligned}
S_{\text{diag}}^{2} = S_{z} + S_{z}^{2} + (S_{-}S_{+})_{\text{diag}},
    \end{aligned}
\end{equation}
where:
\begin{equation} \label{eq:Spm_diag}
    \begin{aligned}
(S_{-}S_{+})_{\text{diag}} = \sum_{j}  \alpha_{j \beta}^{\dagger} \alpha_{j \beta} -  \alpha_{j \beta}^{\dagger} \alpha_{j \beta}\alpha_{j \alpha}^{\dagger} \alpha_{j \alpha}.
    \end{aligned}
\end{equation}

To target the $\langle S^{2}\rangle=0$ sector, the symmetry $S_{\text{diag}}^{2}$ (equation \ref{eq:S2_op3}) can be incorporated into the QUBO cost function with an appropriate energy shift $\lambda$: $H\mapsto H + \lambda S_{\text{diag}}^{2}$ as discussed in \cite{ralli2025bridging}. This is because under the Jordan-Wigner transform $S_{\text{diag}}^{2}$ is a quadratic boolean function and so the problem will remain a QUBO. The shift value $\lambda$ can always be chosen to be equal to the 1-norm of the QUBO cost function, which will shift the energy of all states with $\langle S^{2}\rangle>0$ above the $\langle S^{2}\rangle=0$ states. We note that such a modification of the QUBO cost function can actually help reduce the number of marked states (as the energy of states with $\langle S^{2}\rangle >0$ are all raised by the shift $\lambda$). 

When targeting arbitrary values of $S^{2}(>0)$, the problem must be modified as: $H \mapsto H + \lambda (S_{\text{diag}}^{2}- \omega)^{2}$, where $\omega$ denotes the target value. Squaring $S_{\text{diag}}^{2}$ introduces higher order  terms into the cost function as previously discussed in \cite{ralli2025bridging} in this case four-local terms. This transforms the quadratic boolean function into a quartic boolean function. These higher-order contributions can be implemented directly in quantum circuit within GAS-SCF; however, will require more controls acting on the system register (in this case gates controlled by at most four qubits) when accumulating values into the $f(\vec{x})$ (cost-function storage) qubit register.

Alternatively, the total spin  $S^{2}$ can be enforced in the same manner as the particle number was enforced in this work (without introducing penalty terms in to the cost function). In this approach, an additional ancilla register is introduced to store the  $S_{\text{diag}}^{2}$ value associated with each Fock state in the system register. Fortunately, under the Jordan-Wigner transform $S_{\text{diag}}^{2}$ is a quadratic boolean function and so the circuit to store $S_{\text{diag}}^{2}$ value in the ancilla has the same construction as how the QUBO function was stored. After this has been done, the same projection circuit used to select a particular particle-number sector (Figure~\ref{fig:QUBO-proj}) can then be employed to select states within the correct symmetry sector. The control logic must be modified to flag the desired spin sector instead, so here $\gamma$ would be the desired value of $S^{2}$ in binary.

We note that the spin-squared operator has eigenvalues given by $S^{2} = S(S+1)$. Consequently, singlet, doublet, triplet, and quintet states correspond to $(S=0, \; S^2=0)$, $(S=\frac{1}{2}, \; S^2=0.75)$, $(S=1, \; S^2=2)$, $(S=\frac{3}{2}, \; S^2=3.75)$, and $(S=2, \; S^2=6)$, respectively. More generally, $S^{2}$ can take fractional values, which is inconvenient in the present GAS-SCF framework, where values are required to be integers (as they are stored in binary). To address this, we observe that:

\begin{equation}
    S^{2}= S(S+1) = \frac{v}{2}\bigg(\frac{v}{2}  +1 \bigg) = \frac{v(v+2)}{4} \qquad \forall v \in \{0,1,2,\dots \} \equiv \mathbb{N}_{0},
\end{equation}
where $v = 2S$ and $\mathbb{N}_{0}$ is the set of Natural Numbers including zero. Multiplying the operator by four therefore yields:

\begin{equation}
    \Omega = 4S^{2} = v(v+2) \qquad \forall v \in \mathbb{N}_{0},
\end{equation}
whose eigenvalues are always integers. This transformation preserves the ordering of the spin sectors while replacing fractional $S^{2}$ eigenvalues with integer labels that can be used directly within the GAS-SCF framework (where the $\Omega$ values will be stored on the ancilla register rather than $S^{2}$).

A single Fock state (Slater determinant) is always an eigenfunction of $S_{z}$, with $M_{S} = \frac{1}{2}(N_{\alpha} - N_{\beta})$. However, it is a configuration state function (CSF) only if it is also an eigenfunction of $S^{2}$, i.e. only if it has a
definite total spin $S$. All single Fock states have a definite $M_{S}$. however it only has a definite value of $S$ if it is an eigenfunction of $S^{2}$. We will see this next.

For a determinant built from orthonormal, spin-restricted orbitals, the expectation value of $S^{2}$ is (equation \ref{eq:S2_op3}):
\begin{equation}
    \langle S^{2} \rangle = M_{S}\left(M_{S}+1\right) + N_{\beta} - N_{\text{pairs}},
    \label{eq:S2_det}
\end{equation}
where $N_{\text{pairs}}$ is the number of doubly occupied orbitals, so that $N_{\beta} - N_{\text{pairs}}$ is the number of open-shell $\beta$ electrons.

For example, $\ket{(1,1),(1,0),(1,0),(1,0)}$ has three singly occupied $\alpha$ orbitals, so $M_{S} = \frac{3}{2}$ and
\begin{equation}
    \langle S^{2} \rangle = \frac{3}{2}\left(1+\frac{3}{2}\right) + 1 - 1 = \frac{15}{4} = 3.75 .
\end{equation}
This is exactly $S(S+1)$ for $S = \frac{3}{2}$, so this determinant is a valid
single-determinant CSF (a quartet). By contrast, $\ket{(1,1),(1,0),(1,0),(0,1)}$ has singly occupied orbitals of both $\alpha$ and $\beta$ spin. It has a well-defined $M_{S} = \frac{3-2}{2} = \frac{1}{2}$, but
\begin{equation}
    \langle S^{2} \rangle = \frac{1}{2}\left(1+\frac{1}{2}\right) + 2 - 1 = \frac{7}{4} = 1.75 ,
\end{equation}
which is not of the form $S(S+1)$ for any allowed $S$ (here $S^{2}=1.75 \implies S = 0.914$, which means there is spin-contamination). Hence this determinant is not an eigenfunction of $S^{2}$ and has no definite total spin.

To see why, recall that for a given total spin $S$ the allowed projections are $M_{S} = -S, -S+1, \dots, S-1, S$. Three singly occupied orbitals correspond to coupling
three spin-$\frac{1}{2}$ particles,
\begin{equation}
    \frac{1}{2} \otimes \frac{1}{2} \otimes \frac{1}{2}
    = \frac{3}{2} \oplus \frac{1}{2} \oplus \frac{1}{2},
\end{equation}
so the only possible total spins are $S \in \{\frac{1}{2}, \frac{3}{2}\}$; for instance, $S = \frac{5}{2}$ is impossible because there are not enough unpaired electrons. The corresponding projections are
\begin{subequations}
    \begin{align}
        S = \frac{1}{2} &\implies M_{S} \in \left\{-\frac{1}{2}, \frac{1}{2}\right\}, \\
        S = \frac{3}{2} &\implies M_{S} \in \left\{-\frac{3}{2}, -\frac{1}{2}, \frac{1}{2}, \frac{3}{2}\right\}.
    \end{align}
\end{subequations}
Importantly, $M_{S} = \frac{1}{2}$ is accessible from both $S = \frac{1}{2}$ and $S = \frac{3}{2}$. The $M_{S} = \frac{1}{2}$ space of the three open shells is spanned by the three determinants $\ket{\uparrow\uparrow\downarrow}$, $\ket{\uparrow\downarrow\uparrow}$ and $\ket{\downarrow\uparrow\uparrow}$, and contains
one quartet and two linearly independent doublets. The quartet component is
\begin{equation}
    \ket{S=\frac{3}{2}, M_{S}=\frac{1}{2}}
    = \frac{1}{\sqrt{3}} \left( \ket{\uparrow\uparrow\downarrow}
    + \ket{\uparrow\downarrow\uparrow} + \ket{\downarrow\uparrow\uparrow} \right).
\end{equation}
Our determinant (open shells $\ket{\uparrow\uparrow\downarrow}$) can therefore be written in terms of CSFs as
\begin{equation}
    \ket{\psi} = c_{d} \ket{S=\frac{1}{2}, M_{S}=\frac{1}{2}}
               + c_{q} \ket{S=\frac{3}{2}, M_{S}=\frac{1}{2}},
\end{equation}
where $\ket{S=\frac{1}{2}, M_{S}=\frac{1}{2}}$ denotes the particular normalized combination of the two doublet CSFs that appears in $\ket{\psi}$. Normalization requires $|c_{d}|^{2} + |c_{q}|^{2} = 1$, and the expectation value of $S^{2}$ gives
\begin{equation}
    \langle S^{2} \rangle = \frac{7}{4}
    = |c_{d}|^{2} \left(\frac{3}{4}\right) + |c_{q}|^{2} \left(\frac{15}{4}\right).
\end{equation}
Substituting $|c_{d}|^{2} = 1 - |c_{q}|^{2}$:
\begin{equation}
\begin{aligned}
    \frac{7}{4} &= \left(1 - |c_{q}|^{2}\right)\frac{3}{4} + |c_{q}|^{2}\,\frac{15}{4} \\
    \implies\quad 7 &= 3\left(1 - |c_{q}|^{2}\right) + 15\,|c_{q}|^{2} = 3 + 12\,|c_{q}|^{2} \\
    \implies\quad |c_{q}|^{2} &= \frac{4}{12} = \frac{1}{3}.
\end{aligned}
\end{equation}
This agrees with the direct projection $|c_{q}|^{2} = |\; \langle [S=\frac{3}{2}, M_{S}=\frac{1}{2}] \Big|  \uparrow\uparrow\downarrow \rangle \; |^{2} = \bigg(\frac{1}{\sqrt{3}} \bigg)^{2} = \frac{1}{3}$.
The $\ket{(1,1),(1,0),(1,0),(0,1)}$ Fock state is therefore $33.3\%$ quartet and $66.7\%$ doublet.

This leads to a useful rule. Let $n_{\text{open}}$ be the number of singly occupied orbitals, so that the maximum possible spin is $S_{\max} = n_{\text{open}}/2$. If $|M_{S}| = S_{\max}$, the determinant is a pure spin state with $S = S_{\max}$, because no other $S$ can produce that projection. Consequently, a determinant of orthonormal, spin-restricted orbitals (doubly occupied orbitals plus singly occupied orbitals) is automatically an $S^{2}$ eigenfunction when all singly occupied orbitals carry the same spin. This is the high-spin case treated by conventional restricted (RHF/ROHF) SCF solvers. (Unrestricted determinants are generally spin-contaminated even in the high-spin case, because their $\alpha$ and $\beta$ spatial orbitals differ.) Conversely, if $|M_{S}| < S_{\max}$, several values of $S$ contribute, and a single determinant is not necessarily an eigenfunction of $S^{2}$. A linear combination of determinants, a configuration state function (CSF), is then
required to obtain a state of definite spin.

\subsection{Point Group Symmetries}

In this subsection, we demonstrate how to enforce the largest Abelian point-group symmetry available for a molecular system within the GAS-SCF framework. Exploiting point-group symmetries is a standard strategy in computational chemistry, as it enables the partitioning of the Hilbert space into symmetry sectors and can significantly reduce computational cost. While many molecular point groups are non-Abelian, calculations are often performed within their largest Abelian subgroup. For example, $D_{4h}$ is a non-Abelian point group; however, by performing a symmetry descent, one may instead work within its Abelian subgroup $D_{2h}$. We focus on these symmetries here and use the $D_{2h}$ point group as a representative example. The $D_{2h}$ group is generated by the symmetry operations $C_{2}$, $C_{2}'$, and inversion $inv$, from which all other group elements can be constructed.

After a Hartree-Fock calculation is performed using some classical chemistry software, such as PySCF, it is possible to obtain the symmetry labels for each molecular orbital (if the symmetry functionality is turned on during the computation). This means the molecular orbitals will be constructed as a symmetry adapted linear combinations (SALC) of atomic orbitals. Let us consider an arbitrary $D_{2h}$ example with the following labels obtained from a SCF calculation:

\begin{equation}\label{eq:SALC_example}
    \ket{q_{0}^{a_{g}} q_{1}^{a_{g}} q_{2}^{b_{2u}} q_{3}^{b_{2u}} q_{4}^{b_{3u}} q_{5}^{b_{3u}}q_{6}^{a_{g}} q_{7}^{a_{g}}},
\end{equation}
where the superscript gives the symmetry labels of each molecular spin orbital labeled $q_{i}$, where $i$ denotes the spin orbital index. These symbols denote how the orbital behaves when operated on by a symmetry element of the point group.
 
Given such information, we can construct the $\mathbb{Z}_{2}$ symmetries that stem from the group generators. First we need to look up the action of each symmetry generator operation on each molecular orbital. Table \ref{table:symm} summarises the information needed here ($D_{2h}$  point group generators).

\begin{table}[h!]
\begin{tabular}{cccc}
\hline
Group Generator ($\mathtt{G}$) & $A_{g}$ & $B_{2u}$ & $B_{3u}$ \\ \hline
$C_{2}$              & 1  & -1  & -1  \\
$C_{2}'$           & 1  & -1  & 1   \\
$inv$               & 1  & -1  & 1   \\ \hline
\end{tabular}
 \caption{Character table elements of $D_{2h}$ point group generators. \label{table:symm}}
\end{table}

We denote a lookup of Table \ref{table:symm} as follows $\text{CHAR\_TAB}[\mathtt{G}][\mathtt{M}]$, e.g. $\text{CHAR\_TAB}[C_{2}][B_{2u}] = -1$. We can use this to define the symmetries as follows:

\begin{equation}
    F_{\mathtt{G}} = \prod_{i=0}^{N-1}  \begin{cases}
\big(I_{i} - 2 \alpha_{i}^{\dagger} \alpha_{i}  \big), & \text{ if } \text{CHAR\_TAB}[\mathtt{G}][\mathtt{M}_{i}]=-1,\\
I, & \text{ otherwise }
\end{cases} 
\end{equation}
where $\mathtt{M}_{i}$ is the orbital symmetry label for spin orbital $i$. For the example labels given in equation \ref{eq:SALC_example} we obtain the following:

\begin{equation}
    F_{C_{2}} = I_{0} \cdot I_{1} \cdot \big(I_{2} - 2 \alpha_{2}^{\dagger} \alpha_{2} \big) \cdot \big(I_{3} - 2 \alpha_{3}^{\dagger} \alpha_{3} \big)  \cdot \big( I_{4} - 2 \alpha_{4}^{\dagger} \alpha_{4} \big) \cdot \big(I_{5} - 2 \alpha_{5}^{\dagger} \alpha_{5}\big) \cdot I_{6}  \cdot I_{7},
\end{equation}
\begin{equation}
    F_{C_{2}'} = I_{0} \cdot I_{1} \cdot \big(I_{2} - 2 \alpha_{2}^{\dagger} \alpha_{2} \big) \cdot \big(I_{3} - 2 \alpha_{3}^{\dagger} \alpha_{3} \big)  \cdot I_{4} \cdot I_{5}  \cdot I_{6}  \cdot I_{7},
\end{equation}

\begin{equation}
    F_{inv} = I_{0} \cdot I_{1} \cdot \big(I_{2} - 2 \alpha_{2}^{\dagger} \alpha_{2} \big) \cdot \big(I_{3} - 2 \alpha_{3}^{\dagger} \alpha_{3} \big)  \cdot \big( I_{4} - 2 \alpha_{4}^{\dagger} \alpha_{4} \big) \cdot \big(I_{5} - 2 \alpha_{5}^{\dagger} \alpha_{5}\big) \cdot I_{6}  \cdot I_{7} .
\end{equation}
We re-iterate, that the orbitals have been labelled as: $[a_{g}, a_{g}, b_{2u}, b_{2u}, b_{3u}, b_{3u}, a_{g}, a_{g}]$ and this ordering is used when determining each $\mathtt{M}_{i}$ in $F_{\mathtt{G}}$. We observe  that $F_{C_{2}} = F_{inv}$. Under the Jordan-Wigner tranform, these symmetries would be written as $F_{C_{2}} \mapsto IIZZZZII$ and $F_{C_{2}'} \mapsto IIZZIIII$. Since these are symmetry operators, they commute with the molecular Hamiltonian. Consequently, fixing their eigenvalues restricts the space to a particular symmetry sector, thereby reducing the accessible Hilbert space to states possessing the desired symmetry.

The symmetry of an electronic state is determined by taking the direct product of the irreducible representations of all electrons participating in the state, where each electron is assigned the irreducible representation of the molecular orbital it occupies. Continuing our example (equation \ref{eq:SALC_example}), for a state $\ket{11110000}$ we obtain $a_{g} \cdot a_{g} \cdot b_{2u}\cdot b_{2u} = A_{g} \cdot A_{g} \cdot B_{2u}\cdot B_{2u} = A_{g}$. Fixing the expectation values of the symmetry generator according to this state will enforce all states to have $A_{g}$ symmetry. Whereas, for the state $\ket{11100100}$ we find: $a_{g} \cdot a_{g} \cdot b_{2u}\cdot b_{3u} =A_{g} \cdot A_{g} \cdot B_{2u}\cdot B_{3u} = A_{g} \cdot B_{1g}= B_{1g}$, which is in a different symmetry sector. Note we use the usual convention that lower case letters are used for orbital irreducible representations (irreps) and upper case letters are used for the irrep of a given molecular state.

We note that these symmetries can be enforced in two distinct ways in GAS-SCF. (1)The first is directly analogous to marking Fock state with the correct particle-number as done in this work. In this approach, an additional ancillary register is introduced to label Fock states according to their point-group symmetry, enabling the selective marking of configurations belonging to the desired irreducible representation. Here controls from the ancilla register would select the $+1$ or $-1$ eigenvalue of each $F_{\mathtt{G}}$ in the marking operation. (2)Alternatively, instead of using another ancilla register a symmetry penalty may be incorporated into the cost function, as discussed in Section 3.1 of ~\cite{ralli2025bridging}. However, this approach generally introduces higher-order terms into the QUBO, thereby converting it into a higher-order unconstrained binary optimization (HUBO) problem. Implementing these higher-order contributions within the cost-function evaluation would require multi-controlled operations acting on the system register (more than two control qubits) when accumulating values into the $f(\vec{x})$ (cost-function storage) qubit register.

% \clearpage

% \begin{figure*}[t]
% \centering
% \includegraphics[width=1.75in,height=1.75in]{figures/toc_image_full_size.png}
% % \caption{Table of Contents (TOC) graphic. }
% %     \label{fig:TOC}
% \end{figure*}

\end{document}